\documentclass[onecolumn,numberedappendix,appendixfloats]{emulateapj}
\usepackage{epsfig}
\usepackage{multirow}
\usepackage{graphicx}
\usepackage{epstopdf}
\usepackage{amsmath, amsthm, amssymb}
\usepackage{natbib}
\usepackage{rotating}
\tightenlines
\newcommand \lsim{\mathrel{\rlap{\lower4pt\hbox{\hskip1pt$\sim$}}
    \raise1pt\hbox{$<$}}}
\newcommand \gsim{\mathrel{\rlap{\lower4pt\hbox{\hskip1pt$\sim$}}
    \raise1pt\hbox{$>$}}}

\newcommand     \kms    {\,{\rm km\:s^{-1}}}
\newcommand     \kpc    {\,{\rm kpc}}

\newcommand{\sm}{{M_\odot}}

\newcommand{\beq}{\begin{equation}}
\newcommand{\eeq}{\end{equation}}
\newcommand{\beqa}{\begin{eqnarray}}
\newcommand{\eeqa}{\end{eqnarray}}

\newcommand{\htwo}      {H$_2$}

\newcommand{\thco}	{^{13}{\rm CO}}

\newcommand{\cotwo}     {^{12}{\rm CO}}

\newcommand{\tex}	{T_{\rm ex}}

\def\HI{\ion{H}{1}}
\def\HII{\ion{H}{2}}

\newlength{\figwidth}
\countdef\decade=200
\advance\decade by \year
\countdef\hours=201
\advance\hours by \time
\divide\hours by 60
\countdef\mins=202
\advance\mins by \hours
\multiply\mins by 60
\multiply\hours by 100
\countdef\miltime=203
\advance\miltime by \hours
\advance\miltime by \time
\advance\miltime by -\mins

\begin{document}

\title{The Giant Molecular Cloud Environments of Infrared Dark Clouds}

%\centerline{DRAFT: \today}
\author{Audra K. Hernandez}
\affil{Department of Astronomy, University of Wisconsin, 475 North Charter Street, Madison,  WI 53706, USA;\\ hernande@astro.wisc.edu}
\author{Jonathan C. Tan}
\affil{Departments of Astronomy \& Physics, University of Florida, Gainesville, FL 32611, USA;\\ jt@astro.ufl.edu}

\begin{abstract}
We study Giant Molecular Cloud (GMC) environments surrounding 10 Infrared Dark Clouds (IRDCs), using $\thco(1-0)$ emission from the Galactic Ring Survey. We measure physical properties of these IRDCs/GMCs on a range of scales extending to radii, $R$, of 30~pc. By comparing different methods for defining cloud boundaries and for deriving mass surface densities and velocity dispersions, we settle on a preferred ``CE,$\tau$,G'' method of ``Connected Extraction'' in position-velocity space plus Gaussian fitting to opacity-corrected line profiles for velocity dispersion and mass estimation. We examine how cloud definition affects measurements of the magnitude and direction of line-of-sight velocity gradients and velocity dispersions, including associated dependencies on size scale. CE,$\tau$,G-defined GMCs show velocity dispersion versus size relations $\sigma\propto{s}^{1/2}$, which are consistent with the large-scale gradients being caused by turbulence. However, IRDCs have velocity dispersions that are moderately enhanced above those predicted by this scaling relation.  We examine the dynamical state of the clouds finding mean virial parameters $\bar{\alpha}_{\rm{vir}}\simeq 1.0$ for GMCs and 1.6 for IRDCs, broadly consistent with models of magnetized virialized pressure-confined polytropic clouds, but potentially indicating that IRDCs have more disturbed kinematics. CE,$\tau$,G-defined clouds exhibit a tight correlation of $\sigma/R^{1/2}\propto\Sigma^n$, with $n\simeq0.7$ for GMCs and 1.3 for IRDCs (c.f., a value of  0.5 expected for a population of virialized clouds). We conclude that while GMCs show evidence for virialization over a range of scales, IRDCs may be moderately super virial. Alternatively, IRDCs could be virialized but have systematically different $\thco$ gas phase abundances, i.e., due to freeze-out, affecting mass estimations.

\end{abstract}

\section{Introduction}

Galactic star formation resides mostly within Giant Molecular Clouds
(GMCs), conventionally defined to have masses $\geq 10^4\:\sm$ and
observed to extend up to several~$\times 10^6\:\sm$
\citep[e.g.,][]{Blitz1993,Williams2000,McKeeOstriker2007}. With
typical mass surface densities of $\sim 100\:M_\odot\:{\rm pc}^{-2}$,
GMCs have mean radial sizes of $\sim 6 - 100$~pc, assuming simple
circular symmetry. However, GMCs are highly irregular and hierarchical
structures. Their dense clumps can spawn stellar clusters and
associations, creating the bulk of the Galactic field star
population. The efficiency and rate of star formation from these
clumps is relatively low, i.e., a few percent per local free-fall time
\citep[][]{ZuckermanEvans1974,KrumholzTan2007}. This appears to be
mostly because much of the GMC material is stable with respect to
gravitational collapse, especially material below a threshold
$A_V\sim10$~mag \citep[e.g.,][]{Lada2010}. Higher total star formation
efficiencies, $\sim10-50\%$, appear to be possible in the star-forming
clumps that form at least moderately bound clusters \citep{Lada2003}.

Different theoretical models of the processes that create star-forming
clumps within GMCs or prevent overdensities in the bulk of the cloud
are actively debated.  These processes include the regulation of star
formation and stabilization of gas by magnetic fields
\citep{McKee1989,Mouschovias2001} or turbulence
\citep{KrumholzMcKee2005,PadoanNordlund2011}, and/or the initiation of
star formation by discrete triggering events, such as converging
atomic flows \citep[e.g.,][]{Heitsch2006}, cloud collisions
\citep[e.g.,][]{Tan2000} or stellar feedback
\citep[e.g.,][]{Samal2014}.

Infrared dark clouds (IRDCs) are likely to be examples of early stage
star-forming clumps
\citep[e.g.,][]{Perault1996,egan1998,Carey2000,Rathborne2006,ButlerTan2009,PerettoFuller2009,Battersby2010}. Thus
their study may help us understand the processes that initiate star
formation in GMCs. There have been many investigations of the internal
properties of IRDCs, including their temperatures
\citep[e.g.,][]{Pillai2006,Peretto2010,Ragan2011,Chira2013}, mass
surface density structure \citep[e.g.,
][]{ButlerTan2009,PerettoFuller2009,Ragan2011,ButlerTan2012,KainulainenTan2013,Butler2014},
kinematics \citep[e.g.,][]{Henshaw2013,JimenezSerra2014} and dynamics
\citep[e.g.,][]{HernandezTan2011,Hernandez2012}, CO depletion
\citep{Fontani2006,Hernandez2011}, chemistry \citep{Sanhueza2013}.
See \citet{Tan2014} for a review.

However, there have been fewer studies connecting IRDCs to their
larger-scale environments, such as the morphology, kinematics and
dynamics of their parent clouds. Theories involving production of
dense gas in shocks have been supported by 
detection of large-scale SiO emission along IRDCs \citep{JimenezSerra2010,NguyenLuong2013}.
However, these studies focus only on a few individual clouds and are
still confined to a few-parsec scales in and around the
filamentary molecular clouds.

Here we study the $^{13}$CO(1-0)-emitting gas in and around 10 well-studied
IRDCs, utilizing data from the BU-FCRAO Galactic Ring Survey
\citep[GRS;][]{Jackson2006}. We consider a range of scales out to
30~pc projected radius, expected to encompass the potential GMC
environment of the IRDC.  While our study connects to scales typical
of other large-sample GMC studies
\citep[e.g.,][]{Heyer2009,RomanDuval2009,RomanDuval2010}, by focusing
on just 10 regions
we are able to investigate their kinematic properties in much greater
detail. Other studies done on such a range of scales, from clump to
GMCs, have been performed on nearby GMCs, such as Orion A 
\citep[e.g.,][]{Shimajiri2011}, Taurus \citep[e.g.,][]{Goldsmith2008},
and Perseus \citep{Ridge2006,Foster2009,Kirk2010}. However, these local 
GMCs do not seem to give rise to the more extreme range of star-forming 
clumps that is found in IRDCs.

The main focus of this paper is to ``bridge the gap'' between IRDC and
GMC studies. The questions we aim to address include: Are IRDCs
typically found within GMCs?  Are IRDCs found within specific
locations with respect to GMCs?  Are IRDCs and their surrounding GMCs
virialized, and does their degree of virialization vary as a function
of cloud physical scale?

The IRDC/GMC sample is presented in \S2. Methods for defining
cloud boundaries and estimating masses and kinematic properties are
described in \S3. Results are presented in \S4, including derived physical properties of
the clouds (\S\ref{S:physprop}), the location of IRDCs within their
respective GMCs (\S\ref{S:loc}), and the kinematic and dynamical
analysis of the IRDC and GMCs (\S\ref{S:kinematics}).  We
conclude in \S5.
    
\section{The IRDC Sample and Molecular Line Data}
We utilize the $\thco (J=1\rightarrow0)$ data of the GRS survey
\citep[][]{Jackson2006}, which covers $18^{\circ}\le l \le
56^{\circ}$, $b\pm1^{\circ}$ and $-5 <v_{\rm LSR}<135~\kms$ (for $l
\le 40^{\circ}$) and $-5<v_{\rm LSR}<85~\kms$ (for $ l \ge
40^{\circ})$. The GRS had a spatial resolution of 46\arcsec, with
22\arcsec\ sampling, and a spectral resolution of $0.212~\kms$.  The
typical rms noise is $\sigma \rm (T_A^*)=0.13$~K (or $\sigma \rm
(T_{B,\nu})\sim 0.26$~K with a main beam efficiency of 0.48).

Our selected IRDCs are the 10 clouds from \citet[][hereafter
  BT09]{ButlerTan2009} \citep[see
  also][]{ButlerTan2012,KainulainenTan2013}.  This sample, a subset of
that of \citet{Rathborne2006}, was chosen while considering the
8$\mu$m (IRAC band 4) images from the \textit{Spitzer} Galactic Legacy
Mid-Plane Survey Extraordinaire
\citep[GLIMPSE;][]{Benjamin2003}. These IRDCs were selected for being
relatively nearby, massive, dark (i.e., relatively high contrast
against the surrounding diffuse emission), and surrounded by
relatively smooth diffuse emission within the 8$\mu$m GLIMPSE images.
Characteristic sizes and boundaries for each IRDC were taken from
\citet{Simon2006a}, where ellipses were fitted based on extinction in
MSX images.  Although these fitted ellipses are not necessarily
accurate of IRDC shapes, they provide a convenient measure of the
approximate cloud structure. The catalog coordinates and sizes for the
10 IRDCs (including ``single component'' sub-classifications, labelled
``-s'', see below) are listed in Table \ref{tab1}.

\begin{table}[h]
\centering
\scriptsize
\caption{The IRDC Sample}
\begin{tabular}{lccccccccccccc}
\hline
IRDC & $l_{\rm IRDC}$\footnotemark[1]& $b_{\rm IRDC}$\footnotemark[1] &  $r_{\rm maj}\footnotemark[1]$ & $r_{\rm min}\footnotemark[1]$  &  PA$_e$\footnotemark[1] & $l_{0}$ & $b_{0}$ & $v_{0}$  & $l_{0,\tau}$ & $b_{0,\tau}$ & $v_{0,\tau}$  & $d$\footnotemark[2]  & $\tex$\footnotemark[3]  \\
          & ($^{\circ}$)& ($^{\circ}$)   &   (\arcmin)   & (\arcmin)     & ($^{\circ}$)  & ($^{\circ}$)& ($^{\circ}$)   &$\kms$ & ($^{\circ}$)& ($^{\circ}$)   &$\kms$ & $(\kpc)$  & (K)\\
\hline
\hline
A & 18.822 & -0.285 & 14.1 & 3.90 & 74.0 & 18.850 & -0.276 & 64.3 & 18.862 & -0.276 & 64.5 & 4.80 & 8.90 \\
B & 19.271 & 0.074 & 8.40 & 1.80 & 88.0 & 19.287 & 0.075 & 26.5 & 19.287 & 0.075 & 26.7 & 2.40 & 7.11 \\
C & 28.373 & 0.076 & 12.0 & 9.30 & 78.0 & 28.370 & 0.062 & 79.0 & 28.364 & 0.062 & 79.2 & 5.00 & 7.86 \\
D & 28.531 & -0.251 & 20.4 & 5.10 & 60.0 & 28.567 & -0.233 & 78.6 & 28.567 & -0.239 & 80.7 & 5.70 & 7.00 \\
D-s &	   &        &      &      &      & 28.555 & -0.251 & 87.9 & 28.549  &  -0.257  &    87.7 &      &      \\
E & 28.677 & 0.132 & 14.6 & 4.10 & 103 & 28.672 & 0.130 & 80.7 & 28.684 & 0.130 & 81.3 & 5.10 & 7.00 \\
F & 34.437 & 0.245 & 5.30 & 2.00 & 79.0 & 34.434 & 0.240 & 57.3 & 34.428 & 0.240 & 57.5 & 3.70 & 6.48 \\
F-s &	   &       &      &      &      & 34.434 & 0.240 & 58.0 & 34.428  &   0.240  &    58.0 &      &      \\
G & 34.771 & -0.557 & 6.60 & 2.00 & 95.0 & 34.764 & -0.559 & 43.9 & 34.764 & -0.559 & 44.1 & 2.90 & 8.69 \\
G-s &	   &        &      &      &      & 34.770 & -0.559 & 44.4 & 34.764  &  -0.559  &    44.4 &      &      \\
H & 35.395 & -0.336 & 20.6 & 6.40 & 59.0 & 35.422 & -0.319 & 47.5 & 35.416 & -0.319 & 48.0 & 2.90 & 6.61 \\
H-s &	   &        &      &      &      & 35.391 & -0.337 & 43.9 & 35.391  &  -0.337  &    44.4 &      &      \\
I & 38.952 & -0.475 & 7.50 & 3.00 & 64.0 & 38.942 & -0.473 & 43.5 & 38.942 & -0.473 & 43.2 & 2.70 & 7.24 \\
J & 53.116 & 0.054 & 1.60 & 1.30 & 50.0 & 53.114 & 0.056 & 21.4 & 53.120 & 0.056 & 21.6 & 1.80 & 7.05 \\
\hline
\label{tab1}
\end{tabular}
\footnotetext[1]{IRDC coordinates and elliptical sizes adopted from \citet{Simon2006a}}
\footnotetext[2]{IRDC kinematic distances adopted from \citet{Rathborne2006}}
\footnotetext[3]{The $\thco$ excitation temperatures adopted from \cite{RomanDuval2010}}
\end{table}

\begin{figure}[ht!]
\begin{center}$
\begin{array}{c}
\includegraphics[width=7.2in, angle=0, trim=0 0.8in 0 0.5in]{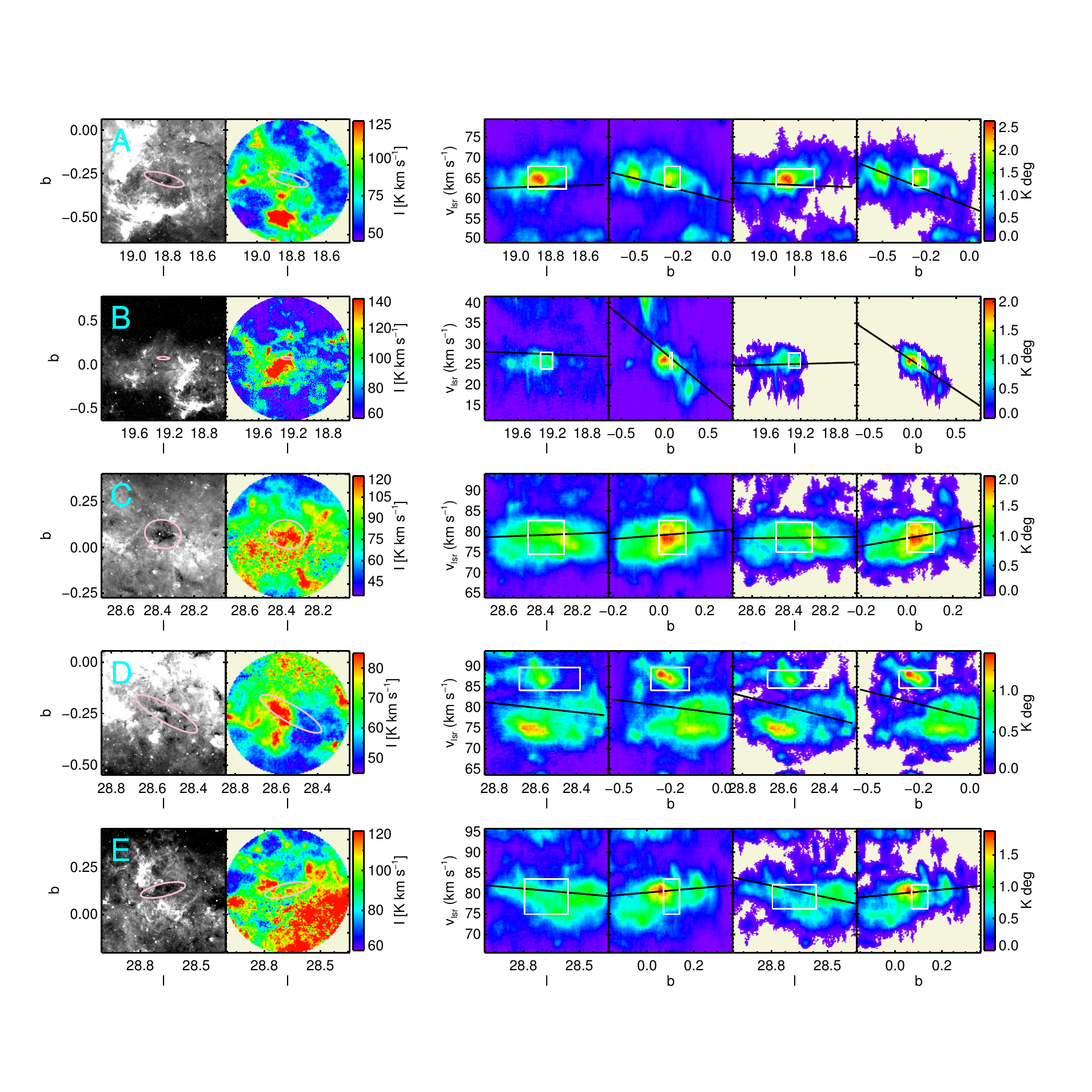}  
\end{array}$
\end{center}
\caption{
The IRDC/GMC sample: each row presents a single cloud (A through E)
showing data out to a radius of 30 pc. \textit{Left to right:} GLIMPSE
$8\:{\rm \mu m}$ image with the IRDC boundary marked with a pink
ellipse (see Figs. \ref{mapA} and \ref{mapB}-\ref{mapE} for intensity scales);
$\thco$(1-0) GRS integrated intensity map (over the range $v_0\pm
15\:\kms$) in units of $\rm K ~\kms$; Full projections of the cloud in 
position-velocity space: simple-extraction (SE) $\thco$(1-0) 
$v_{\rm lsr}$ vs $l$ map, SE $\thco$(1-0) $v_{\rm lsr}$
vs $b$ map; connected extraction (CE) $\thco$(1-0) $v_{\rm lsr}$ vs
$l$ map; CE $\thco$(1-0) $v_{\rm lsr}$ vs $b$ map.  
The IRDC is marked with a white box of width equal to the extent of
the elliptical boundary and height equal to the FWHM of the IRDC
co-added spectra. The solid black line shows the mass-weighted linear velocity gradient ($dv_0/ds$) across each cloud at $R=30$ pc.}
\label{PV1}
\end{figure}
 
Kinematic distances were adopted from \citet{Rathborne2006} and
\citet{Simon2006b}, where the IRDC central velocity was matched
morphologically between the mid-infrared (MIR) extinction estimated
from MSX and the $\thco$ emission from the GRS.  These distances were
estimated assuming the rotation curve of \citet{Clemens1985}. We
assume uncertainties of 20\%, given the size of streaming motions of
$\sim 10 - 20\:\kms$ of clouds with trigonometric parallax
measurements \citep[e.g.,][]{Brunthaler2009,Reid2014}.

\section{Methods: Derivations of Molecular Cloud Properties}

\subsection{Definition of Cloud Boundaries}

\begin{figure}[t!]
\begin{center}$
\begin{array}{c}
\includegraphics[width=7.2in, angle=0, trim=0 0.8in 0 0.5in]{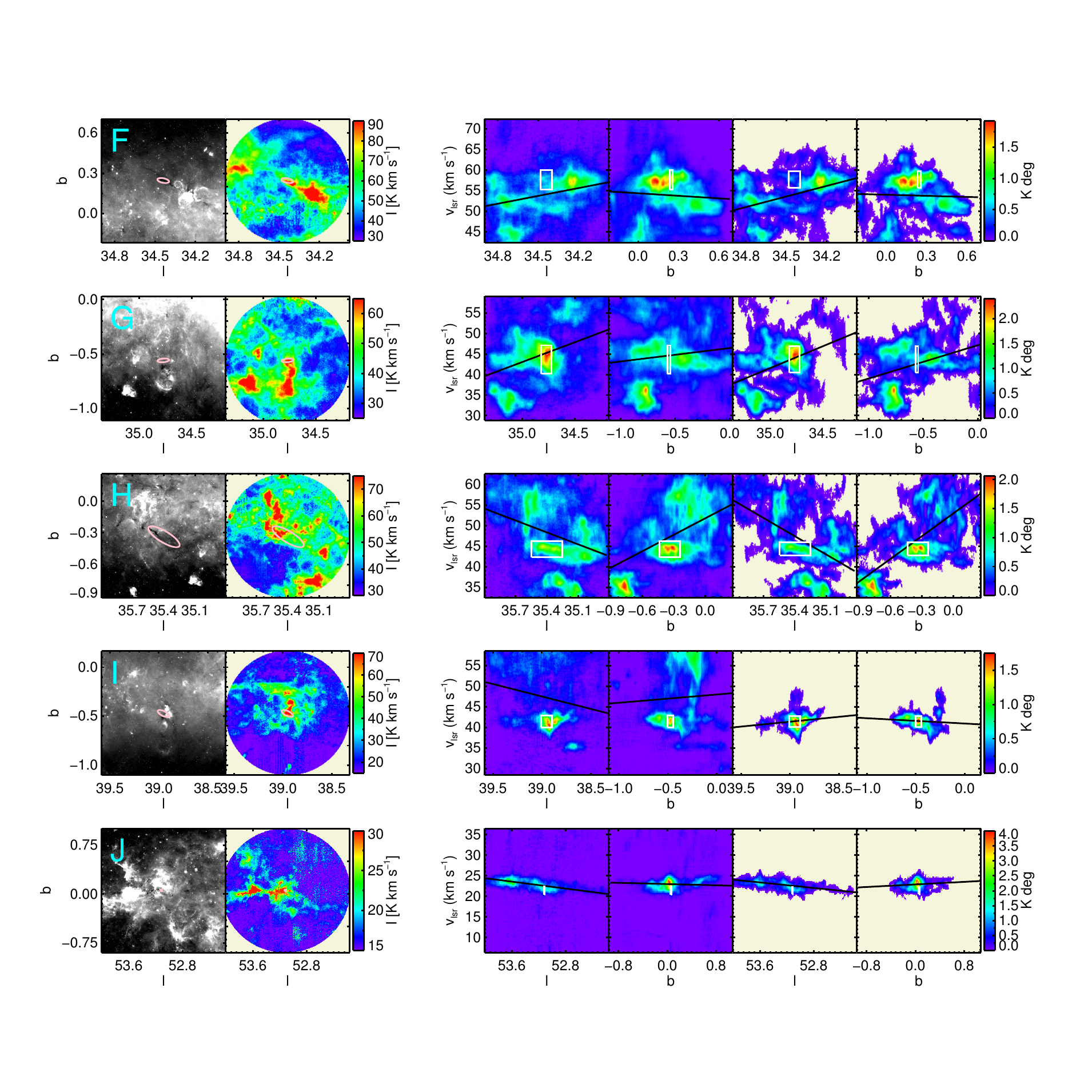}  
\end{array}$
\end{center}
\caption{The same as Figure \ref{PV1}, but for Clouds F through J.}
\label{PV2}
\end{figure}

We first determined the cloud center-of-mass in position-velocity
space. To do this, we co-added the $\thco$ spectra within an
elliptical boundary defined by twice the $r_{\rm maj}$ and $r_{\rm min}$
values, using the full GRS velocity range. Then, a velocity interval
of $30~\kms$ was considered, centered on the peak of the emission
profile. The cloud's center-of-mass was evaluated for both optically
thin and opacity corrected gas, $(l_0,b_0,v_0$) and
$(l_{0,\tau},b_{0,\tau},v_{0,\tau}$), respectively.  We then use this
as the reference point around which to search for GMC-scale gas, out
to a projected radius of 30~pc and over a new velocity range of
$v_0\pm15\:\kms$. This range is wide enough to sample the
  velocities of GMC gas, even those that are potentially undergoing
  collisional interactions at velocities set by Galactic shear at the
  tidal radii of the clouds \citep{Tan2000, TaskerTan2009}.
The GLIMPSE $8\mu$m and GRS $\thco$ integrated intensity maps centered
on these locations are presented in Figures \ref{PV1} and \ref{PV2}.

We examined $\thco$ emission within the $v_0\pm15$~km/s velocity range
and within apertures of varying radii from the cloud center-of-mass
coordinate. To explore how boundary definitions affect estimated
physical properties, we used two different methods to select $\thco$
line emission associated with the cloud. First, ``Simple Extraction
(SE)'' selected all the $\thco$ emission within radii $R=5, 10, 20$,
and 30~pc and $v_0\pm15$~km/s. Second, ``Connected Extraction (CE)''
defined a cloud as a connected structure in $l$-$b$-$v$-space with all
cloud voxels required to be above a given threshold intensity: we
defined a given voxel as ``molecular cloud gas'' if its $\thco$(1-0)
line intensity $T_{B,\nu}\geq 1.35$~K (i.e., the GRS $5\sigma_{\rm rms}$
noise level). The CE search was also limited to within a 30~pc radius
and $\pm15\:\kms$ of the IRDC center-of-mass. Position-velocity maps 
for each cloud, defined by both SE and CE, are shown in Figures 
\ref{PV1} and \ref{PV2}.

Clouds defined by CE do not have simple radial sizes. Therefore, we
estimated three circular boundaries, centered on the extracted cloud's
center-of-mass: (1) mass-weighted radius, $R_M$, defined as mean
projected radial distance of cloud mass from the center of mass; (2)
areal radius, $R_A$, defined by the total projected area $A = \pi R_A^2
= N_p A_p$, where $N_p$ is the total number of pixels subtended by the
cloud and $A_p$ is the area of one image pixel; (3) half-mass radius, 
$R_{1/2}$, defined as the radius from cloud center that contains half 
of the total mass.

For the IRDCs, we also selected ``cloud'' material via SE and CE using
the spatial coordinates and elliptical boundaries from
\citet{Simon2006a}, along with the $v_0\pm15~\kms$ velocity intervals.

\subsection{Column Densities and Masses from $\thco$ emission}\label{S:13COmass}

We estimated the $\thco$ column density of each molecular cloud voxel,
$dN_{\rm 13CO}$ from their $J=1\rightarrow 0$ line emission assuming a
partition function with a thermal distribution described by an
excitation temperature $T_{\rm ex}$ via:
\beq \frac{dN_{\rm 13CO}}{dv} = \frac{8\pi Q_{\rm
    rot}}{A \lambda_0^3} \frac{g_l}{g_u} \frac{\tau_{\nu} }{1-{\rm
    exp} (-h\nu /[kT_{\rm ex}])},
\label{eq:dN}
\eeq 
where $Q_{\rm rot}$ is the partition function, $A = 6.294\times
10^{-8}{\rm s^{-1}}$ is the Einstein coefficient, $\lambda_0=
0.27204$~cm, $g_l = 1$ and $g_u = 3$ are the statistical weights of
the lower and upper levels and $\tau_\nu$ is the optical depth of the
line at frequency $\nu$, i.e., at velocity $v$. 
Each GRS voxel has a velocity width of ${\rm
  d}v=0.212~\kms$.  For linear molecules, the partition function is
$Q_{\rm rot}=\sum_{J=0}^{\infty} (2J+1) {\rm exp}(-E_J/kT_{\rm ex})$
with $E_J =J(J+1)h B$ where $J$ is the rotational quantum number and
$B=5.5101\times 10^{10}\:{\rm s}^{-1}$ is the rotational constant. For
$\thco$(1-0) we have $E_J/k = 5.289$~K.

Many studies of the physical properties of GMCs and IRDCs have
accounted for line optical depth
when estimating their physical properties \citep[e.g.,][]{Heyer2009,
  RomanDuval2010, HernandezTan2011, Hernandez2011, Hernandez2012}.  In
\citet{HernandezTan2011}, we showed that optical depth correction
factors can increase the $\thco$ column density by a factor of $\sim2$
in the densest, sub-parsec scale clumps of IRDCs.  However, for the
more diffuse GMCs the optical depth correction factors are expected to
be smaller. To gauge the importance of this effect, we carry out
column density estimates for both the optically thin assumption and
accounting for opacity corrections.

The optical depth is evaluated via
\beq 
T_{B,\nu}=\frac{h\nu}{k}[f(T_{\rm ex})-f(T_{\rm bg})][1-e^{-\tau_{\nu}}],
\label{eq:tau}
\eeq where $T_{B,\nu}$ is the brightness temperature at frequency
$\nu$, $f(T)\equiv [{\rm exp}(h\nu/[kT])-1]^{-1}$, and $T_{\rm
  bg}=2.725$~K is the background temperature. $T_{B,\nu}$ is derived
from the antenna temperature, $T_A$, via $T_A\equiv \eta f_{\rm
  clump}T_{B,\nu}$, where $\eta$ is the main beam efficiency
($\eta=0.48$ for the GRS) and $f_{\rm clump}$ is the beam dilution
factor of the $\thco$ emitting gas, which we assume to be unity due to
the large scale extent of GMCs. Smaller scale structures are
undoubtedly present, e.g., as revealed in the BT09 MIR extinction
maps, but to gauge the effects of these on the CO emission requires
higher resolution molecular line maps of the clouds. For $\tau \ll 1$,
Equation (\ref{eq:tau}) can be simplified to express $\tau_{\nu}$ by:
$\tau_{\nu}=(T_{B,\nu}k / [h\nu] )[f(T_{\rm ex}) - f(T_{\rm
    bg})]^{-1}$.  With this simplification, for an observed voxel
$T_{B,\nu}$ and an assumed $T_{\rm ex}$, the optically thin $\thco$
column density per voxel is given by combining Equations (\ref{eq:dN})
and (\ref{eq:tau}): 
\beq \frac{dN_{\rm 13CO}}{dv} = 1.251 \times 10^{14} \frac{Q_{\rm
    rot}}{f(T_{\rm ex}) - f(T_{\rm bg})} \frac{T_{B,\nu}}{1-{\rm
    exp}(-{h\nu/[kT_{\rm ex}])}} ~\frac{\rm{cm}^{-2}}{~\kms}.
\label{eq:dN2}
\eeq

Early studies of IRDCs estimated typical gas kinetic temperatures
$T_{\rm gas}\sim20$~K \citep{Carey1998, Carey2000, Pillai2006}. IRDC F
was estimated to have a temperature of 19~K based on $\rm NH_3$~(1,1)
and (2,2) VLA observations \citep{Devine2009}. However, as discussed
below, CO excitation temperatures appear to be significantly lower.

In our previous study of IRDC H, we used IRAM~30m observations of $\rm
C^{18}O$(2-1) and (1-0) emission from around the IRDC filament to
estimate a mean $T_{\rm ex}\sim7$~K \citep[][hereafter
  H11]{Hernandez2011}. Here, for a uniform analysis of the 10 IRDCs,
we now use $\cotwo$ $T_{\rm ex}$ estimates from \citet[][hereafter
  RD10]{RomanDuval2010}.  In their study of 580 molecular clouds,
brightness temperatures from $\cotwo$(1-0) emission line data
(Univ. of Massachusetts-Stony Brook (UMSB) Galactic Plane Survey),
were used to derive proxy $\thco$ excitation temperatures, assuming
$\cotwo$ emission was optically thick and that 
$\thco$ and $\cotwo$ excitation temperatures are
equal. Ultimately, RD10 cited a mean excitation temperature for all
their molecular clouds based on all cloud voxels above $4\sigma_{\rm rms}$. The
RD10 clouds were extracted from the GRS data using a modified version
of CLUMPFIND \citep{Williams1994}, which allowed for varying
thresholds (i.e., contour increment and minimum brightness, see
\citet{Rathborne2009} for details). 

Eight of our IRDCs overlapped with at least one of their molecular
clouds, and in these cases we adopted $T_{\rm ex}$ from the RD10 value
from the overlapping cloud(s).  For the remaining two IRDCs (D and E)
we set $T_{\rm ex}=7$~K, similar to the mean values of 7.2~K of H11
and 6.32~K of RD10. Our adopted values of $T_{\rm ex}$ are
  listed in Table \ref{tab1}.  For our column density estimates that
  assume optically thin $^{13}$CO emission, these $T_{\rm ex}$ values
  are assumed to be constant throughout the cloud.  These
temperatures are slightly lower (by a few~K) than those used in
previous studies (e.g., \citealt{Simon2001,Simon2006b} who assumed a
fixed value of 10~K).  However, the results from
\citet{Heyer2009} indicate that CO gas throughout GMCs is mostly sub-thermally excited.
Note that, in this optically thin limit, varying $T_{\rm ex}$
from 5~K to 10~K would change the derived column density of the cloud
by only $\sim 20\%$.

For the opacity-corrected case, the use of a single mean excitation
temperature of relatively low value ($\sim 7$~K) can lead to
non-physical results in certain regions of the cloud.
Equation \ref{eq:tau} implies
$\tau_{\nu}=-\ln{(1-[(T_{B,\nu}k / [h\nu] )[f(T_{\rm ex}) - f(T_{\rm bg})]^{-1}])}$.
Thus, for a given observed $T_{B,\nu}$, $\tau_{\nu}$ will become
undefined if $T_{\rm ex} \le T_{\rm ex,crit}$ (i.e., when
$[(T_{B,\nu}k / [h\nu] )[f(T_{\rm ex}) - f(T_{\rm bg})]^{-1}] >
1$).  For example, a voxel with an observed brightness temperature of
$T_{B,\nu}=5$~K will have a numerically undefined opacity at $T_{\rm
  ex,crit}<8.2$~K.
H11 showed modest temperature variations were present in IRDC H, with a peak
temperature of $T_{\rm ex} \sim10$~K within the densest clumps and $T_{\rm ex}
\sim 7$~K in the more diffuse gas within the filament envelope (see
H11, Fig. 1).  Hence, we expect that the voxels containing the largest
brightness temperatures have excitation temperatures which are a few K
larger than the constant $T_{\rm ex}$ values adopted from RD10.

To estimate the opacity-corrected column density, we first apply our
adopted $T_{\rm ex}$ to estimate $\tau_{\nu}$ in each voxel. Then, for
each voxel with an undefined $\tau_{\nu}$, we specified a new
excitation temperature of $T_{\rm ex,crit}+1$~K, given the voxel's
observed $T_{B,\nu}$. This revised excitation temperature allows us to
estimate a real solution for $dN_{\rm 13CO}/dv$ throughout the
cloud. After considering a range of possible temperature offsets from
$T_{\rm ex,crit}$, we estimate that the uncertainty in $dN_{\rm
  13CO}/dv$ in an individual voxel by this method is at a level of
$\sim 20\%$.

Ideally, excitation temperatures would be estimated locally from
$^{12}$CO $J=1-0$ observations of the clouds, assuming that the
$^{12}$CO line is optically thick. However, there are currently no
other $^{12}$CO surveys that match both the resolution and spatial
extent of the GRS. The widely used Columbia-CfA $^{12}$CO survey
covers the whole Galactic plane, but with a low angular resolution of
8\arcmin\ \citep{Dame2001}. The RD10 temperatures measurements are
based on the UMSB survey, which has a 44\arcsec\ spatial resolution
with 3\arcmin\ sampling.

For the clouds defined by SE extraction and with $R=30$ pc, we find
that on average $0.7\%$ of the cloud voxels require higher
temperatures than those listed in Table \ref{tab1}. Cloud D has the
highest percentage, 1.5\%, and for these voxels
$T_{\rm ex,crit}$ peaks at 16.4 K with a mean $T_{\rm ex,crit}=8.3$~K.
For the clouds defined by CE extraction and $R_A$, we find that on
average $10\%$ of voxels require higher excitation temperatures. Here,
Cloud J has the highest percentage, 27\%, with a peak $T_{\rm ex,crit}$ of 
25.0~K and a mean $T_{\rm ex,crit}=9.2$~K.

The $\thco$-based mass per voxel, 
$dM$ was then calculated by
assuming a $n_{\rm 12CO}/n_{\rm 13CO}=54$ \citep{Milam2005} and
$n_{\rm 12CO}/n_{\rm H2}=2.0 \times 10^{-4}$ \citep{Lacy1994}.  Hence, the
assumed abundance of $\rm ^{13}CO$ to $\rm H_2$ is $3.70\times
10^{-6}$. The mass located at a given voxel is then given by 
\beq
dM=1.45\times10^{-4} \frac{dN_{\rm
    13CO}}{10^{13}~\rm{cm}^{-2}} \frac{dv}{0.212~\kms} \left(\frac{d}{\rm
  kpc}\right)^2 \frac{\Delta l}{22\arcsec} \frac{\Delta b}{22\arcsec}
~\rm{M_{\odot}},
\label{eqn:M}
\eeq
where $d$ is the cloud distance, $\Delta l$ and $\Delta b$ are the
angular sizes of the GRS pixels, and assuming a mass per H nucleus of
$\mu_{\rm H}=2.34\times 10^{-24}\:{\rm g}$.  The total $\thco$-derived
cloud mass, $M$, is simply the total mass of all cloud voxels within
radius $R$ and velocity range $v_0\pm15~\kms$. For clouds defined by
SE, any pixels with total integrated intensities below $5\sigma_{\rm
  rms}$ were omitted from further analysis.  

We estimate an uncertainty of $20\%$ in $T_{\rm ex}$ due to its
intrinsic variation, which in addition to uncertainties due to
intrinsic abundance variations, leads to $\sim30\%$ uncertainty in the mass
surface density ($\Sigma$). We thus estimate $\sim$50\% random errors in
$M$, after accounting for these uncertainties in 
$\Sigma$ and the cloud kinematic distance estimates (assumed to be
$\sim 20\%$). However, we also anticipate that there could be
global systematic uncertainties in $M$ (of the whole cloud sample) of up to a factor of 2, given the
uncertainties in overall absolute $\thco$ abundance.

\subsection{Cloud Kinematics}

\begin{figure}[ht!]
\begin{center}$
\begin{array}{c}
\includegraphics[width=7.61in, angle=0]{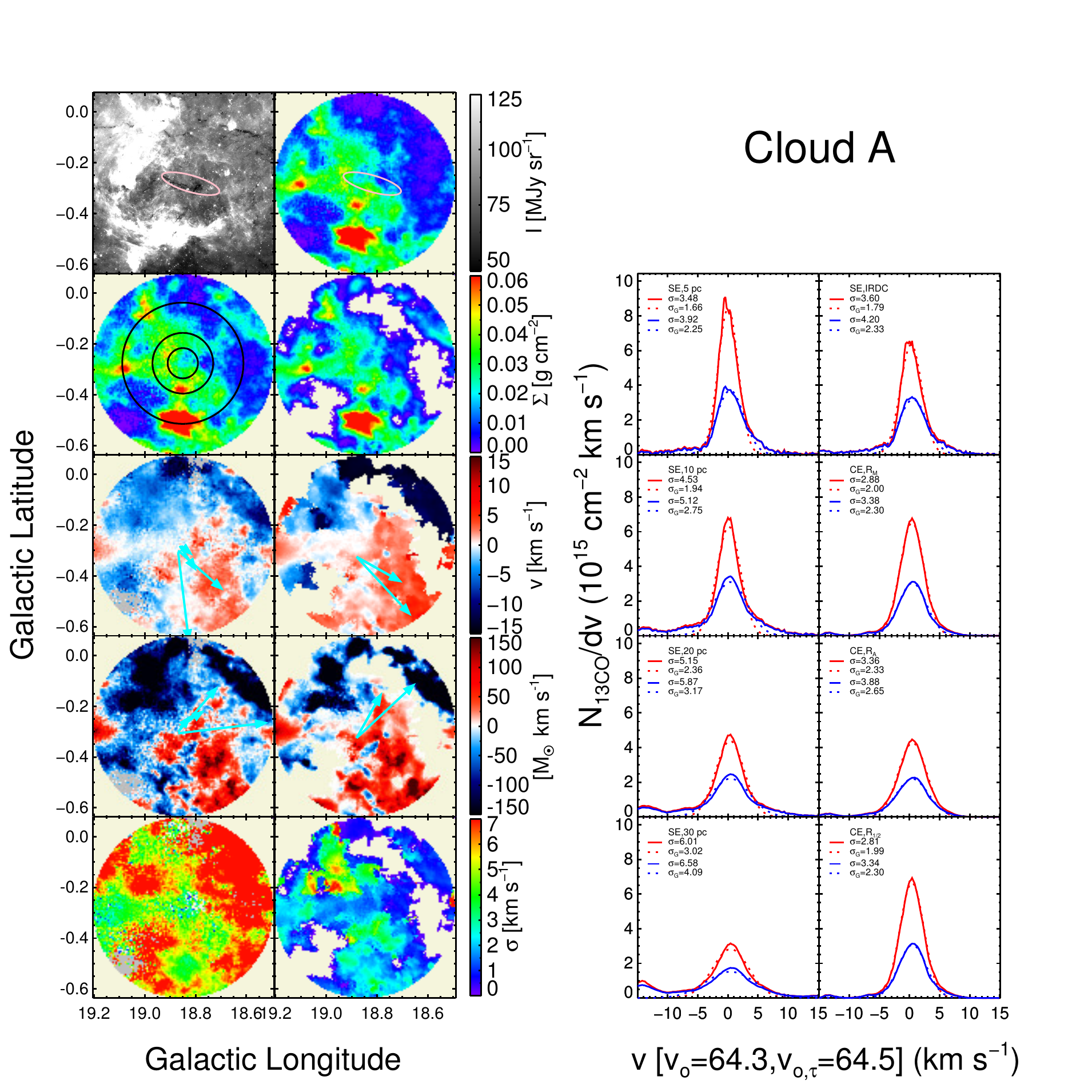} 
\end{array}$
\end{center}
\caption{
Cloud A: \textit{Left panel maps:} {\it Top row:} {\it Left:} GLIMPSE
8$\rm \mu$m image. {\it Right:} $\thco$ mass surface density
($\Sigma$) map over the velocity range of the IRDC emission profile
$v_{\rm lsr}=54.6-77.1~\kms$ (ellipse denotes IRDC boundary). {\it 2nd
  row:} {\it Left:} SE $\thco$-derived $\Sigma$ map (circles denote 5,
10, 20~pc boundaries). {\it Right:} CE $\thco$-derived $\Sigma$
map. {\it 3rd row:} {\it Left:} SE $\thco$ mean velocity (first
moment) map. {\it Right:} CE $\thco$ mean velocity (first moment)
map 
relative to $v_{0}$ defined from the IRDC. 
The arrows show the direction of total (mass-weighted) line-of-sight
velocity gradients, $dv_0/ds$. An arrow is shown for each extraction
radius (defining the length of the arrow) and is centered on either
the extracted cloud's center of mass (for SE cases) or the $R=30$ pc cloud's
center of mass (for CE cases).
{\it 4th row:} {\it Left:} SE $\thco$-derived (optically thin) linear
momentum map. {\it Right:} CE $\thco$-derived (optically thin) linear
momentum map. 
The arrows show the position angle, $\theta_v$, of the cloud's
rotation axis, i.e., orthogonal to the direction of the velocity
gradient, for each extracted cloud (with arrow length showing cloud
radius) {\it 5th row:} {\it Left:} SE $\thco$ velocity dispersion
(second moment; optically thin) map. {\it Right:} CE $\thco$ velocity
dispersion (second moment; optically thin) map. All maps in Rows 2-5
cover the velocity interval of $v_0\pm15~\kms$, with $v_0$ defined by
the emission of the IRDC.  Omitted pixels, those that fall below the
5$\sigma_{\rm rms}$ level, are shown in gray.
\textit{Right panel velocity distributions:} 
In each panel, the column density distributions with velocity are
derived assuming optically thin conditions (blue solid line) or with
opacity correction (red solid line) from the $\thco$(1-0) co-added
spectra. Gaussian fits are shown by the dotted blue and red lines.  In all panels,
$\sigma$ and $\sigma_G$ are noted in the upper left corner of the
plot. {\it Left column:} {\it Top to bottom:} SE out to $R=5, 10, 20,
30$~pc. {\it Right column:} {\it Top to bottom:} SE of IRDC ellipse,
CE out to $R_M$, $R_A$, $R_{1/2}$.
}
\label{mapA}
\end{figure}

We used co-added $\thco$ column density velocity distributions (e.g.,
Fig. \ref{mapA}: \textit{Right}) to determine the mean velocity, $v_0$
of each extracted cloud.
The velocity dispersion was estimated using two standard methods: 1)
the rms 1D velocity dispersion, $\sigma$;
2) the width of a fitted Gaussian profile, $\sigma_G$. We estimate an mean uncertainty in $\sigma$ 
of $10\%$. Additionally, we used these Gaussian fitted profiles to estimate a Gaussian profile
mass, $M_G$, of each cloud. To visualize how $v_0$ and $\sigma$ vary
throughout the cloud, we show mass-weighted first and second moment
maps of each GMC (e.g., Fig. \ref{mapA}).

Using the first moment maps, we derive the velocity gradients in each
spatial direction.  For example, the longitudinal velocity gradient,
$dv_0/d l$, was derived by first estimating the mean velocity at each
longitudinal position, via a mass-weighted sum along the perpendicular
($b$) direction, then finding the best (mass-weighted) linear fit to
these velocities. This method was repeated for $dv_0/db$. The
magnitude of the total linear velocity gradient across the cloud,
$dv_0/ds$, and its position angle direction, $\theta_v$, were then
calculated. Note, we choose to use mass-weighted velocity
  gradients to prevent the results being unduly affected by tenuous
  wisps of cloud material. Also, in the context of interpreting
  velocity gradients as due to rotation and thus measuring rotational
  energies of the cloud (below), this mass-weighted gradient is the
  appropriate one to use.

Many studies of the dynamics of molecular clouds have interpreted
total GMC linear velocity gradients as due to solid body rotation
\citep[e.g.,][]{Phillips1999,Rosolowsky2003,ImaraBlitz2011}.
However, the possibility remains that the identified single ``cloud''
actually consists of spatially independent structures.  For a cloud
undergoing solid body rotation, the line of sight velocity gradient in
the plane of the sky, $dv_{0}/ds$, is equal to the projected angular
velocity, i.e., $\Omega_0 = dv_{0}/ds$. The true angular velocity is
$\Omega = \Omega_0/{\rm sin} i$, where $i$ is the angle between the
rotation axis and the line of sight.

The position angle of the projected rotation axis of the cloud
contains information that may constrain theories of GMC formation and
evolution. For example, if a GMC forms rapidly from atomic gas in the
Galactic plane, then the GMC rotation is expected to be prograde with
respect to Galactic rotation \citep[e.g.,][]{TaskerTan2009}.
If strong gravitational encounters and collisions are frequent between
gravitationally bound GMCs (Tan 2000), then a more random set of
orientations of the positional angles of projected rotation axes are
expected, including both pro- and retrograde rotating clouds.  
For clouds observed in the Galactic plane,
$-90^{\circ}<\theta_v<+90^{\circ}$ represents retrograde rotation and
$-90^{\circ}<\theta_v<-180^{\circ}$ and
$90^{\circ}<\theta_v<180^{\circ}$ represent prograde rotation.

We then also estimated the projected moment of inertia, $I_0 = \sum
dM_{i} s_{i}^2$, of each cloud using the sky projected rotation
axis, defined by $\theta_v$ and the cloud center-of-mass coordinate,
where $s_i$ is the shortest distance to the rotation axis. This allows
estimation of the projected rotational energy of the cloud, $E_{\rm
  rot,0}=(1/2) I_0 \Omega_0^2$.

\begin{sidewaystable}[h]
%\emulate
\centering
%\rotate
%\begin{deluxetable*}{lccccc}
%\tablecolumns{5}
%\tablewidth{0pt}
\tiny
\caption{GMC Physical Parameters Estimates}
\begin{tabular}{lcccccccccccccc}
\hline
IRDC  &       &                     &                      &                          &                        & &  &                  &  &  &  & & & \\ 
Case &  $l_{\rm CM}$,$l_{\rm CM,\tau}$   &  $b_{\rm CM}$,$b_{\rm CM,\tau}$ &  $ v_{\rm CM}$,$v_{\rm CM,\tau}$ & $\Delta s$ & $\Delta v$ &R &$M$,$M_{\tau}$,$M_{G}$ & $\sigma$,$\sigma_{\tau}$,$\sigma_{G}$  &  $\frac{dv_0}{ds}$,$\frac{dv_{0,\tau}}{ds}$  & $\theta_v$,$\theta_{v,\tau}$ &  $I_0$,$I_{0,\tau}$   &  $E_{\rm rot,0}$,$E_{\rm rot,0,\tau}$ & $\beta,\beta_{\tau}$ & $\log(\alpha_{\rm vir})$,$\log(\alpha_{\rm vir,\tau})$, \\

                      &                       &                     &        &              &             &           &  $\times10^4 $ &                   &  $\times \rm 10^{-2}$ &                & $\times \rm 10^{3}$   & $\times 10^{46}$ & ($10^{-2}$)   & $\log(\alpha_{\rm vir,G})$ \\
    (pc)                  &  ( $^{\circ}$)   & ($^{\circ}$)  & ($\kms$)& (pc) & ($\kms$)&    (pc)  &  $(\sm)$           & ($\kms$)   &  (km/s/pc)       & ($\deg$) & ($M_{\odot}$~pc$^2$)  & (erg) & &  \\

 (1)   & (2)       & (3)                    &(4)                      & (5)                             & (6)                       & (7) & (8) &              (9)    & (10) & (11) & (12) &(13) & (14) &(15) \\       
\hline
\hline
A&       &                     &                      &                          &                        & &  &                  &  &  &  & & & \\ 
5&18.856,18.868&-0.282,-0.276&64.3,64.3&2.06,2.12&0.0,0.0&5.00&1.10,1.87,1.62&3.92,3.48,1.66&36.0,32.3&141,124&0.0814,0.131&1.37,4.01&7.62,3.39&0.912,0.575,-0.00704\\
10&18.862,18.875&-0.282,-0.282&63.9,64.1&2.58,2.57&-0.4,-0.2&10.0&4.83,7.20,5.73&5.12,4.53,1.94&15.2,16.1&124,123&1.36,1.97&13.3,29.6&2.34,1.72&0.802,0.521,-0.115\\
20&18.868,18.881&-0.301,-0.313&63.5,63.7&3.46,4.02&-0.9,-0.6&20.0&16.1,24.0,19.2&5.87,5.15,2.36&16.8,15.6&139,142&16.8,25.7&74.2,165&6.38,3.78&0.696,0.411,-0.170\\
30&18.868,18.881&-0.307,-0.319&62.6,63.1&3.72,4.37&-1.7,-1.3&30.0&31.0,44.1,34.9&6.58,6.01,3.02&12.7,12.1&96.8,96.3&68.2,96.7&183,370&5.97,3.80&0.688,0.457,-0.0389\\
$R_M$&18.881,18.893&-0.325,-0.337&63.3,63.5&5.10,6.60&-1.7,-1.3&17.4,17.0&9.63,16.7,15.7&3.38,2.88,2.00&17.5,14.0&151,153&7.63,12.9&30.5,94.2&7.67,2.69&0.380,-0.00984,-0.298\\
$R_A$&18.881,18.893&-0.325,-0.337&63.3,63.5&5.10,6.60&-1.7,-1.3&26.6,26.6&16.6,28.7,26.9&3.88,3.36,2.33&16.5,14.0&133,134&32.3,54.3&59.3,177&14.8,5.97&0.447,0.0862,-0.204\\
$R_{1/2}$&18.881,18.893&-0.325,-0.337&63.3,63.5&5.10,6.60&-1.7,-1.3&17.1,16.2&9.38,15.5,14.7&3.34,2.81,1.99&17.2,13.5&151,153&7.18,11.0&29.4,84.9&7.18,2.36&0.374,-0.0177,-0.291\\
\hline
B&       &                     &                      &                          &                        & &  &                  &  &  &  & & & \\ 
5&19.299,19.299&0.050,0.050&27.4,26.7&1.65,1.46&0.9,0.6&5.00&0.799,1.38,1.19&5.27,4.39,1.59&34.1,34.3&105,97.6&0.0347,0.0547&0.729,2.18&5.50,2.94&1.31,0.909,0.0934\\
10&19.323,19.330&0.032,0.032&26.9,26.5&2.94,2.94&0.4,0.4&10.0&2.30,3.68,2.88&5.88,5.09,1.87&67.5,64.0&114,112&0.428,0.603&3.01,7.75&64.3,31.7&1.24,0.912,0.150\\
20&19.342,19.354&0.025,0.025&27.4,26.9&3.72,3.93&0.9,0.9&20.0&5.35,7.64,5.21&6.67,6.18,2.21&58.2,57.4&89.3,89.1&2.72,3.62&8.16,16.6&112,71.2&1.29,1.07,0.340\\
30&19.317,19.323&0.013,0.019&27.4,27.1&3.30,3.10&0.9,1.1&30.0&7.70,10.7,7.26&7.14,6.79,2.80&42.3,41.7&87.3,86.4&6.91,9.34&11.3,21.9&109,73.6&1.36,1.18,0.577\\
$R_M$&19.447,19.434&0.062,0.050&24.8,25.0&7.23,6.78&-1.1,-0.9&7.79,7.34&1.26,2.35,2.27&2.20,2.13,1.78&36.3,36.5&109,112&0.195,0.327&1.17,4.30&21.9,10.1&0.541,0.219,0.0764\\
$R_A$&19.447,19.434&0.062,0.050&24.8,25.0&7.23,6.78&-1.1,-0.9&12.2,12.2&1.87,3.42,3.26&2.19,2.08,1.62&31.5,31.1&100.,101&0.353,0.556&1.64,5.47&21.2,9.77&0.559,0.256,0.0577\\
$R_{1/2}$&19.447,19.434&0.062,0.050&24.8,25.0&7.23,6.78&-1.1,-0.9&7.20,6.60&1.12,2.02,1.96&2.25,2.19,1.84&40.8,41.5&110,113&0.159,0.253&0.987,3.54&26.6,12.2&0.581,0.259,0.124\\
\hline
C&       &                     &                      &                          &                        & &  &                  &  &  &  & & & \\ 
5&28.370,28.364&0.062,0.062&78.4,78.4&0.54,0.76&-0.2,-0.2&5.00&1.75,3.23,3.10&4.18,3.49,2.76&11.6,11.5&-118,-114&0.110,0.191&3.51,11.9&0.423,0.212&0.763,0.341,0.154\\
10&28.370,28.364&0.055,0.055&78.6,78.6&1.07,1.20&0.0,0.0&10.0&5.77,9.31,8.82&4.67,4.09,3.08&6.40,6.77&-74.0,-86.7&1.31,1.86&19.0,49.5&0.282,0.171&0.643,0.320,0.0965\\
20&28.370,28.358&0.049,0.049&79.0,79.2&1.61,1.94&0.4,0.6&20.0&19.7,28.0,25.6&5.50,5.16,3.82&6.61,7.05&-121,-119&18.7,25.2&111,224&0.735,0.556&0.553,0.344,0.122\\
30&28.370,28.364&0.031,0.025&79.0,79.0&3.22,3.79&0.4,0.4&30.0&36.7,50.2,45.9&5.77,5.50,4.15&4.54,4.75&-116,-117&68.0,88.2&256,479&0.544,0.414&0.501,0.324,0.117\\
$R_M$&28.364,28.358&0.018,0.012&78.4,78.4&4.33,4.95&0.2,0.0&17.1,16.6&11.4,17.2,17.2&3.45,3.27,3.07&4.50,4.22&-75.6,-66.2&6.60,9.71&43.5,102&0.306,0.169&0.318,0.0792,0.0259\\
$R_A$&28.364,28.358&0.018,0.012&78.4,78.4&4.33,4.95&0.2,0.0&29.0,29.0&24.4,36.9,35.3&4.21,4.21,3.46&7.72,7.12&-91.4,-90.5&34.7,51.6&117,268&1.75,0.970&0.389,0.209,0.0588\\
$R_{1/2}$&28.364,28.358&0.018,0.012&78.4,78.4&4.33,4.95&0.2,0.0&17.8,17.2&12.1,18.2,18.2&3.48,3.27,3.09&5.09,4.43&-82.9,-75.5&7.38,10.6&47.3,110&0.402,0.189&0.314,0.0725,0.0235\\
\hline
D&       &                     &                      &                          &                        & &  &                  &  &  &  & & & \\ 
5&28.573,28.567&-0.233,-0.239&80.7,82.0&0.61,0.86&0.2,0.4&5.00&1.95,3.23,2.09&7.59,7.03,1.83&20.2,19.9&-130,-126&0.113,0.176&4.32,11.9&1.06,0.583&1.24,0.949,-0.0280\\
10&28.579,28.573&-0.233,-0.239&79.6,80.9&1.22,0.61&-0.9,-0.6&10.0&5.96,9.23,4.66&7.61,7.31,1.88&18.4,20.1&118,112&1.42,2.15&20.3,48.6&2.35,1.78&1.05,0.828,-0.0534\\
20&28.586,28.592&-0.227,-0.233&79.4,80.3&1.93,1.84&-1.1,-1.3&20.0&18.5,27.6,27.9&7.45,7.47,11.7&17.9,19.3&62.4,59.4&19.6,28.8&97.6,218&6.37,4.92&0.845,0.673,1.06\\
30&28.573,28.567&-0.196,-0.202&79.6,80.3&3.72,3.12&-0.9,-1.3&30.0&36.1,54.8,55.1&7.07,7.11,9.24&8.59,10.4&50.0,47.3&81.5,123&249,571&2.41,2.34&0.684,0.509,0.733\\
$R_M$&28.567,28.567&-0.177,-0.184&79.6,80.5&5.05,4.45&-1.7,-1.9&17.6,18.0&10.5,18.3,10.7&6.57,6.74,2.46&19.9,23.7&67.6,63.1&9.31,16.8&35.8,106&10.3,8.86&0.926,0.716,0.0741\\
$R_A$&28.567,28.567&-0.177,-0.184&79.6,80.5&5.05,4.45&-1.7,-1.9&28.2,28.2&22.1,37.5,37.1&6.23,6.53,7.21&17.4,20.9&46.3,44.1&51.4,87.3&99.1,285&15.7,13.3&0.761,0.571,0.662\\
$R_{1/2}$&28.567,28.567&-0.177,-0.184&79.6,80.5&5.05,4.45&-1.7,-1.9&18.1,18.5&11.0,19.2,11.3&6.57,6.76,2.51&19.9,23.9&67.4,62.3&10.3,18.9&38.2,114&10.7,9.35&0.917,0.709,0.0795\\
\hline
D-s&       &                     &                      &                          &                        & &  &                  &  &  &  & & & \\ 
5&28.567,28.567&-0.233,-0.239&87.1,86.9&0.61,0.86&-0.2,-0.2&5.00&0.991,1.94,1.96&1.97,1.66,1.66&4.22,5.23&172,-177&0.0874,0.355&1.12,4.31&0.138,0.224&0.358,-0.0818,-0.0853\\
10&28.573,28.573&-0.239,-0.245&87.5,87.3&0.61,1.22&0.2,0.2&10.0&2.44,4.45,4.41&2.06,1.78,1.72&11.2,10.8&124,121&0.547,0.968&3.41,11.3&2.02,0.992&0.305,-0.0813,-0.109\\
20&28.604,28.604&-0.257,-0.270&88.1,87.9&3.92,4.78&0.9,0.9&20.0&6.14,10.0,10.2&2.39,2.12,2.31&14.1,14.2&64.8,65.0&5.13,7.54&10.8,28.7&9.39,5.27&0.334,0.0174,0.0848\\
30&28.616,28.610&-0.245,-0.257&88.3,87.9&4.46,4.41&1.1,0.9&30.0&10.4,15.9,16.2&2.57,2.27,2.52&11.2,10.5&46.5,50.2&20.4,25.8&20.6,47.9&12.4,5.90&0.345,0.0551,0.137\\
$R_M$&28.635,28.629&-0.264,-0.270&87.9,87.7&6.29,6.12&1.1,0.9&13.5,13.4&3.58,7.14,7.08&1.96,1.90,1.88&22.3,22.1&74.4,78.3&1.28,2.56&5.43,21.7&11.6,5.72&0.224,-0.102,-0.107\\
$R_A$&28.635,28.629&-0.264,-0.270&87.9,87.7&6.29,6.12&1.1,0.9&20.5,20.5&4.91,9.35,9.32&2.09,2.03,2.06&21.5,20.6&51.6,60.8&3.15,5.32&6.71,24.4&21.7,9.20&0.328,0.0204,0.0361\\
$R_{1/2}$&28.635,28.629&-0.264,-0.270&87.9,87.7&6.29,6.12&1.1,0.9&11.7,11.1&3.00,5.76,5.73&1.86,1.75,1.72&21.3,21.4&69.5,75.4&0.890,1.64&4.40,17.1&9.11,4.37&0.194,-0.164,-0.176\\
\hline
E&       &                     &                      &                          &                        & &  &                  &  &  &  & & & \\ 
5&28.672,28.684&0.130,0.130&79.4,79.4&0.55,0.54&-0.4,-0.6&5.00&0.907,1.27,0.966&6.67,6.16,2.71&31.3,11.8&144,128&0.0571,0.0742&0.939,1.84&5.92,0.559&1.46,1.24,0.646\\
10&28.672,28.690&0.118,0.118&80.7,81.1&1.23,1.55&0.9,1.1&10.0&3.52,5.03,4.20&6.45,6.28,3.66&18.3,23.5&74.8,79.6&0.747,1.01&7.08,14.5&3.52,3.87&1.14,0.960,0.568\\
20&28.653,28.666&0.105,0.099&80.3,80.5&3.10,2.95&0.4,0.4&20.0&15.0,21.1,18.6&6.52,6.40,4.32&5.69,6.81&-41.2,-28.0&15.6,22.7&63.8,127&0.787,0.822&0.820,0.655,0.369\\
30&28.641,28.647&0.081,0.075&80.3,80.5&5.47,5.64&0.4,0.4&30.0&32.4,46.4,40.4&6.93,6.87,4.86&6.34,7.14&-44.8,-41.1&68.8,101&199,409&1.38,1.25&0.714,0.551,0.310\\
$R_M$&28.616,28.622&0.056,0.050&79.8,80.3&8.21,8.21&0.0,0.4&17.7,17.6&9.01,14.8,13.7&5.47,5.76,3.86&14.3,15.9&-35.2,-35.6&7.29,12.1&26.2,71.3&5.66,4.26&0.834,0.662,0.349\\
$R_A$&28.616,28.622&0.056,0.050&79.8,80.3&8.21,8.21&0.0,0.4&27.5,27.5&20.2,34.2,31.4&5.44,5.62,4.02&10.5,11.3&-8.71,-3.01&40.1,71.4&84.7,242&5.16,3.77&0.671,0.472,0.216\\
$R_{1/2}$&28.616,28.622&0.056,0.050&79.8,80.3&8.21,8.21&0.0,0.4&18.1,17.9&9.41,15.3,14.1&5.47,5.76,3.86&14.2,15.6&-37.2,-36.3&7.93,12.8&27.9,74.2&5.67,4.18&0.826,0.655,0.344\\
\hline
F&       &                     &                      &                          &                        & &  &                  &  &  &  & & & \\ 
5&34.428,34.422&0.240,0.240&56.9,56.9&0.89,1.25&-0.6,-0.6&5.00&0.897,1.51,1.30&4.09,3.55,2.09&20.2,25.8&75.7,64.8&0.0373,0.0535&0.919,2.60&1.65,1.36&1.03,0.688,0.289\\
10&34.422,34.409&0.234,0.228&56.0,56.3&1.43,2.31&-1.5,-1.3&10.0&2.87,4.32,3.68&4.99,4.50,2.93&25.6,24.1&109,103&0.642,0.861&4.70,10.7&8.92,4.66&1.00,0.737,0.434\\
20&34.415,34.397&0.215,0.209&54.6,55.0&2.54,3.66&-3.0,-2.6&20.0&10.6,16.2,14.8&5.64,5.25,4.18&18.0,18.6&161,159&12.6,19.5&32.0,74.9&12.8,8.91&0.845,0.597,0.438\\
30&34.465,34.458&0.197,0.197&53.7,53.9&3.55,3.39&-3.8,-3.6&30.0&21.3,31.7,30.6&6.07,5.79,5.46&10.3,11.2&162,160&58.7,88.6&86.1,191&7.21,5.73&0.782,0.568,0.531\\
$R_M$&34.471,34.458&0.215,0.215&53.5,53.7&2.81,2.32&-3.4,-3.4&18.9,18.6&5.03,9.23,7.44&4.76,4.45,2.44&25.5,25.3&-171,-172&5.81,10.3&7.66,26.1&49.1,25.2&0.996,0.667,0.240\\
$R_A$&34.471,34.458&0.215,0.215&53.5,53.7&2.81,2.32&-3.4,-3.4&26.2,26.1&8.91,16.2,14.8&4.95,4.84,3.85&15.7,16.7&-176,-178&18.8,33.9&17.3,57.5&26.7,16.4&0.923,0.643,0.482\\
$R_{1/2}$&34.471,34.458&0.215,0.215&53.5,53.7&2.81,2.32&-3.4,-3.4&19.6,19.2&5.39,9.79,7.91&4.79,4.51,2.54&24.5,24.9&-172,-173&6.67,11.7&8.48,28.5&47.2,25.4&0.987,0.667,0.261\\
\hline
\label{tab2}
%\end{deluxetable}
\end{tabular}
%\footnotetext[1]{$\pm10^{\circ}$}
\end{sidewaystable}

\begin{sidewaystable}[h]
%emulate
\centering
%\rotate
%\begin{deluxetable*}{lccccc}
%\tablecolumns{5}
%\tablewidth{0pt}
\tiny
\caption{Cont. Table \ref{tab2}}
\begin{tabular}{lcccccccccccccc}
\hline
IRDC  &       &                     &                      &                          &                        & &  &                  &  &  &  & & & \\ 
Case &  $l_{\rm CM}$,$l_{\rm CM,\tau}$   &  $b_{\rm CM}$,$b_{\rm CM,\tau}$ &  $ v_{\rm CM}$,$v_{\rm CM,\tau}$ & $\Delta s$ & $\Delta v$ &R &$M$,$M_{\tau}$,$M_{G}$ & $\sigma$,$\sigma_{\tau}$,$\sigma_{G}$  &  $\frac{dv_0}{ds}$,$\frac{dv_{0,\tau}}{ds}$  & $\theta_v$,$\theta_{v,\tau}$ &  $I_0$,$I_{0,\tau}$   &  $E_{\rm rot,0}$,$E_{\rm rot,0,\tau}$ & $\beta,\beta_{\tau}$ & $\log(\alpha_{\rm vir})$,$\log(\alpha_{\rm vir,\tau})$, \\

                      &                       &                     &        &              &             &           &  $\times10^4 $ &                   &  $\times \rm 10^{-2}$ &                & $\times \rm 10^{3}$   & $\times 10^{46}$ & ($10^{-2}$)   & $\log(\alpha_{\rm vir,G})$ \\
    (pc)                  &  ( $^{\circ}$)   & ($^{\circ}$)  & ($\kms$)& (pc) & ($\kms$)&    (pc)  &  $(\sm)$           & ($\kms$)   &  (km/s/pc)       & ($\deg$) & ($M_{\odot}$~pc$^2$)  & (erg) & &  \\

 (1)   & (2)       & (3)                    &(4)                      & (5)                             & (6)                       & (7) & (8) &              (9)    & (10) & (11) & (12) &(13) & (14) &(15) \\       
\hline
\hline
F-s&       &                     &                      &                          &                        & &  &                  &  &  &  & & & \\ 
5&34.428,34.422&0.240,0.240&57.7,57.5&0.89,1.25&0.0,0.0&5.00&0.664,1.21,1.18&2.51,2.21,1.77&8.85,9.76&-24.7,-21.8&0.0464,0.0838&0.503,1.68&0.718,0.472&0.742,0.368,0.192\\
10&34.415,34.403&0.221,0.221&58.2,58.0&2.25,2.86&0.4,0.4&10.0&1.89,3.13,2.79&2.89,2.67,1.88&1.87,2.17&9.94,-3.59&0.414,0.783&2.04,5.58&0.0706,0.0655&0.711,0.423,0.170\\
20&34.360,34.335&0.191,0.184&58.2,58.0&6.28,7.83&0.4,0.4&20.0&5.74,9.75,9.02&2.68,2.48,1.95&1.05,1.32&-105,-88.7&2.45,3.99&9.41,27.1&0.0287,0.0255&0.463,0.167,-0.00943\\
30&34.379,34.366&0.178,0.178&58.0,57.7&5.90,6.47&0.2,0.2&30.0&9.42,15.3,14.3&2.61,2.41,1.88&0.480,0.504&-129,-122&8.39,9.78&16.9,44.4&0.0114,0.00557&0.401,0.122,-0.0638\\
$R_M$&34.366,34.354&0.178,0.178&57.3,57.3&6.47,7.07&0.0,0.0&14.9,14.0&3.35,6.67,6.27&2.26,2.13,1.67&4.86,4.73&-84.3,-86.3&0.786,1.34&4.31,18.1&0.428,0.164&0.422,0.0453,-0.138\\
$R_A$&34.366,34.354&0.178,0.178&57.3,57.3&6.47,7.07&0.0,0.0&20.5,20.5&4.06,8.21,7.84&2.14,2.02,1.64&4.84,4.96&-80.3,-77.5&1.56,3.27&4.60,18.8&0.792,0.425&0.431,0.0733,-0.0877\\
$R_{1/2}$&34.366,34.354&0.178,0.178&57.3,57.3&6.47,7.07&0.0,0.0&12.3,10.9&2.84,5.44,5.04&2.33,2.20,1.65&2.60,2.20&-91.7,-104&0.458,0.602&3.73,15.5&0.0827,0.0187&0.437,0.0511,-0.162\\
\hline
G&       &                     &                      &                          &                        & &  &                  &  &  &  & & & \\ 
5&34.758,34.758&-0.565,-0.571&43.9,44.1&0.44,0.88&0.4,0.9&5.00&1.05,1.61,1.58&3.58,3.23,2.85&47.1,46.0&47.0,50.8&0.0575,0.0804&1.25,2.96&10.2,5.71&0.853,0.577,0.475\\
10&34.764,34.764&-0.577,-0.590&44.6,44.6&0.93,1.59&1.1,1.3&10.0&3.16,4.47,4.14&4.57,4.14,3.08&12.1,12.7&-2.58,3.59&0.623,0.698&5.69,11.4&1.59,0.988&0.887,0.650,0.425\\
20&34.807,34.813&-0.614,-0.620&44.8,44.6&3.55,3.80&1.3,1.3&20.0&11.1,14.6,13.1&6.47,6.29,4.99&21.8,22.2&-161,-160&9.22,11.3&35.1,60.5&12.4,9.14&0.944,0.802,0.645\\
30&34.844,34.862&-0.571,-0.590&44.4,43.7&4.09,4.92&0.9,0.4&30.0&21.3,28.0,27.7&7.22,7.21,8.46&20.1,22.9&-161,-158&41.2,53.6&86.7,149&19.0,18.8&0.931,0.812,0.955\\
$R_M$&34.936,34.948&-0.602,-0.620&42.0,41.6&8.98,9.84&-1.5,-1.7&15.2,14.8&5.65,9.28,6.13&5.56,5.91,2.61&42.4,56.3&-128,-131&3.38,6.07&12.0,33.3&50.4,57.5&0.984,0.810,0.280\\
$R_A$&34.936,34.948&-0.602,-0.620&42.0,41.6&8.98,9.84&-1.5,-1.7&23.9,23.9&9.07,14.6,15.1&5.59,5.66,6.94&31.7,35.8&-134,-136&9.70,16.4&19.6,51.2&49.6,40.8&0.982,0.784,0.948\\
$R_{1/2}$&34.936,34.948&-0.602,-0.620&42.0,41.6&8.98,9.84&-1.5,-1.7&13.9,13.3&4.91,7.66,5.30&5.48,5.92,2.65&42.2,59.9&-124,-128&2.53,4.18&9.91,25.2&45.3,59.3&0.995,0.851,0.313\\
\hline
G-s&       &                     &                      &                          &                        & &  &                  &  &  &  & & & \\ 
5&34.764,34.764&-0.571,-0.571&44.4,44.4&0.69,0.69&0.6,0.9&5.00&0.929,1.47,1.51&2.57,2.40,2.66&28.7,29.1&57.9,63.1&0.0487,0.0714&0.985,2.47&4.05,2.44&0.617,0.357,0.435\\
10&34.776,34.770&-0.583,-0.596&44.8,44.8&1.28,1.87&1.1,1.3&10.0&2.57,3.81,3.89&2.75,2.56,2.77&18.0,17.8&14.6,15.2&0.408,0.507&3.77,8.27&3.47,1.94&0.534,0.302,0.361\\
20&34.838,34.832&-0.590,-0.602&45.2,45.2&3.76,3.80&1.5,1.7&20.0&7.01,9.59,9.70&2.95,2.81,2.97&8.12,8.92&-18.9,-20.1&4.79,6.32&14.0,26.2&2.24,1.91&0.461,0.282,0.326\\
30&34.862,34.862&-0.528,-0.540&45.0,45.0&4.92,4.76&1.3,1.5&30.0&11.7,15.2,15.4&3.03,2.91,3.11&1.19,1.35&-26.1,-23.6&16.8,20.9&26.0,44.0&0.0909,0.0859&0.437,0.290,0.339\\
$R_M$&34.887,34.881&-0.547,-0.559&44.8,44.8&5.95,5.60&1.1,1.3&14.5,14.1&3.64,5.54,5.58&2.58,2.50,2.59&13.9,15.3&-7.89,-13.1&1.84,2.90&5.22,12.4&6.76,5.41&0.490,0.266,0.295\\
$R_A$&34.887,34.881&-0.547,-0.559&44.8,44.8&5.95,5.60&1.1,1.3&21.8,21.8&5.36,7.95,7.94&2.57,2.48,2.50&2.38,2.57&35.9,14.1&2.41,3.91&7.55,16.6&0.180,0.154&0.493,0.291,0.298\\
$R_{1/2}$&34.887,34.881&-0.547,-0.559&44.8,44.8&5.95,5.60&1.1,1.3&13.1,12.5&3.25,4.79,4.87&2.62,2.53,2.70&16.1,17.0&-4.82,-7.47&1.48,2.16&4.59,10.5&8.32,5.91&0.508,0.288,0.337\\
\hline
H&       &                     &                      &                          &                        & &  &                  &  &  &  & & & \\ 
5&35.428,35.422&-0.313,-0.319&48.6,48.2&1.12,0.88&1.1,1.3&5.00&0.970,1.52,0.789&6.53,6.18,1.12&51.8,57.5&-87.4,-81.5&0.0500,0.0761&1.07,2.65&12.4,9.46&1.41,1.16,-0.0351\\
10&35.422,35.422&-0.288,-0.288&48.8,48.4&1.97,2.27&1.3,1.5&10.0&3.27,5.02,2.72&6.82,6.53,1.54&25.0,28.7&-68.9,-68.6&0.662,1.04&6.10,14.4&6.73,5.91&1.22,0.996,0.00640\\
20&35.403,35.397&-0.251,-0.257&49.2,49.0&3.75,3.79&1.7,2.1&20.0&9.90,14.5,6.60&7.16,6.95,2.12&31.6,33.1&-52.4,-53.3&9.80,14.7&27.9,59.6&34.8,26.8&1.08,0.890,0.198\\
30&35.367,35.354&-0.282,-0.307&47.8,46.9&3.08,3.06&0.2,0.0&30.0&20.9,31.0,31.0&7.92,8.05,13.2&32.7,35.7&-53.9,-54.6&53.7,84.0&82.8,183&69.0,58.3&1.02,0.863,1.29\\
$R_M$&35.348,35.336&-0.337,-0.362&46.5,45.6&3.48,4.34&-0.2,-0.6&18.6,18.6&4.52,8.30,4.91&5.95,5.90,1.60&40.4,40.5&-56.2,-60.0&4.55,8.61&6.27,21.1&118,66.4&1.23,0.958,0.0548\\
$R_A$&35.348,35.336&-0.337,-0.362&46.5,45.6&3.48,4.34&-0.2,-0.6&25.7,25.7&8.22,15.8,7.07&7.41,7.53,1.94&48.5,49.9&-52.9,-54.4&18.5,37.6&15.0,55.1&288,169&1.30,1.03,0.204\\
$R_{1/2}$&35.348,35.336&-0.337,-0.362&46.5,45.6&3.48,4.34&-0.2,-0.6&20.0,19.9&5.04,9.23,5.14&6.19,6.30,1.62&43.7,45.9&-56.1,-59.7&5.89,11.1&7.25,24.4&154,95.4&1.25,0.998,0.0716\\
\hline
H-s&       &                     &                      &                          &                        & &  &                  &  &  &  & & & \\ 
5&35.428,35.422&-0.319,-0.325&43.9,43.9&1.56,1.12&0.2,0.0&5.00&0.478,0.848,0.781&1.50,1.37,1.10&13.9,15.5&-51.8,-46.3&0.0321,0.0616&0.260,0.820&2.38,1.80&0.436,0.108,-0.0466\\
10&35.416,35.410&-0.294,-0.301&43.9,43.9&2.27,1.89&0.2,0.0&10.0&1.50,2.64,2.62&1.57,1.47,1.46&16.2,16.6&-57.8,-56.0&0.397,0.740&1.29,3.99&8.03,5.09&0.283,-0.0205,-0.0250\\
20&35.330,35.330&-0.288,-0.294&43.7,43.7&4.49,4.32&0.0,-0.2&20.0&3.36,5.54,5.60&1.69,1.60,1.68&7.84,7.66&-47.5,-48.3&2.79,4.42&3.22,8.74&5.31,2.95&0.298,0.0293,0.0677\\
30&35.244,35.244&-0.276,-0.282&43.7,43.7&8.67,8.56&0.0,-0.2&30.0&5.99,9.34,9.53&1.81,1.72,1.86&2.84,2.66&-11.7,-14.5&9.29,14.4&6.82,16.6&1.09,0.612&0.279,0.0457,0.105\\
$R_M$&35.231,35.231&-0.294,-0.301&43.7,43.7&8.98,8.61&-0.2,-0.2&13.4,13.2&1.83,3.38,3.47&1.40,1.36,1.50&5.55,4.94&-14.3,-9.80&0.994,1.83&1.44,4.93&2.12,0.901&0.223,-0.0746,-0.00303\\
$R_A$&35.231,35.231&-0.294,-0.301&43.7,43.7&8.98,8.61&-0.2,-0.2&19.2,19.3&2.98,5.54,5.60&1.48,1.44,1.52&3.96,3.86&-8.14,-17.6&3.25,6.19&2.64,9.06&1.92,1.01&0.216,-0.0739,-0.0344\\
$R_{1/2}$&35.231,35.231&-0.294,-0.301&43.7,43.7&8.98,8.61&-0.2,-0.2&13.1,12.9&1.78,3.23,3.32&1.40,1.35,1.49&5.61,5.11&-15.5,-10.2&0.929,1.64&1.38,4.62&2.10,0.921&0.224,-0.0705,0.00264\\
\hline
I&       &                     &                      &                          &                        & &  &                  &  &  &  & & & \\ 
5&38.936,38.936&-0.460,-0.460&44.3,42.6&1.04,1.04&0.6,0.4&5.00&0.814,1.47,1.27&5.92,4.47,1.21&1.76,14.9&-100.,130&0.0423,0.0671&0.756,2.47&0.0172,0.600&1.40,0.898,-0.173\\
10&38.930,38.930&-0.454,-0.460&44.9,43.2&1.45,1.30&1.3,1.1&10.0&2.25,3.58,2.71&6.33,5.26,1.42&11.9,6.46&-39.0,-10.1&0.452,0.588&2.90,7.32&2.21,0.333&1.32,0.955,-0.0629\\
20&38.942,38.942&-0.393,-0.411&46.9,45.4&3.81,2.96&3.2,3.2&20.0&5.15,6.96,3.86&6.87,6.42,1.61&24.3,22.6&-71.4,-67.8&2.69,3.54&7.57,13.8&20.8,13.1&1.33,1.14,0.194\\
30&38.961,38.961&-0.362,-0.380&47.1,45.6&5.22,4.36&3.4,3.4&30.0&6.81,8.96,4.35&7.29,6.94,1.62&13.5,15.1&-18.3,-10.2&9.39,11.1&8.83,15.3&19.4,16.6&1.43,1.27,0.326\\
$R_M$&38.936,38.936&-0.460,-0.467&41.3,41.1&1.04,0.91&0.0,-0.2&6.70,6.18&0.789,1.61,1.59&1.42,1.34,1.27&24.8,25.9&91.4,89.1&0.0849,0.152&0.531,2.40&9.80,4.24&0.297,-0.0932,-0.137\\
$R_A$&38.936,38.936&-0.460,-0.467&41.3,41.1&1.04,0.91&0.0,-0.2&11.6,11.6&1.35,2.71,2.58&1.79,1.67,1.32&3.43,5.61&139,112&0.269,0.696&0.891,3.61&0.353,0.604&0.506,0.145,-0.0381\\
$R_{1/2}$&38.936,38.936&-0.460,-0.467&41.3,41.1&1.04,0.91&0.0,-0.2&6.40,5.70&0.755,1.47,1.46&1.41,1.29,1.24&25.1,25.9&89.1,84.0&0.0754,0.120&0.509,2.17&9.33,3.66&0.289,-0.122,-0.158\\
\hline
J&       &                     &                      &                          &                        & &  &                  &  &  &  & & & \\ 
5&53.139,53.145&0.056,0.050&21.6,21.6&0.77,0.98&0.4,0.4&5.00&0.434,0.794,0.679&5.44,4.61,1.11&48.3,50.5&44.6,41.5&0.0222,0.0403&0.215,0.720&23.9,14.2&1.60,1.19,0.0203\\
10&53.164,53.164&0.025,0.025&22.0,22.0&1.82,1.82&0.9,0.9&10.0&0.881,1.53,1.31&5.16,4.42,1.17&20.0,18.1&33.2,37.5&0.132,0.222&0.443,1.34&11.8,5.37&1.55,1.17,0.0849\\
20&53.250,53.280&0.062,0.050&22.7,22.7&4.25,5.22&1.5,1.5&20.0&1.65,3.05,2.64&4.87,4.17,1.22&12.2,10.9&-21.1,-19.0&1.21,2.26&0.775,2.65&23.2,10.1&1.52,1.12,0.119\\
30&53.293,53.324&0.124,0.105&22.7,22.7&5.99,6.75&1.5,1.5&30.0&2.31,4.09,3.35&5.28,4.59,1.24&6.84,6.59&10.8,9.23&2.83,4.71&1.01,3.18&13.0,6.40&1.62,1.25,0.203\\
$R_M$&53.317,53.342&0.013,0.019&22.7,22.7&6.52,7.24&1.5,1.5&9.89,9.34&0.697,1.47,1.50&1.10,1.06,1.12&12.4,13.1&-37.2,-44.5&0.189,0.353&0.280,1.33&10.3,4.54&0.301,-0.0818,-0.0402\\
$R_A$&53.317,53.342&0.013,0.019&22.7,22.7&6.52,7.24&1.5,1.5&13.1,13.1&0.939,2.00,2.04&1.11,1.08,1.15&8.78,8.75&-14.4,-14.9&0.483,1.03&0.385,1.74&9.62,4.53&0.301,-0.0541,-0.00341\\
$R_{1/2}$&53.317,53.342&0.013,0.019&22.7,22.7&6.52,7.24&1.5,1.5&9.20,8.60&0.629,1.28,1.30&1.08,1.06,1.12&13.3,15.1&-40.7,-50.1&0.137,0.227&0.246,1.09&9.85,4.75&0.301,-0.0549,-0.0138\\
\hline
\label{tab2c}
%\end{deluxetable}
\end{tabular}
%\footnotetext[1]{$\pm10^{\circ}$}
\end{sidewaystable}

\begin{sidewaystable}[h]
%emulate
\centering
%\rotate
%\begin{deluxetable*}{lccccc}
%\tablecolumns{5}
%\tablewidth{0pt}
\tiny
\caption{IRDC Physical Parameters Estimates}
\begin{tabular}{cccccccccccccc}
\hline
IRDC &  Case &  $l_{\rm CM}$,$l_{\rm CM,\tau}$   &  $b_{\rm CM}$,$b_{\rm CM,\tau}$ &  $ v_{\rm CM}$,$ v_{\rm CM,\tau}$ & $R$ &$M$,$M_{\tau}$,$M_{G}$ & $\sigma$,$\sigma_{\tau}$,$\sigma_{G}$  &  $\frac{dv_0}{ds}$,$\frac{dv_{0,\tau}}{ds}$   & $\theta_v$,$\theta_{v,\tau}$ &  $I_0$,$I_{0,\tau}$   &  $E_{\rm rot,0}$,$E_{\rm rot,0,\tau}$ & $\beta,\beta_{\tau}$ & $\log(\alpha_{\rm vir})$,$\log(\alpha_{\rm vir,\tau})$, \\

                      &                       &                     &        &                   &           &  $\times10^4 $ &                   &  $\times \rm 10^{-2}$ &                & $\times \rm 10^{3}$   & $\times 10^{46}$ & ($10^{-2}$)   & $\log(\alpha_{\rm vir,G})$ \\

&              &  ( $^{\circ}$)   & ($^{\circ}$)  & ($\kms$)&    (pc)  &  $(\sm)$           & ($\kms$)   &  (km/s/pc)      & ($\deg$) & ($M_{\odot}$~pc$^2$)  & (erg) &  & \\

 (1)   & (2)       & (3)                    &(4)                      & (5)                             & (6)                       & (7) & (8) &              (9)    & (10) & (11) & (12) &(13) & (14)\\         
\hline
\hline
A&SE&18.831,18.844&-0.282,-0.282&64.3,64.3&5.18&1.07,1.65,1.41&4.20,3.60,1.79&24.6,27.9&117,117&143,214&123,293&6.97,5.66&0.998,0.675,0.136\\
&CE&18.838,18.844&-0.282,-0.276&65.0,64.8&&0.802,1.36,1.30&2.18,1.97,1.63&27.1,28.7&122,121&114,187&69.4,199&12.0,7.73&0.553,0.236,0.0890\\
\hline
B&SE&19.268,19.274&0.075,0.075&26.5,26.1&1.36&0.0891,0.178,0.170&3.96,3.06,1.51&37.6,35.2&-93.6,-93.5&0.102,0.204&3.20,12.8&0.446,0.197&1.44,0.920,0.327\\
&CE&19.274,19.274&0.075,0.075&25.9,25.9&&0.0742,0.161,0.165&1.41,1.33,1.46&22.8,22.5&40.5,40.3&0.712,1.52&2.22,10.5&1.66,0.727&0.625,0.239,0.307\\
\hline
C&SE&28.370,28.370&0.068,0.068&78.6,78.6&7.68&3.58,6.19,5.96&4.09,3.52,2.83&10.5,10.6&-98.4,-102&371,599&947,2830&0.429,0.238&0.621,0.254,0.0788\\
&CE&28.370,28.370&0.068,0.068&78.1,78.4&&3.07,5.63,5.68&2.78,2.53,2.63&14.7,15.0&-92.7,-93.8&309,527&698,2340&0.956,0.506&0.351,0.00859,0.0370\\
\hline
D&SE&28.567,28.573&-0.233,-0.233&80.5,81.5&8.46&3.54,5.67,3.27&7.46,7.09,1.90&5.66,5.40&-4.15,37.0&1490,528&822,2110&0.577,0.0727&1.19,0.941,0.0380\\
&CE&28.579,28.579&-0.227,-0.227&81.3,82.4&&2.37,4.29,2.93&7.07,6.53,1.71&22.3,32.3&101,107&387,727&368,1210&5.21,6.25&1.32,0.990,-0.00652\\
\hline
D-s&SE&28.573,28.573&-0.233,-0.233&87.3,87.1&8.46&1.63,3.05,3.05&2.06,1.75,1.72&6.34,7.73&113,114&363,584&174,612&0.834,0.568&0.408,-0.00586,-0.0187\\
&CE&28.579,28.579&-0.233,-0.233&86.9,86.9&&1.35,2.77,2.84&1.58,1.49,1.63&24.3,24.5&121,120&259,481&120,502&12.7,5.71&0.260,-0.0996,-0.0343\\
\hline
E&SE&28.678,28.678&0.130,0.130&79.8,80.1&5.74&1.29,1.86,1.52&6.38,5.97,3.05&40.7,41.9&71.4,69.5&120,196&163,336&12.2,10.2&1.32,1.11,0.611\\
&CE&28.672,28.678&0.130,0.124&79.8,79.8&&0.770,1.26,1.21&3.58,3.60,2.41&20.0,18.5&57.4,53.4&136,251&57.8,155&9.32,5.53&1.05,0.838,0.506\\
\hline
F&SE&34.440,34.440&0.246,0.246&57.5,57.5&1.75&0.160,0.294,0.255&3.74,3.12,1.67&26.1,26.9&26.5,23.8&2.38,4.67&8.25,28.0&1.95,1.20&1.25,0.831,0.348\\
&CE&34.440,34.440&0.246,0.246&56.9,57.1&&0.122,0.250,0.243&1.82,1.69,1.55&46.8,46.2&-29.6,-31.9&2.29,4.58&4.82,20.3&10.4,4.79&0.742,0.367,0.305\\
\hline
F-s&SE&34.440,34.440&0.246,0.246&57.7,57.5&1.75&0.127,0.252,0.247&2.10,1.81,1.58&20.7,21.1&-23.4,-23.5&2.63,5.21&5.20,20.6&2.15,1.12&0.851,0.424,0.315\\
&CE&34.440,34.440&0.246,0.246&57.3,57.3&&0.111,0.235,0.239&1.44,1.40,1.50&32.7,34.6&-26.5,-25.9&2.18,4.66&4.01,17.9&5.80,3.10&0.580,0.230,0.286\\
\hline
G&SE&34.764,34.770&-0.559,-0.559&43.5,43.3&1.53&0.139,0.249,0.244&3.20,2.75,2.36&54.8,49.7&61.4,60.4&0.882,1.50&7.06,22.7&3.73,1.62&1.12,0.734,0.610\\
&CE&34.764,34.764&-0.559,-0.559&43.5,43.3&&0.127,0.236,0.239&2.41,2.17,2.28&70.5,67.1&67.0,66.5&0.628,1.09&5.94,20.4&5.22,2.40&0.912,0.550,0.590\\
\hline
G-s&SE&34.770,34.770&-0.559,-0.559&43.7,43.5&1.53&0.126,0.234,0.236&2.42,2.16,2.24&33.2,33.9&72.1,71.1&0.483,0.861&5.87,20.0&0.902,0.490&0.915,0.551,0.579\\
&CE&34.770,34.770&-0.559,-0.559&43.7,43.5&&0.122,0.228,0.233&2.23,2.03,2.20&36.0,36.0&69.9,69.5&0.514,0.902&5.42,19.1&1.22,0.610&0.865,0.507,0.569\\
\hline
H&SE&35.410,35.410&-0.325,-0.331&47.5,46.9&4.84&0.924,1.56,1.05&6.12,5.64,1.42&81.4,83.0&-60.2,-60.5&134,232&99.3,282&88.9,56.3&1.36,1.06,0.0358\\
&CE&35.416,35.416&-0.325,-0.331&46.7,46.3&&0.641,1.24,0.940&5.05,4.79,1.24&105,106&-61.6,-61.5&94.5,187&47.7,179&218,116&1.35,1.02,-0.0359\\
\hline
H-s&SE&35.403,35.403&-0.337,-0.337&43.7,43.9&4.84&0.560,1.06,1.04&1.54,1.43,1.40&30.6,29.4&-56.0,-56.2&95.8,183&36.5,130&24.6,12.2&0.380,0.0397,0.0277\\
&CE&35.403,35.397&-0.337,-0.337&43.9,43.9&&0.481,0.966,0.940&1.34,1.30,1.24&41.9,40.8&-56.2,-56.2&84.5,171&26.9,108&55.0,26.1&0.322,-0.00712,-0.0359\\
\hline
I&SE&38.954,38.954&-0.473,-0.473&43.7,42.2&1.86&0.187,0.396,0.361&5.33,3.56,1.06&17.2,28.6&150,137&3.17,5.27&10.6,47.7&0.871,0.896&1.52,0.842,-0.169\\
&CE&38.954,38.954&-0.473,-0.473&41.3,41.3&&0.143,0.353,0.356&1.10,1.01,1.04&27.6,30.0&-23.1,-23.3&2.20,4.93&6.24,38.1&2.68,1.16&0.260,-0.206,-0.181\\
\hline
J&SE&53.114,53.114&0.056,0.056&21.2,21.2&0.378&0.00333,0.00734,0.00718&2.30,1.75,0.728&217,208&5.45,5.68&0.0398,0.0780&0.0166,0.0809&1130,416&1.84,1.26,0.512\\
&CE&53.114,53.114&0.056,0.056&21.2,21.2&&0.00298,0.00693,0.00707&0.698,0.654,0.711&1.18,7.40&66.9,171&0.000768,0.0319&0.0133,0.0721&0.000795,0.241&0.857,0.434,0.497\\
\hline
\label{tabirdc}
%\end{deluxetable}
\end{tabular}
%\footnotetext[1]{$\pm10^{\circ}$}
\end{sidewaystable}

\section{Results}

\subsection{Molecular Cloud Physical Properties}
\label{S:physprop}

Table \ref{tab2} presents the derived physical properties for all 10
clouds. The global properties of each cloud were evaluated for data
extracted out to seven different radii, shown on different lines in
the table: 5, 10, 20, and 30 pc using SE; and out to mass weighted
($R_M$), areal ($R_A$), and half-mass ($R_{1/2}$) radii using CE. Each
cloud definition case is noted in column 1. The center-of-mass
coordinates of each cloud are listed in columns 2 and 3, evaluated for
the optically thin and opacity-corrected cases. 
We note that the CE cloud definitions with $R_M$, $R_A$ and $R_{1/2}$
share the same center. Mean velocities are reported in column 4.
Columns 5 and 6 list the separations between the GMC centers and the
IRDC centers in plane of sky distance and velocity offset,
respectively, again for both optically thin and opacity-corrected methods.
Column 7 lists the cloud radii (with $R_M$, $R_A$ and $R_{1/2}$ cases
having two values for the optical thin and opacity-corrected
cases). Column 8 lists the three cloud mass estimates: optically thin
mass ($M$), opacity-corrected mass ($M_{\tau}$), and the mass
estimated from a gaussian fit of the co-added opacity corrected column
density spectra ($M_G$). Column 9 lists the three velocity dispersions
measured: the standard deviations of the cloud's optically thin and
opacity corrected column density spectra ($\sigma$, $\sigma_{\tau}$)
and the dispersion estimated from the gaussian fitted opacity
corrected column density spectra ($\sigma_{G}$). Column 10 presents
the two velocity gradient magnitudes, one estimated from the optically
thin column density weighted first-order moment map ($dv_0/ds$) and
one estimated from the opacity-corrected column density weighted
first-order moment map ($dv_{0,\tau}/ds$).
Column 11 presents the position angles of the projected cloud rotation
axes estimated from each cloud's angular momentum for both optically
thin and opacity corrected masses ($\theta_v$ and $\theta_{v,\tau}$). The
projected moments of inertia estimated using the optically thin mass
($I_0$) and opacity corrected mass ($I_{0,\tau}$) are presented in column
12.  Column 13 shows projected rotational energies, $E_{\rm rot,0}$ and
$E_{\rm rot,0,\tau}$, and column 14 shows the ratios of these to the
gravitational energy, i.e., $\beta$ and $\beta_\tau$. Finally, column
15 shows log$_{10}$ of the virial parameters of the clouds using the three
measures of mass, i.e., $\alpha_{\rm vir}$, $\alpha_{\rm vir,\tau}$, $\alpha_{\rm vir,G}$.

A visual overview of each GMC-scale region is given in Figure
\ref{mapA} and Figures \ref{mapB}-\ref{mapJ}. The left side of the
figures display a series of images of the molecular gas out to a
radius of 30~pc. The first row displays the GLIMPSE $8\micron$ image
and the $\thco$ mass surface density map, $\Sigma_{\rm 13CO}$,
generated using the velocity interval given solely by the IRDC. The
IRDC elliptical boundary is also shown \citep[][]{Simon2006a}. The
subsequent rows display the $\Sigma_{\rm 13CO}$ maps assuming our
defined velocity interval of $v_0\pm15~\kms$, the column-density
weighted mean velocity (first moment), the linear momentum map derived
from each pixel's column density and velocity with respect to the
cloud center of mass, and the velocity dispersion map (second moment)
for the SE cloud (\textit{Left column}) and CE cloud (\textit{Right
  column}).  The right side of the figures display the optically thin
and opacity-corrected co-added column density velocity spectra at
various extraction radii for the SE and CE clouds and the SE IRDC.

Four of the clouds (D, F, G, and H) were found to contain multiple
velocity components, based on the co-added $\thco$ column density
spectra of their SE 60 pc diameter GMCs (e.g., Cloud D; Fig. \ref{mapD}). We
used a series of integrated intensity maps to isolate the velocity
range of the molecular gas associated with the IRDC
(Fig. \ref{chanmaps}).  In general, these velocity ranges span
$\sim10~\kms$ (Cloud D: $84.1-93.7~\kms$; Cloud F: $54.3-65.5~\kms$;
Cloud G: $39.7-51.8~\kms$; Cloud H: $39.7-49.9~\kms$). To investigate
how the cloud's derived physical properties were affected by isolating
only these velocity ranges, we include these four cases as separate
entries in Table \ref{tab2} (noted with ``-s''). For all subsequent 
analyses, the ``-s'' values for these four clouds are used in place of those 
derived using the full velocity range of $v_0\pm15~\kms$.
   
%\newpage   
\subsection{IRDC and GMC Morphologies}
\label{S:loc}
\subsubsection{Description of Individual IRDCs/GMCs}
   
%\noindent
\underbar{\textit{Cloud A:}} We estimate a cloud areal radius of
$R_A\simeq 27$~pc and a mass of $M_{G,A}\simeq 2.7\times10^5~\rm
M_{\odot}$. The GRS $\thco$ GMC study by RD10 identified five molecular clouds within the extent of Cloud A, with a
total $\thco$ mass of $2.1\times10^5~\rm M_{\odot}$ (note, we scaled
all $\thco$ masses of RD10 by a factor 0.49 to
convert to our assumed $\thco$/\htwo ~abundance ratio; see
\S\ref{S:13COmass}). IRDC A is located about 20~pc north of its parent
GMC's highest $\Sigma$ clump, which appears bright at $\rm 8\:\mu m$
and is thus likely to be star-forming.
The WISE \HII~ region catalog \citep{Anderson2014} lists the W39 \HII~
region to be within $\sim3\arcmin$ of this clump's center, and with a
similar LSR velocity ($v_{\rm LSR}=65.5~\kms$). Another \HII~ region
lies in a smaller dense clump at $l=19.07^\circ$, $b=-0.28^\circ$ with
velocity and distance consistent with those of Cloud A. The isolated,
simple $\rm ^{13}CO$ emission profiles, along with the spatial
connection between the IRDC and the \HII~ regions, demonstrated
through CE, indicates that GMC A contains clumps at a variety of
stages of star formation activity.

\underbar{\textit{Cloud B:}} We estimate a radius of $R_{A}\simeq 12$~pc
and a mass of
$M_{G,A}\simeq3.3\times10^4\:M_{\odot}$. RD10
estimated a $\thco$ mass of $2.4\times10^4~\rm M_{\odot}$ and a
smaller projected areal radius of 6.3 pc.
The IRDC is located within the densest region of the GMC, which 
extends $\sim10$ pc to the south-east of the IRDC and accounts
for most of the cloud's mass.  Spatially, Cloud B is
well defined and completely isolated at an extraction radius of 20~pc
in CE.  Although the cloud is isolated in velocity space, the
column density profile indicates a low level of $\thco$ emission at
neighboring velocities, both lower and higher. 
Although the GLIMPSE $8\mu$m emission map indicates a large bright
emission structure to the west of the cloud, this feature is not at
velocities consistent with Cloud B ($v_{\rm LSR}\sim 50\kms$).

\begin{figure}[ht!]
\begin{center}$
\begin{array}{cc}
\includegraphics[width=2.51in, angle=-90]{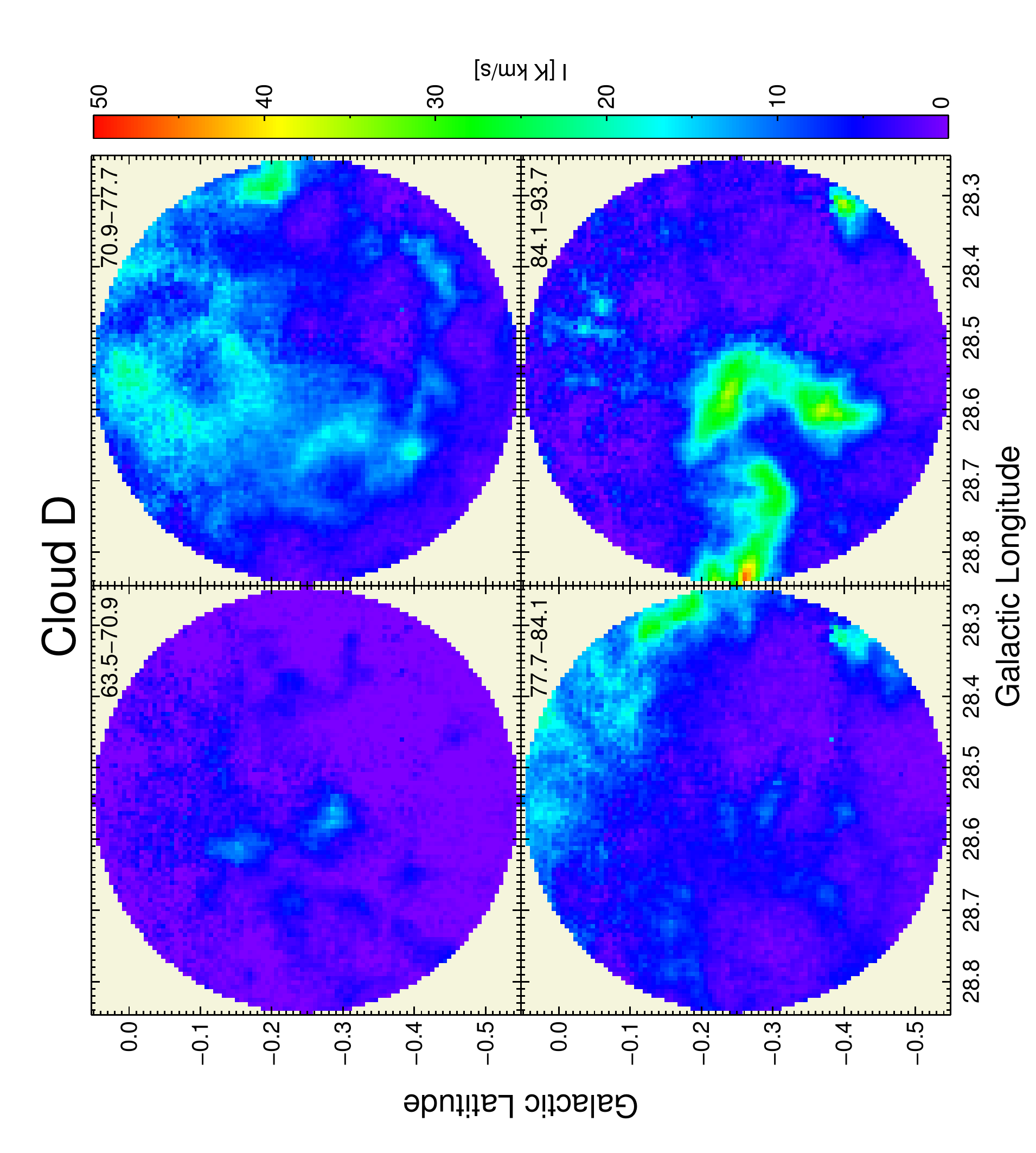} & \includegraphics[width=2.51in, angle=-90]{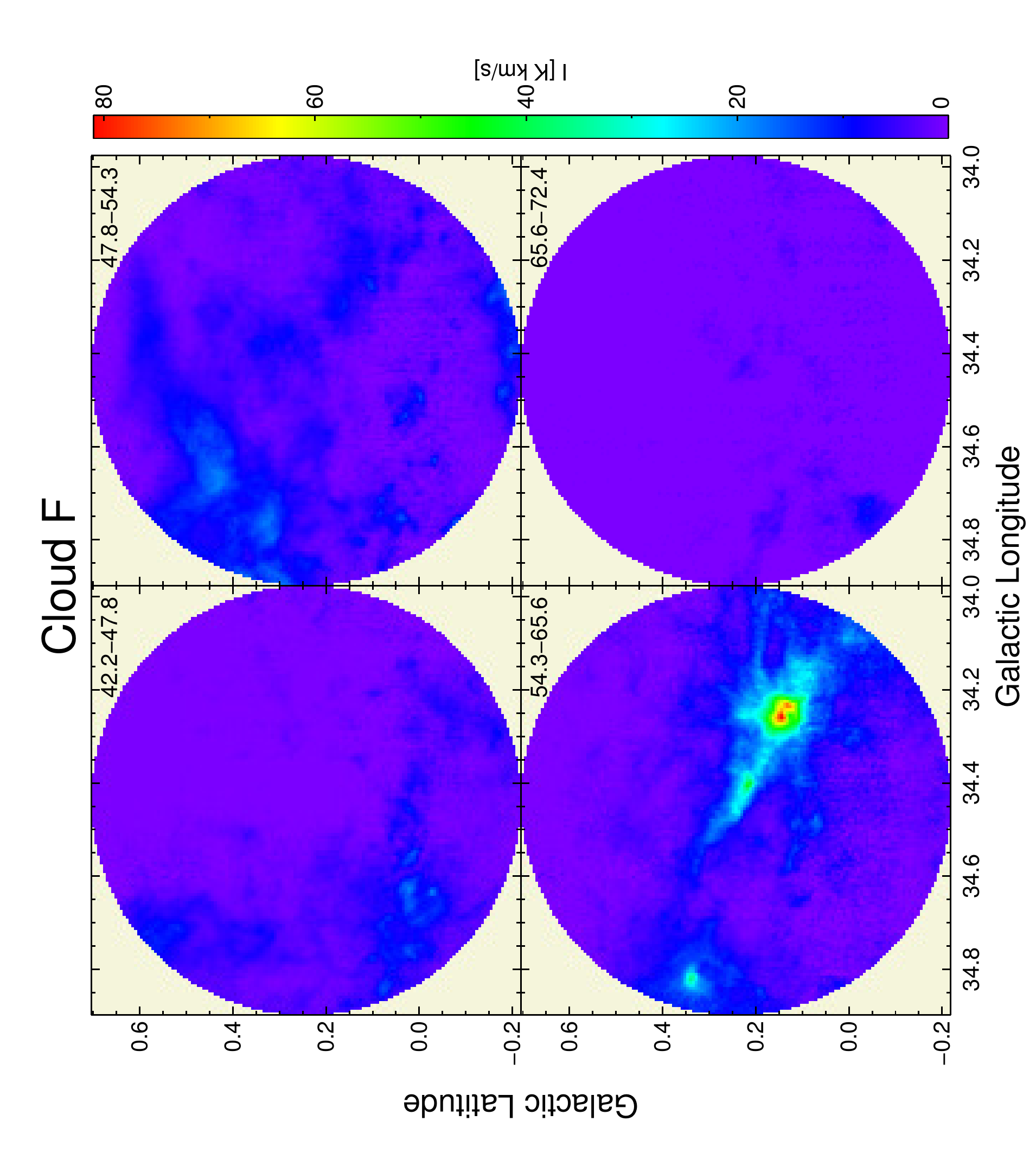} \\
\includegraphics[width=2.51in, angle=-90]{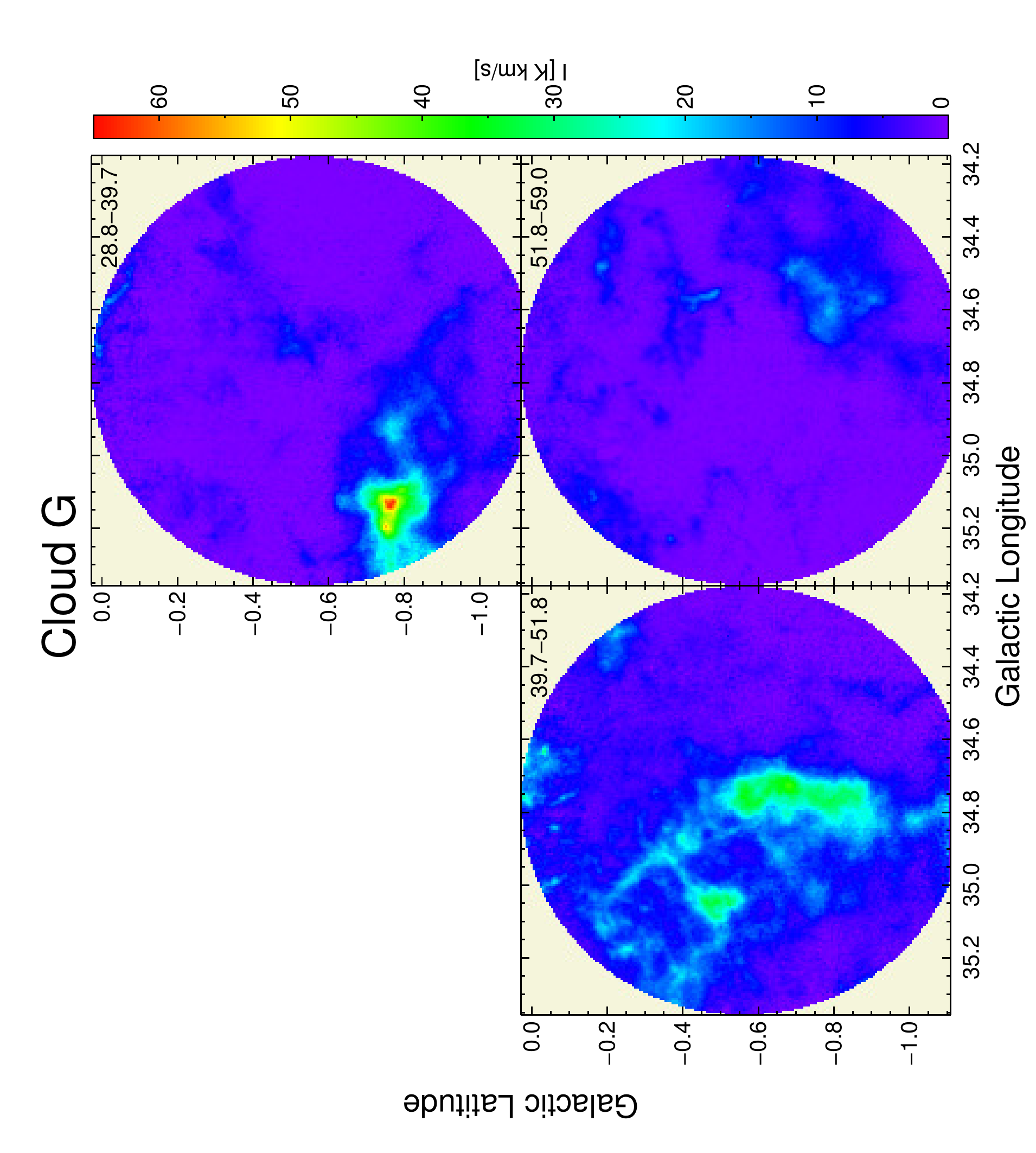} & \includegraphics[width=2.51in, angle=-90]{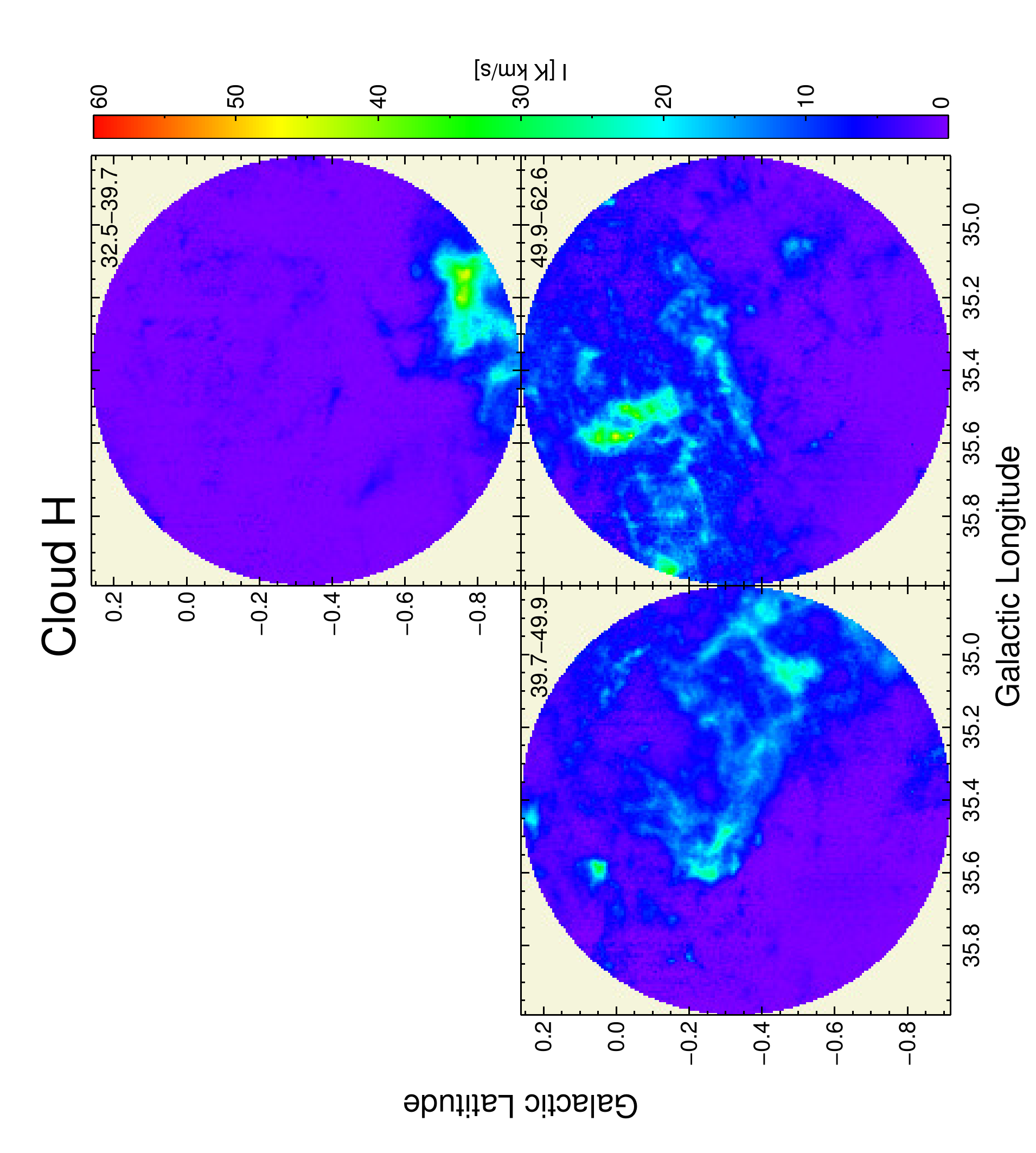} 
\end{array}$
\end{center}
\caption{$\thco$ channel maps of the multi-component clouds.  \textit{top left}: Cloud D; \textit{top right}: Cloud F; \textit{bottom left} Cloud G; \textit{bottom right}: Cloud H.  The velocity interval of each panel is shown in the upper-right corner of each map. }
\label{chanmaps}
\end{figure} 

\underbar{\textit{Cloud C:}} We estimate a GMC radius of $R_A\simeq
29$~pc and a mass of $M_{G,A}\simeq
3.5\times10^5\:M_{\odot}$.  RD10 derived a radius of
20.9~pc and a mass of $\sim3\times10^5~\rm M_{\odot}$.  IRDC C is
located within the dense central region of its parent GMC.  The
integrated intensity maps indicate dense and clumpy structure
throughout the cloud (Fig. \ref{mapC}).  The velocity distribution
shows a broad centrally-peaked profile, but with potential
substructure.  Additionally, there is a low level of $\thco$ emission
at higher velocities. \citet{Butler2014} studied the IRDC C using
near- and mid-infrared extinction mapping to probe the high-dynamic
range of its mass surface density (down to $A_V\sim1$ mag), finding an
IRDC mass of $\sim7\times10^4\:M_{\odot}$. This is one of the most
massive, high-$\Sigma$ IRDCs known, with the potential to form a
massive star cluster.  Using the same IRDC elliptical boundary from
\citet{Simon2006a}, we estimate a $\thco$ mass of
$5.7\times10^4\:M_{\odot}$. Our slightly lower mass estimate may be
the result of abundance uncertainties and/or CO depletion.

\underbar{\textit{Cloud D:}} For the GMC at $v_{\rm LSR}=84.1-93.7~\kms$,
we estimate a radius of $R_A\simeq 21$~pc and a mass of $M_{G,A} \simeq
9.3\times10^4\:M_{\odot}$. RD10 identified two
clouds within the Cloud D-s region with a total $\thco$ mass of
$\sim5.2\times10^4~\rm M_{\odot}$. The IRDC is located in the dense
central region of the GMC. However, the $\thco$ spectra show four
cloud components within $v_0\pm15~\kms$.  Although the $\thco$
emission corresponding to the IRDC is evident within a radius of 5~pc,
a lower-velocity gas component begins to dominate on scales out to
about 20~pc. The position-velocity map (Fig. \ref{PV1}) shows there is a
significant amount of $\thco$ emission at lower velocities,
$v_{\rm LSR}=63.5$-$84.1~\kms$, that extends largely over the northern
region of the cloud. In the same velocity interval of the IRDC,
$v_{\rm LSR}=84.1-93.7~\kms$, a substantial molecular gas clump contains
the largest integrated peak intensity ($\sim50$~K~$\kms$) at the
western edge of the 30~pc extraction radius. Figure \ref{chanmaps}
suggests that this emission is only tenuously connected to the $\thco$
emission associated with the main GMC.
  
\underbar{\textit{Cloud E:}} We estimate a GMC radius of $R_A\simeq
28$~pc and a mass of $M_{G,A}\simeq 3.1\times10^5\:M_{\odot}$. No clouds
within the Cloud E spatial and velocity ranges were listed in the
RD10 catalog.  The integrated emission map indicates
that the GMC is highly substructured.  The IRDC is contained within a
small dense clump central to its parent GMC, but more massive clumps
are seen on scales beyond 5~pc. CE finds an extended region of clumpy
$\thco$ gas in the cloud's south-west region, beginning at 10~pc and
extending beyond the 30~pc boundary.

\underbar{\textit{Cloud F:}} For the cloud in the velocity range
$v_{\rm LSR}=54.3-65.6~\kms$, we estimate a radius of $R_A\simeq21$~pc
and a mass of $M_{G,A}\simeq7.8\times10^4\:M_{\odot}$, assuming a
distance of 3.7~kpc \citep{Rathborne2006}. Note, \citet{Kurayama2011}
estimated a parallax distance $1.56\pm0.12$~kpc, however this result
has been called into question by
\citet{Foster2012}. RD10 found two GMCs associated
with Cloud F: one cloud matches the location and velocity interval of
Cloud F's western massive clump, having a mass of
$\simeq6.9\times10^4\:M_{\odot}$ and radius of 12.3~pc. The other
molecular cloud is associated with the smaller eastern clump, for
which they estimate a mass of $\sim2.2\times10^4~\rm M_{\odot}$.  Our
analysis recovers this larger total mass, $\sim 10^5\:M_\odot$, if we
include all the $\thco$ emission within $v_0\pm15\:\kms$.
The larger molecular cloud complex surrounding Cloud F was recently
cataloged within the Giant Molecular Filament (GMF) study of
\citet{Ragan2014}. They found that their filament GMF38.1-32.4 is
actually the projection of two overlapping filamentary structures.
Cloud F is within the further filament (3.0-3.7~kpc), which has a
total length of 232~pc. The GMF contains multiple infrared dark
regions (identified in both GLIMPSE $8\mu$m and HiGAL $250\mu$m
images) and several dense gas clumps (identified by the
\citet{Wienen2012} $\rm NH_3$ and Bolocam Galactic Plane surveys),
demonstrating that overall it is in an early phase of star
formation. However, in Cloud F alone, there are two \HII~regions
within the two densest parts of the cloud with consistent LSR
velocities and distances, at $l=34.256, b=0.136$ and $l=34.821,
b=0.351$ \citep{Anderson2014}. While IRDC~F, which is also highly
filamentary, contains some of the highest column densities in the GMC,
the highest $\Sigma$ clump is actually another feature extending west
of the IRDC between 5 and 25~pc (Fig. \ref{mapF}). The velocity
distribution of the GMC for material out to 30~pc indicates that there
are four cloud components within our defined velocity interval
$v_0\pm15~\kms$. Figure \ref{PV2} shows the PV maps of the cloud out
to 30 pc and indicates that the majority of the $\thco$ emission
resides within the velocity interval of the IRDC
($v_{\rm LSR}=54.3-65.6~\kms$) and is connected to the weaker molecular
emission at surrounding velocities.  Additionally, the massive clump
and IRDC are kinematically associated with another dense clump on the
eastern edge of the cloud at $\sim25$ pc. The cloud defined by CE out
to a radius of 30 pc indicates that the regions above and below the
IRDCs parent cloud are filled with clumpy molecular material related
to the other three emission profiles.

\underbar{\textit{Cloud G:}} For the material in the velocity range of
$v_{\rm lsr}=39.7-51.8~\kms$ we estimate a cloud radius of $R_A\simeq
22$~pc and a mass of $M_{G,A}\simeq8.0\times10^4\:M_{\odot}$.
RD10 identified two clouds within the Cloud G
boundary with a total $\thco$ mass of
$\sim5.7\times10^4\:M_{\odot}$. 
At small cloud radii, $<10$~pc, Cloud G's velocity distribution is
reasonably well described by a single peak.  However, at larger radii
we clearly see three separate components within $v_0\pm15~\kms$. The
IRDC, and its parent GMC, are confined to the central velocity
interval of $v=39.7-51.8~\kms$.  In the GMC defined by SE, the IRDC
lies at the northern edge of the cloud's densest region, which extends
south $\sim18$ pc.  For the cloud defined by CE, this same cloud
region breaks up into three smaller clumps. The lower velocity
component of the cloud contains a dense molecular clump at
$l\simeq35.2,b\simeq-0.7$, a region which also contains a high level
of mid-infrared emission at $8\mu$m (Fig. \ref{mapG}). The WISE \HII~region
database lists a candidate \HII~region near the center of the
low-velocity clump with an estimated parallax distance of 2.2 kpc,
versus the Cloud G distance of 2.9~kpc.

\underbar{\textit{Cloud H:}} For gas in the velocity range of $v_{\rm
  lsr}=39.7-47.1~\kms$, we estimate a cloud radius of $R_A\simeq19$~pc
and a mass of $M_{G,A}\simeq5.6\times10^4\:M_{\odot}$. RD10
identified three molecular clouds at the location of Cloud H, with a
total mass of $\simeq2.8\times10^4\:M_{\odot}$. The PV diagrams
indicate that there are three separate cloud components within
$v_0\pm15\:\kms$ (Fig. \ref{PV2}). The IRDC and its parent GMC are
confined to the velocity interval of $v=39.7-47.1~\kms$, in which the
IRDC is the main structure of a larger diffuse cloud extending to the
west.  There are additional dense gas clumps at the neighboring
velocities: a large coherent clump about $25$~pc to the south at
$v=32.5-39.7~\kms$ (part of Cloud G), and a high concentration of
dense, clumpy gas along the entire northern region at velocities of
$v=47.1-62.6\:\kms$. These higher velocity clumps overlap with a
region of bright $8\mu$m dust emission (Fig. \ref{mapH}). The WISE
\HII~ database lists five \HII~regions overlapping these dense clumps
with similar LSR velocities \citep{Anderson2014}.

\underbar{\textit{Cloud I:}} We find that GMC I has a radius of $R_A\simeq
12$~pc and a mass of $M_{G,A}\simeq2.6\times10^4\:M_{\odot}$.
However, RD10 identified two clouds that match the location of Cloud I, and estimated the much
larger mass of $\sim2.4\times10^5~\rm M_{\odot}$. This difference is
due to the surrounding diffuse $\thco$ emission, which is included by
their cloud finding method that uses the CLUMPFIND algorithm
\citep{Rathborne2009}. IRDC I lies within the densest region of the
GMC, which extends roughly 7 pc north of the IRDC.  The cloud's SE
column density spectra indicates a symmetric and well isolated
emission profile out to radii of $\sim10$ pc.  However, at larger
cloud radii, diffuse $\thco$ gas at higher velocities is seen.
However, the spectra of the cloud defined by CE indicates that this
high-velocity emission is at intensities less than the $5\sigma$
threshold, suggesting that the GMCs is surrounded by a reservoir of
low-density molecular gas. For the cloud defined by CE, the central
dense region is broken down into three dense clumps and the GLIMPSE
$8\mu$m image indicates that there are multiple bright dust regions
along these dense $\thco$ clumps.  However, only one \HII~ region is
identified within the WISE \HII~ catalog, with a slightly lower LSR
velocity of $\sim38~\kms$.

\underbar{\textit{Cloud J:}}
We find that Cloud J has a radius of $R_A\simeq 13$~pc and a mass of 
$M_{G,A}\simeq2.0\times10^4\:M_{\odot}$.
However, Cloud J was included in the GMF catalog by
\citet{Ragan2014} in which they estimate a total cloud length of 79
pc.  They estimated a total cloud mass of $6.8\times10^4\:M_{\odot}$
(part of this difference results from the larger considered area; part
from assuming a smaller $n_{\rm 12CO}/n_{\rm H2}$ ratio of
$1.1\times10^{-4}$). Our analysis also finds that at the largest extraction radii, Cloud J 
is a filamentary cloud, extending $\sim60$~pc in the CE
case. Additionally, the co-added distribution of column density with
velocity at all cloud radii indicate that Cloud J is quite symmetric
and well isolated in velocity space. IRDC J, i.e., G053.11+00.05 from
\citet{Simon2006a}, defined the search region, but Cloud J also
includes the IRDC G053.31+00.00 within its 30~pc radius.
The $8\mu$m images of the Cloud J reveal a
significant amount of bright dust emission, which \citet{Ragan2014}
associated with a foreground high-mass star-forming region at
$v_{\rm LSR}\sim4~\kms$.  
Our search through the WISE \HII~ database found one radio quiet
\HII~region at $l=53.09,b=0.12$ with an LSR velocity consistent with
Cloud J \citep{Anderson2014}.

\subsubsection{Spatial and Velocity Offsets between IRDCs and GMCs}

The spatial and velocity offsets between the IRDC and GMC center of
masses are listed in Table~\ref{tab2}, quoted to the nearest
resolution element of the GRS position-velocity grid. In general,
spatial offsets grow as one considers larger GMC scales. At the 30~pc
aperture scale applied to connected extraction, this offset is
typically $\sim 2-10$~pc. All but clouds C and I have this GMC center
located outside the \citet{Simon2006a} IRDC boundary.

Velocity offsets are typically relatively small, $\sim1$~km/s. This
suggests that IRDCs have been formed by contraction/compression of
structures that were already pre-existing within the GMC.
Clouds A, B, C, E, I, J have relatively simple CE CO morphologies in
position velocity space. The remaining clouds exhibit more complex,
multi-component structures.

\begin{figure}[t!]
\begin{center}$
\begin{array}{c}
\includegraphics[width=7in, angle=0,trim=0 0.7in 0 1in]{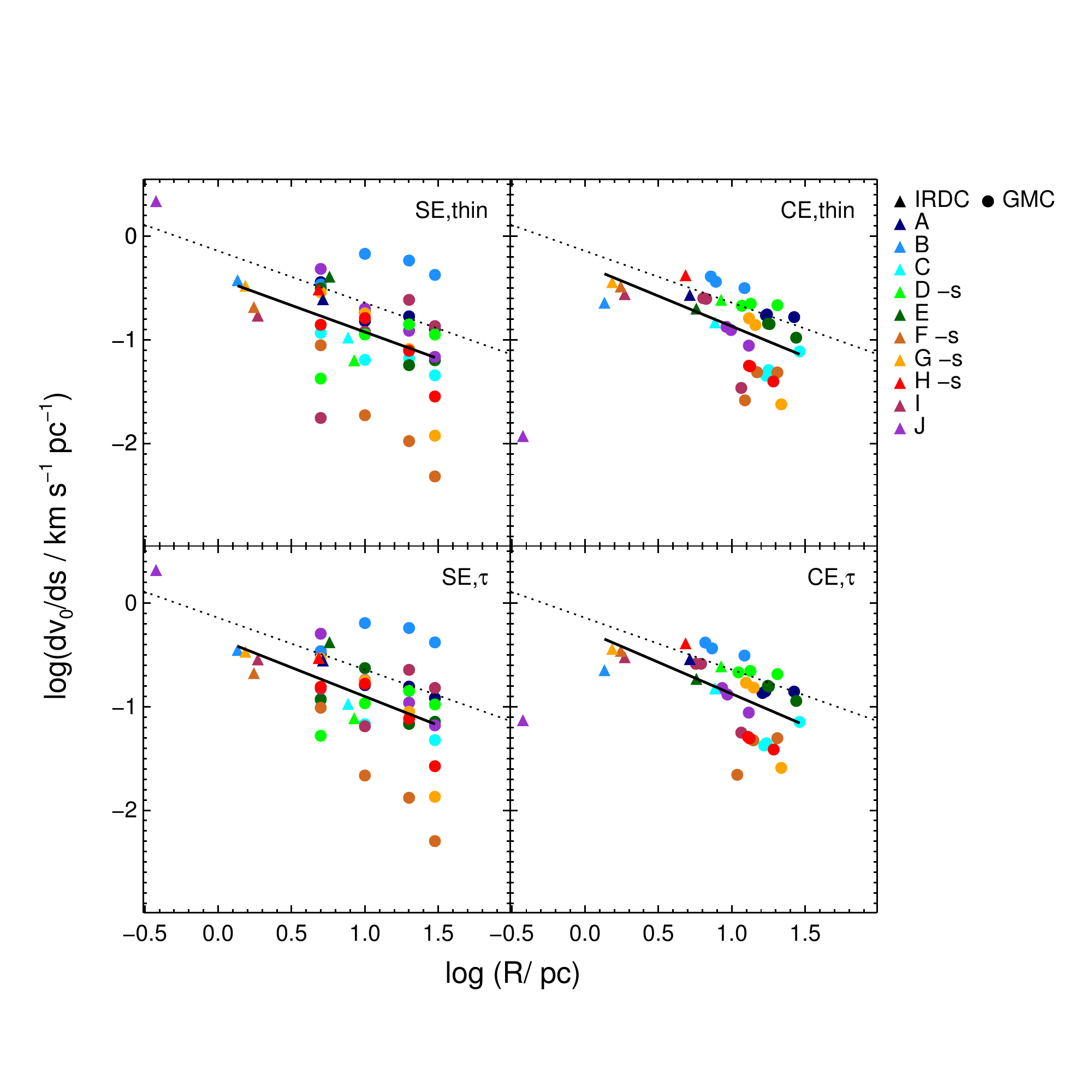}  
\end{array}$
\end{center}
\caption[]{
Velocity gradient $d v_0/ ds$ versus radius $R$ of the clouds over
which the gradient is evaluated. Each panel represents a different
cloud definition: {\it (a) Top left:} simple extraction, optically
thin column density derivation; {\it (b) Bottom left:} simple
extraction, opacity-corrected column densities; {\it (c) Top right:}
connected extraction, optically thin column density derivation; {\it
  (d) Bottom right:} connected extraction, opacity-corrected column
densities.
n each panel, the 10 IRDCs are shown by the colored filled triangles
as defined in the upper right-hand corner.  For Clouds D, F, G, and H,
these estimates are based on the single component emission profile.
GMCs are shown with filled circles.
The best-fit power-law relation from \citet[][]{McKeeOstriker2007}
($dv_0 / d s \simeq \sigma_{\rm pc} (s/{\rm pc})^{\alpha_\sigma - 1}$,
with values of $\alpha_\sigma\simeq 0.5$ and $\sigma_{\rm pc}\simeq
0.72\:\kms$) is shown by the dotted lines.
The global (IRDCs and GMCs, excluding IRDC J) best-fit power-law relation
for each panel is shown by the solid lines.}
\label{Fig:omega}
\end{figure}

\subsection{Kinematics and Dynamics}\label{S:kinematics}

\subsubsection{Velocity Gradients and Implications for Turbulence and Rotation}
The GMCs defined by CE have a mean velocity gradient of
$dv_{0,\tau}/ds\simeq 0.14\:\kms\:\rm pc^{-1}$. A similar value of
$0.13~\kms~\rm pc^{-1}$ was found by \citet{ImaraBlitz2011} for five
GMCs. The $\sim100$-pc-scale filaments have global velocity gradients
that are smaller than these values. \citet{Jackson2010} find ${\rm d}
v/{\rm d} s <0.09\:{\rm km\:s^{-1}\:pc^{-1}}$ in ``Nessie'',
\citet{BattersbyBally2012} find ${\rm d} v/{\rm d} s <0.05\:{\rm
  km\:s^{-1}\:pc^{-1}}$ in their 80-pc long cloud, and
\citet{Ragan2014} measure ${\rm d} v/{\rm d} s \simeq 0.06\:{\rm
  km\:s^{-1}\:pc^{-1}}$ as an average of the 7 filaments in their
sample.

\begin{figure}[ht!]
\begin{center}$
\begin{array}{c}
\includegraphics[width=6.5in, angle=0,trim=0 1in 0 0]{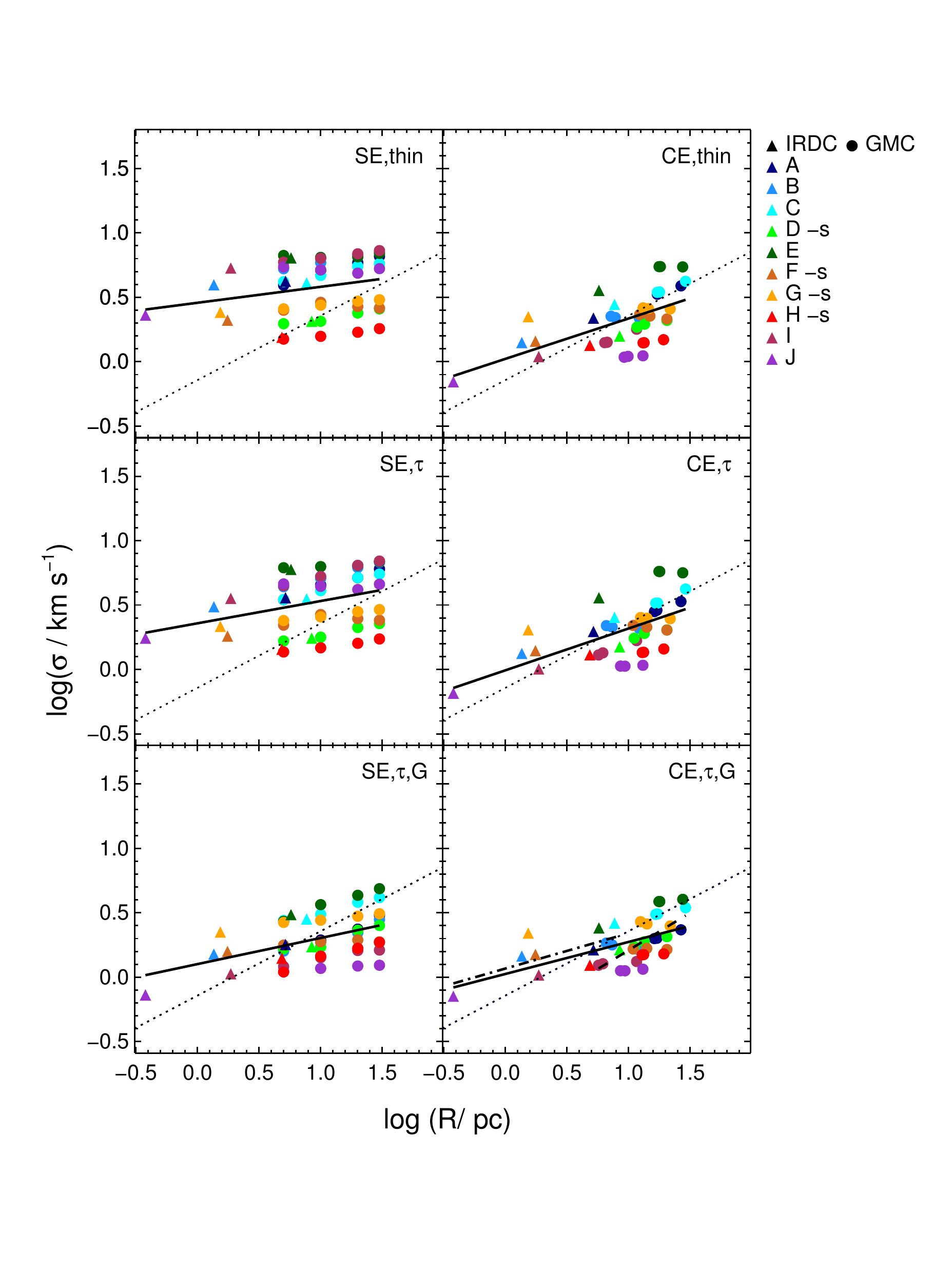}  
\end{array}$
\end{center}
\caption[]{
Velocity dispersion $\sigma$ versus radius $R$ of the clouds over
which the dispersion is evaluated. Each panel represents a different
cloud definition: 
{\it (a) Top left:} simple extraction, optically thin column density
derivation; {\it (b) Middle left:} simple extraction,
opacity-corrected column densities; {\it (c) Bottom left:} simple
extraction, opacity-corrected column densities fitted by a Gaussian
profile; {\it (d) Top right:} connected extraction, optically thin
column density derivation; {\it (e) Middle right:} connected
extraction, opacity-corrected column densities.  {\it (f) Bottom
  right:} connected extraction, opacity-corrected column densities
fitted by a Gaussian profile. 
In each panel, the 10 IRDCs are shown by
the colored filled triangles as defined in the upper right-hand
corner.  For Clouds D, F, G, and H, these estimates are based on the
single component emission profile. GMCs are shown with filled circles.  
The best-fit power-law relation from \citet[][]{McKeeOstriker2007} 
($\sigma =\sigma_{\rm pc} (s/{\rm
  1\:pc})^{\alpha_\sigma}$, with values of $\alpha_\sigma\simeq 0.5$
and $\sigma_{\rm pc}\simeq 0.72\:\kms$) is shown by the dotted
lines. The global (IRDCs and GMCs) best-fit power-law relation for
each panel is shown by the solid lines. In panel \textit{(f)}, the best-fit 
power-law relations are also shown for the IRDCs (dot-dashed line) and the GMCs 
(dashed line).}
\label{Fig:sigmaR}
\end{figure}

For the elliptical IRDC regions also defined by CE, we find a mean
velocity gradient of $dv_{0,\tau}/ds \simeq 0.26~\kms~\rm pc^{-1}$, 
larger than that of the GMCs. For comparison, on smaller scales within
IRDC H, \citet{Henshaw2014} find global velocity gradients of $0.08,
0.07, 0.30\:{\rm km\:s^{-1}\:pc^{-1}}$ in several sub-filaments
measured over $\sim$2 parsec length scales, based on centroid
velocities of the dense gas tracer $\rm N_2H^+(1-0)$. Larger local
gradients $\sim 1.5-2.5\:{\rm km\:s^{-1}\:pc^{-1}}$ can be present on
sub-parsec scales.
\citet{Ragan2012} found velocity gradients of 2.4 and 2.1~$\rm km
s^{-1} pc^{-1}$ within sub-pc regions of IRDCs G5.85-0.23 and
G24.05-0.22, based on observations of $\rm NH_3~(1,1)$.

Observed velocity gradients may be a signature of turbulence
\citep[e.g.,][]{BurkertBodenheimer2000,McKeeOstriker2007}.  For a
turbulent line-width-size (LWS) relation of the form $\sigma =
\sigma_{\rm pc} (s/{\rm 1\:pc})^{\alpha_\sigma}$, where values of
$\alpha_\sigma\simeq 0.5$ and $\sigma_{\rm pc}\simeq 0.72\:\kms$ have
been reported \citep[][]{McKeeOstriker2007}, we expect velocity
gradients measured over some scale $s$ to vary as $dv_0 / d s \simeq
\sigma_{\rm pc} (s/{\rm pc})^{\alpha_\sigma - 1}$.

Figure \ref{Fig:omega} displays our estimates of $dv_0/ds$ as a
function of cloud size for both IRDCs and GMCs 
(we note that the small size of IRDC J makes it difficult for velocity
gradients to be properly resolved with GRS data, so we exclude it from
the following analysis). Fitting the data for clouds defined by CE and
including opacity-corrected column densities (lower-right panel) we
find $\alpha_\sigma= 0.39\pm0.14$ and $\sigma_{\rm pc}=0.54\pm0.19\kms$. 
If each cloud is considered individually,
$\alpha_\sigma$ has mean and median values of $0.45$ and $0.51$,
respectively, and a standard deviation of 0.59. By inspection of the
other panels in Fig.~\ref{Fig:omega} we see that the results for the
slope and normalization of the power law relation are not too much
affected by the choice of cloud definition and mass measurement, but
the CE results show smaller dispersions.

We also examine directly the observed velocity dispersions as a
function of size scale (effective radius) of the clouds. Figure
\ref{Fig:sigmaR} displays estimates of $\sigma$ as a function of cloud
size for both IRDCs and GMCs. Now we see a larger effect on the
velocity dispersion versus size relation depending on how the cloud is
identified (i.e., SE or CE), how its mass is measured (i.e., whether
or not to carry out an opacity correction for the $^{13}$CO-derived
mass), and how to derive velocity dispersion (i.e., via a direct
estimate or via a fitted Gaussian). The CE, $\tau$-corrected,
Gaussian-fitted results yield a relatively tight power-law, with the
smallest dispersion.

Fitting the data for all clouds defined by CE and including
opacity-corrected column densities (lower-right panel) we find
$\alpha_\sigma=0.25\pm0.06$ and
$\sigma_{\rm pc}= 1.07\pm0.15\kms$. If each cloud is considered
individually, $\alpha_\sigma$ has a mean and median value of $0.17$
and $0.16$, respectively, and a standard deviation of 0.10.
If we fit IRDCs and GMCs separately, we find
$\alpha_\sigma=0.27\pm0.11$ and $\sigma_{\rm pc}= 1.17\pm0.16\kms$ for
IRDCs and $\alpha_\sigma=0.58\pm0.13$ and $\sigma_{\rm pc}=0.43\pm0.15\kms$ for GMCs.
For virialized, constant mass surface density clouds, where $\sigma
\propto (GM/R)^{1/2} \propto \Sigma^{1/2} R^{1/2}$ \citep{Larson1981,
  McKeeOstriker2007, McKee2010} one expects $\alpha_\sigma=0.5$ and
this is consistent with the GMC-only results. However, for IRDCs alone
or for the full sample including IRDCs, we see a shallower relation,
which could still be consistent with expectations from virial
equilibrium since IRDCs have higher mass surface densities. 
We will discuss the virialization of the clouds in more detail below
in \S\ref{S:virial}.

Observed velocity gradients within molecular clouds can also be
interpreted as due to cloud rotation
\citep[e.g.,][]{Goodman1993,ImaraBlitz2011}.  With such an
interpretation, we assess the projected rotational energy of the
clouds $E_{\rm rot,0}=(1/2)I_0\Omega_0^2$, as discussed in \S3.3.
The derived values of $I_0$ and $E_{\rm rot,0}$ are listed in Table~2.

We next compare rotational to gravitational energies.  Following
\cite[][, hereafter BM92]{Bertoldi1992}, the gravitational energy of
an ellipsoidal cloud is given by
\begin{equation}
W=\frac{3}{5}a_1 a_2 \frac{GM^2}{R},
\end{equation}
where parameter $a_1$ describes the effects of the internal density
distribution of the cloud, where for a power-law distribution
$\rho\propto r^{-k_\rho}$, $a_1=(1-k_{\rho}/3)/(1-2k_{\rho}/5)$.  
Here, for both GMCs and IRDCs,
we adopt $k_{\rho}=1$, such that $a_1=10/9$, based on the IRDC density
profile study by BT12. Somewhat steeper density profiles with
$k_\rho\simeq 1.5$ are derived by BT12 if envelope subtraction is
carried out, but this would only change $a_1$ by $\sim10\%$. The
parameter $a_2$ accounts for the effect for the cloud's
ellipticity. Following BM92 and our ellipsoidal cloud virial analysis
from HT11, an ellipsoidal cloud with radius $R$ normal to the axis of
symmetry and size $2Z$ along the axis has an aspect ratio of $y=Z/R$,
and semi-major and semi-minor axes of $r_{\rm max}$ and $r_{\rm min}$,
respectively. Since we do not know the inclination angles of the IRDCs
or GMCs, we assume that $R=r_{\rm min}$, $Z=r_{\rm max}$, and that the
observed cloud radius is equivalent to the geometric mean radius,
$R_{\rm obs}=(R_{\rm min}R_{\rm max})^{1/2}$.  These assumptions
differ from those of HT11, where we adopted fiducial inclination angle
values of $60^{\circ}$ for the filamentary IRDCs F and H. Now, with a
larger sample of both clumpy and filamentary IRDCs, we make no
assumption regarding their inclination angles. However, as noted in
HT11, an uncertainty of $15^{\circ}$ in inclination would lead to a
$\sim15\%$ uncertainty in the value of y.  We follow the BM92
definition of $a_2$, where
\begin{equation}
a_2= \frac{R_m}{R} \frac{{\rm arcsinh} (y^2-1)^{1/2}}{(y^2-1)^{1/2}},
\end{equation}
where, $R_m$ is the mean value of $R_{\rm obs}$ averaged over all
viewing angles.  For clouds with low aspect ratios ($y<10$), BM92 give
an approximation for the mean observed radius of $R_m/R\sim
y^{1/2}(1\pm0.27|1-y^{-2}|^{0.81})^{0.31}$, accurate to within
$2\%$. Since all of the IRDCs in this study have $y$ values below 10,
with IRDC B having the largest aspect ratio of $y=4.67$, the BM92
approximation of $a_2$ is valid for use in our gravitational energy
estimates.

We thus evaluate $W$, and then the ratio $\beta = E_{\rm rot}/|W|$,
which we also list in Tables \ref{tab2} and \ref{tabirdc}. We find
that the GMCs defined by CE and with opacity-corrected column density
estimates have mean $\beta_{\tau}\simeq 0.038$.  For the elliptical
IRDCs also defined by CE, we find mean $\beta_{\tau}\simeq
0.051$. These values are larger than the $\beta \sim 3\times 10^{-4}$
values estimated by \citet{Ragan2012} for two IRDCs. On much smaller
scales, the study by \citet{Goodman1993} found that most dense cores
within dark clouds have values of $\beta\le0.18$. The increasing
values of $\beta$ that we see on going from GMC to IRDC scales may be
indicative of slightly more disturbed kinematics within
IRDCs. However, it could also be caused by systematic changes in mass
estimation from $\rm ^{13}CO$, e.g., due to increased CO freeze-out
within IRDCs.

\subsubsection{``Rotation'' Directions}
\begin{figure}[t!]
\begin{center}$
\begin{array}{c}
\includegraphics[width=4.5in, angle=0, trim=0 0.6in 0.6in 0]{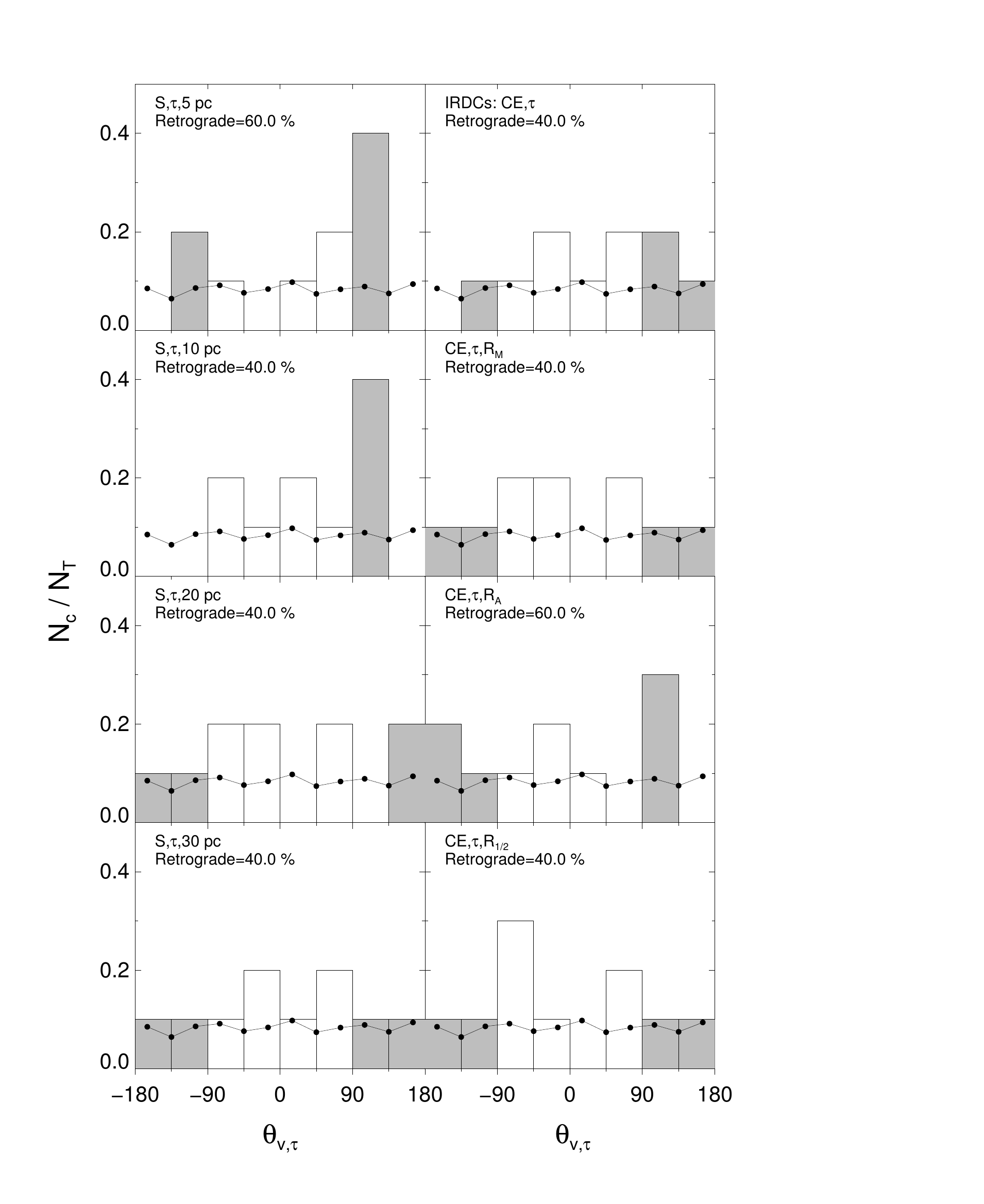}   
\end{array}$
\end{center}
\caption[]{
Distribution of relative position angle of the angular momentum
vectors of the 10 IRDCs/GMCs assuming opacity corrected column
densities with respect to the direction of Galactic rotation,
$\theta_{v,\tau}$. The effect of different cloud definitions is
illustrated in the different panels. In each panel, the black line
shows the results from the numerical simulation of
\citet{TaskerTan2009} for clouds at $\rm t=250~Myr$ and the
appropriate inclination of viewing angle of $0^{\circ}$ with respect
to the galactic plane, i.e., an in-plane view.}
\label{PA}
\end{figure}

Figure~\ref{PA} displays the distribution of position angles of
cloud angular momentum vectors, 
$\theta_{v,\tau}$, with respect to the
direction of Galactic rotation for the cloud definition cases assuming
opacity corrected column densities. Our results show that while a
cloud's position angle can vary depending on cloud definition, the
overall, global distribution of cloud position angles appears
consistent with near equal fractions of clouds with prograde and
retrograde rotation, although based on a relatively small sample of
only 10 analyzed clouds.

Young GMCs, recently formed from the diffuse ISM should possess
position angles which are prograde with Galactic rotation ($\theta_v <
-90^{\circ}$; $\theta_v > 90^{\circ}$). For all of the clouds,
measured at various radii for the SE and CE definitions, we find that
there is a large fraction of clouds with retrograde rotation
($-90^\circ<\theta_v < +90^{\circ}$). We find that $\sim50\%$ of the
CE clouds analyzed with opacity corrected column densities have
retrograde motion.

In the Milky Way, \citet{Phillips1999} compiled a database of
rotational measurements for 156 individual clouds from previously
published studies, finding that there is a random distribution between
the direction of a cloud's angular velocity and the Galactic plane.
However, isolated clouds had orientations towards the Galactic
poles. They concluded that the rotation of large-scale structures was
due to Galactic shear, while the randomized orientation of cloud
position angles was due to dynamical and/or magnetic interactions. By
estimating the velocity gradients for more than 500 clouds within the
GRS, \citet{Koda2006} found that the populations of prograde and
retrograde rotation were nearly equal. Beyond the Milky Way,
\citet{Rosolowsky2003} found that 40\% of the GMC population in M33
are in retrograde rotation with respect to the galaxy. In a subsequent
study of the \HI~ envelopes of 45 GMCs from \citet{Rosolowsky2003},
\citet{Imara2011} found that 53\% of these GMCs are in retrograde
rotation with respect to the local \HI.  Furthermore, they suggest
that if the linear gradients of the \HI ~envelopes are due to
rotation, then 62\% are in retrograde rotation with respect to M33.

In comparison to theoretical models, the disk galaxy simulations of
\citet{TaskerTan2009} found near equal amounts of apparent prograde
and retrograde clouds ($\sim50\%$) for simulated clouds at
$t=250$~Myr and a viewing angle of $0^{\circ}$ inclination to the
galactic plane, i.e., an in-plane view.
A substantial fraction of retrograde clouds results if GMCs live long
enough to undergo mergers and/or collisions with neighboring clouds.
\citet{Dobbs2013} performed 3D smooth particle hydrodynamic (SPH)
simulations of GMCs within a galactic disk, including self-gravity
and stellar feedback.  They found a retrograde GMC population of
$\sim40\%$,
similar to their previous studies \citep[e.g.,
][]{Dobbs2011b}.  However, the most massive GMCs tend to be prograde.

Studying the distribution of angular momentum vectors can test
scenarios of star formation that involve frequent galactic shear
driven cloud-cloud collisions \citep{Tan2000}. At the moment, with
only 10 clouds in our pilot sample, we are not yet in a position to
make detailed statistical comparison with numerical simulations. We
defer such an analysis, using larger samples of Galactic GMCs, to a
future paper.

\subsubsection{Cloud Virialization}\label{S:virial}
Various mechanisms, such as turbulence, magnetic fields, and thermal
gas pressure, may be responsible for supporting molecular clouds
against gravitational collapse \citep[e.g.,][]{McKeeOstriker2007}.
The dimensionless virial parameter, $\alpha_{\rm vir}$, describes a cloud's
dynamical state 
via \citep{Bertoldi1992}:
\begin{equation}
\alpha_{\rm vir} \equiv \frac{5\sigma^2 R}{GM} = 2a \frac{E_K}{|E_G|}
\label{eqn:alpha}
\end{equation}
where $\sigma$ is the 1D mass-averaged velocity dispersion, $a\equiv
|E_G| / (3GM^2/[5R])$ is the ratio of gravitational energy, $E_G$
(assuming negligible external tides), to that of a uniform sphere, and
$E_K$ is the kinetic energy. As discussed above in
\S\ref{S:kinematics}, for spherical clouds with power-law density
distribution, $\rho \propto r^{-k_\rho}$, $a=
(1-k_\rho/3)/(1-2k_\rho/5)$, i.e, rising as 1, 25/24, 10/9, 5/4, 5/3
when $k_\rho$ rises as 0, 0.5, 1, 1.5, 2, respectively.

\begin{figure}[t!]
\begin{center}$
\begin{array}{c}
\includegraphics[width=5in, angle=0, trim=0 1in 0 0.5in]{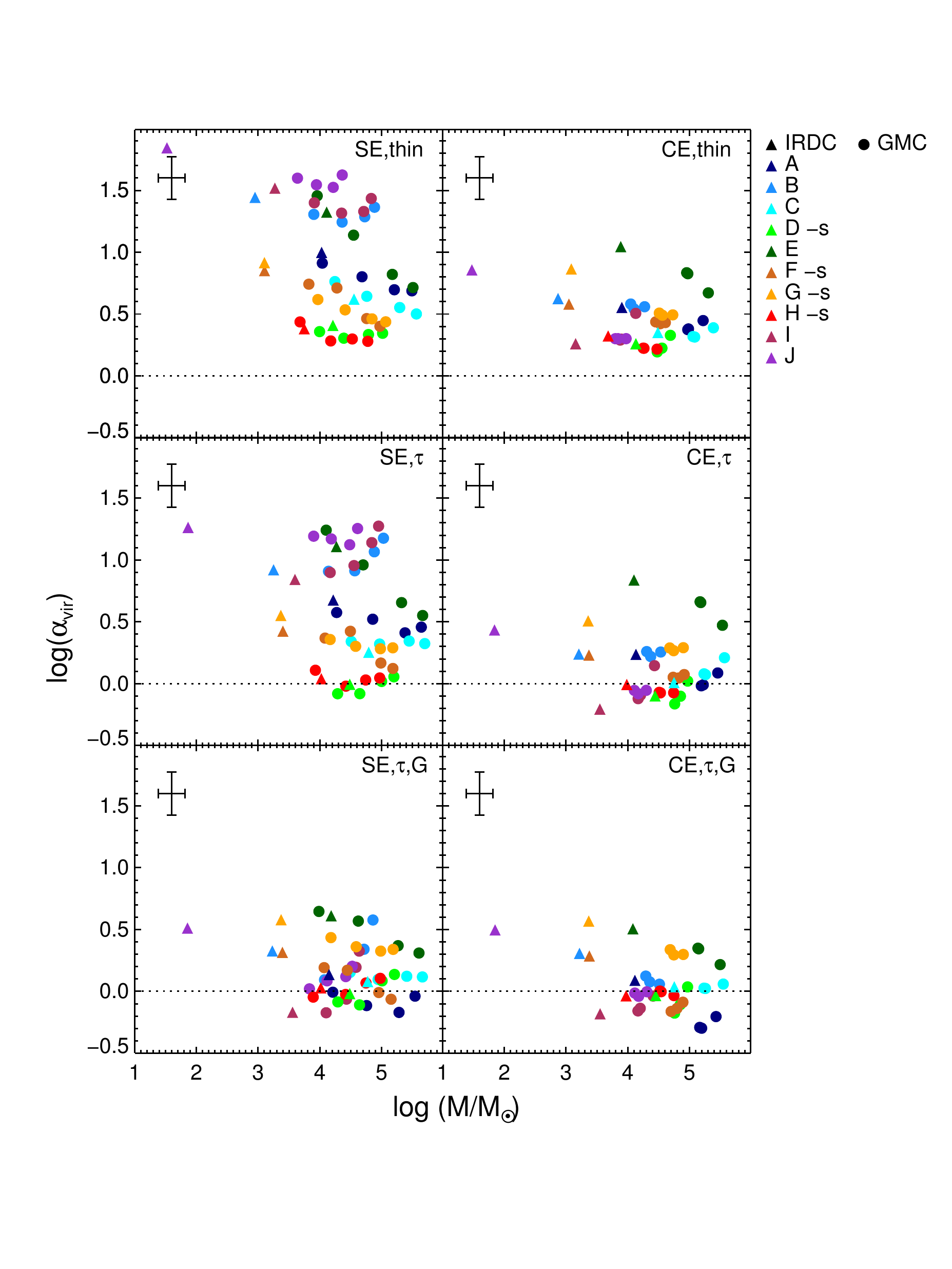} 
\end{array}$
\end{center}
\caption[]{
Comparison of cloud virial parameter, $\alpha_{\rm vir}$, with cloud
mass.  Each panel represents a different cloud definition: {\it (a)
  Top left:} simple extraction, optically thin column density
derivation; {\it (b) Middle left:} simple extraction,
opacity-corrected column densities; {\it (c) Bottom left:} simple
extraction, opacity-corrected column densities fitted by a Gaussian
profile; {\it (d) Top right:} connected extraction, optically thin
column density derivation; {\it (e) Middle right:} connected
extraction, opacity-corrected column densities.  {\it (f) Bottom
  right:} connected extraction, opacity-corrected column densities
fitted by a Gaussian profile. In each panel, the 10 IRDCs are shown by
the colored filled triangles (see legend).  For Clouds D, F, G, and H,
these estimates are based on the isolated single fitted emission
profile. The error bar in the top-left of each panel represents the mean uncertainties 
of $\sim40\%$ in $\alpha_{\rm vir}$ and $\sim50\%$ in $M$.
}
\label{alpha1}
\end{figure}

\begin{figure}[t!]
\begin{center}$
\begin{array}{cc}
\includegraphics[width=3.3in, angle=0, trim=0 1.2in 1in 0]{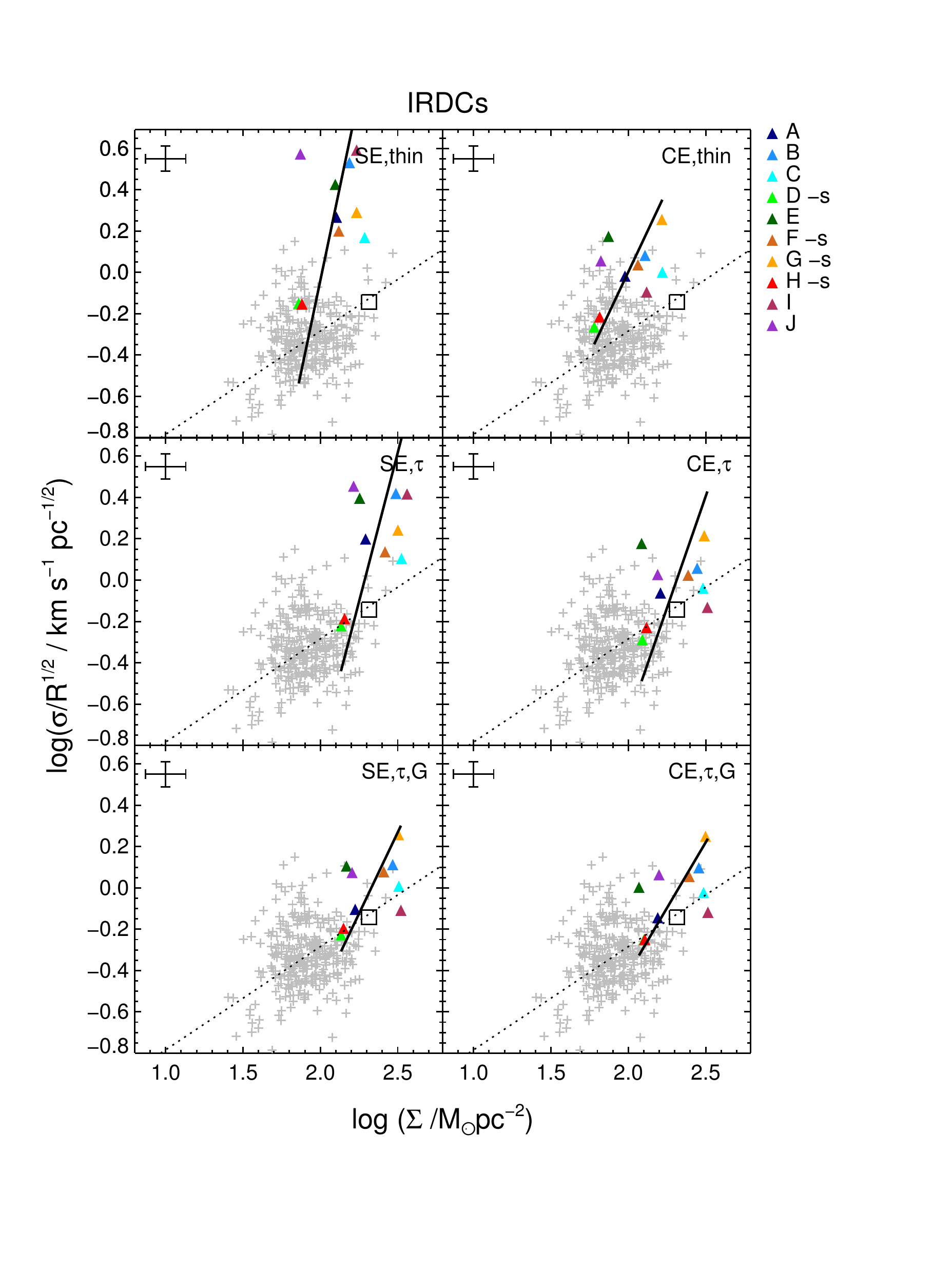}  & \includegraphics[width=3.3in, angle=0, trim=0 1.2in 1in 0]{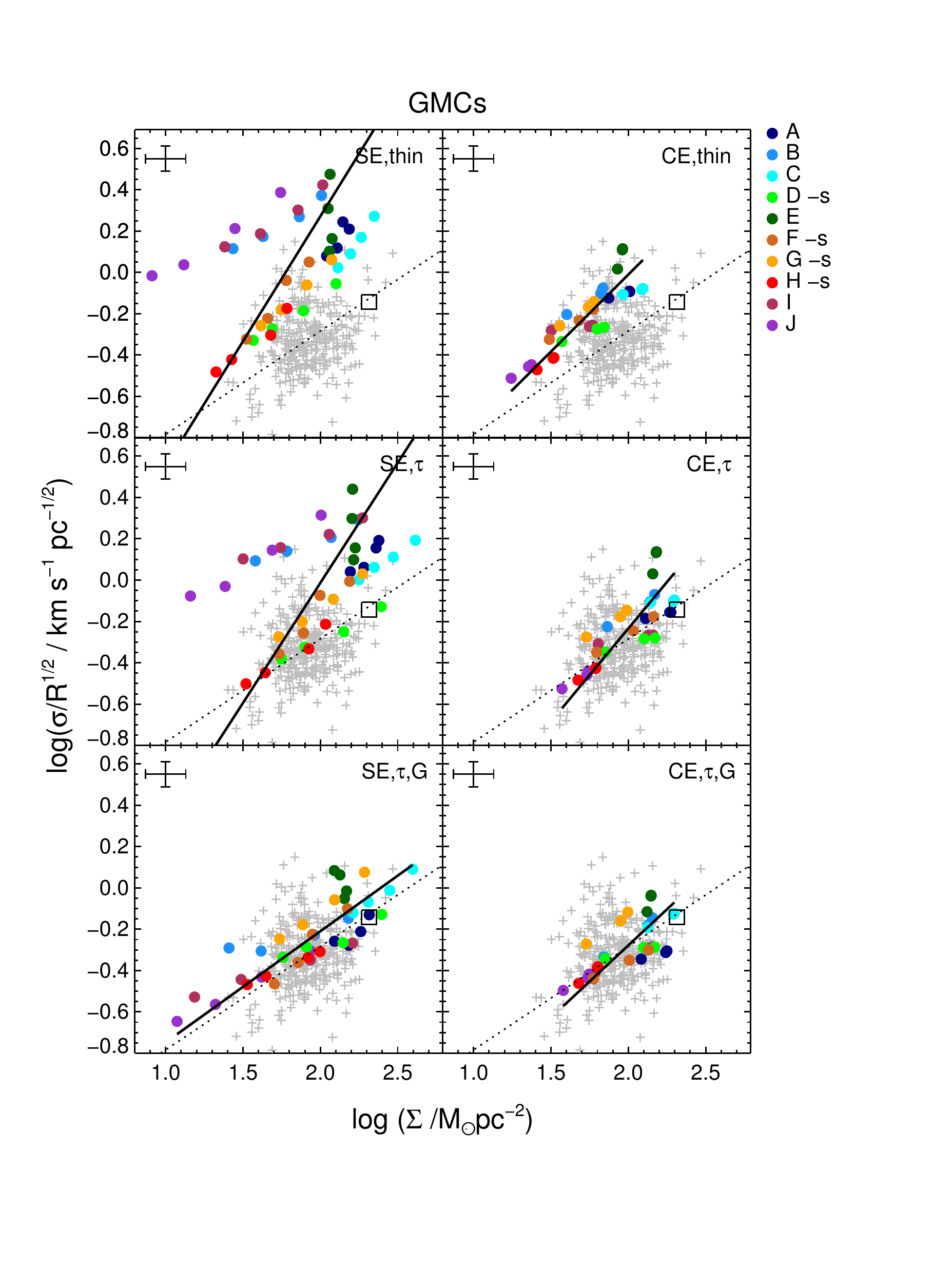}  
\end{array}$
\end{center}
\caption[]{
Dependence of $\sigma/R^{1/2}$ with mass surface density,
$\Sigma$. Results for the IRDCs (\textit{Left}) and GMCs
(\textit{Right}) are shown separately. Each panel represents a
different cloud definition: {\it (a) Top left:} simple extraction,
optically thin column density derivation; {\it (b) Bottom left:}
simple extraction, opacity-corrected column densities; {\it (c) Middle
  right:} connected extraction, optically thin column density
derivation; {\it (d) Middle right:} connected extraction,
opacity-corrected column densities.  {\it (c) Bottom right:} connected
extraction, optically thin column density, Gaussian fitted profile
derivation; {\it (d) Bottom right:} connected extraction,
opacity-corrected column density, Gaussian fitted profile
derivation. In each panel, the IRDCs/GMCs are shown by the colored
filled triangles/circles as defined in the upper right-hand corners.
For Clouds D, F, G, and H, these estimates are based on the single
component emission profiles.
The error bar in the upper-left corner represents the mean
uncertainties of $\sim14\%$ in $\sigma/R^{1/2}$ and $\sim30\%$ in
$\Sigma$.
The results of RD10, a total of 329 clouds ($M>10^{4}~\sm$) scaled by
a factor of 0.49 to reflect our adopted $^{13}$CO abundance
\citep[see][]{Tan2013} are shown by the gray crosses. The dotted line
represents virialized conditions with $\alpha_{\rm vir}=1$, i.e.,
$\sigma/R^{1/2}=(\pi G/5)^{1/2} \Sigma^{1/2}$ (see text). The mean
results of \citet{Solomon1987} are also shown (black open square).
Best-fit power-law relations ($\sigma/R^{1/2} \propto \Sigma^n$) are
shown by the solid lines.}
\label{heyer}
\end{figure}

Effects of nonspherical geometry for ellipsoidal clumps have also been
considered by BM92, and found to be relatively small ($\lesssim 10\%$)
for aspect ratios $Z/R\lesssim6$, where $Z$ is the clump radius along
the axis of symmetry and $R$ the radius normal to this
axis. Therefore, here we will ignore these effects of clump
elongation.

Ignoring surface pressure and magnetic terms, a cloud in virial
equilibrium has $|E_G|=2 E_K$, so that $\alpha_{\rm vir}=a$. Such a
cloud in free-fall has $\alpha_{\rm vir}\rightarrow 2a$ from below as
time progresses. In principle one can imagine a cloud starting with
very small levels of internal motion, so that $\alpha_{\rm vir}\ll1$,
and then crossing through the condition $\alpha_{\rm vir}=a$ as part
of its free-fall collapse. However, such a tranquil and synchronized
initial condition would seem to be difficult to achieve in reality if
GMCs are accumulating from a turbulent ISM, including potential
agglomerating via GMC collisions.

Although it is difficult to differentiate infall motions from
virialized turbulent motions, in cases where infall rates have been
claimed towards dense clumps, including IRDCs,
\citep[e.g.,][]{Peretto2013}, the infall velocities are relatively
small compared to virial motions (equivalent to the mass doubling
times being long ($\sim 10\times$) compared to the free-fall time; see
\citet[][]{Tan2014}, for a review). On the larger scales of GMCs,
\citet{ZuckermanEvans1974} and \citet{KrumholzTan2007} have argued
against fast rates of global collapse and infall, since overall star
formation rates are much smaller. However, converging flows associated
with the collision of independent, self-gravitating clouds may be
relatively common \citep{Tan2000,TaskerTan2009}, and the kinematic
signatures of such flows would be difficult to distinguish from global
infall.

The effect of surface pressure on the cloud increases the velocity
dispersion of the virial equilibrium state and thus raises the value
of its virial parameter. For example, for a cloud with $k_\rho=1.5$
and negligible surface pressure we have $\alpha_{\rm
  vir}=a=5/4$. However, if the cloud is embedded in an ambient medium
with the cloud's surface pressure equal to the ambient pressure, as in
the Turbulent Core model of \citet[][hereafter MT03]{McKeeTan2003},
then $\alpha_{\rm vir}= (5/2)(3-k_\rho)/[(5-2k_\rho)(k_\rho-1)] =
15/4$.
Note as $k_\rho\rightarrow 1$ from above,
$\alpha_{\rm vir}\rightarrow \infty$ (a self-similar singular
polytropic sphere with $k_\rho=1$ would have a spatially constant
pressure, i.e., zero pressure gradient, so there is no self-consistent
physical solution; MT03). These results show that virial equilibrium
pressure confined singular polytropic spherical clouds with $k_\rho
\lesssim 1.5$ have virial parameters $\gg 1$.

Support by large-scale magnetic fields acts to reduce the velocity
dispersion of the virial equilibrium state. For example, in the
fiducial MT03 Turbulent Core model, the mean velocity dispersion of
the virial equilibrium state is reduced by a factor
$\phi_B^{-1/2}\rightarrow 2.8^{-1/2}=0.598$ if allowing magnetic
fields strong enough that the mean Alfv\'en Mach number is unity. This
implies that the value of the virial parameter of the virial
equilibrium state would be reduced by a factor of 0.357.

Magnetic field observations in molecular clouds have been reviewed by 
\citet{Crutcher2012} and \citet{Li2014}. 
Field strengths increase with density approximately as $B\propto
\rho^{2/3}$ and are close to values expected for equipartition
magnetic and turbulent energy densities and also close (within a
factor of $\sim 2$) of the critical field strength that would prevent
collapse. It thus seems likely that magnetic fields can have
significant dynamical effects.  Recently, \citet{Pillai2015} have
found evidence for strong, dynamically important magnetic fields in
two IRDCs, based on relatively ordered directions of polarized sub-millimeter
emission, thought to be caused by dust grains that have aligned with
the $B$-fields. Unfortunately for the IRDCs/GMCs in our sample, we do
not have detailed information on magnetic field strengths.

For the GMCs and IRDCs analyzed in this paper, for each cloud, three
virial parameters were estimated.  First, we estimated $\alpha_{\rm
  vir}$ assuming optically thin conditions to derive $\sigma$ and
$M$. Second, using the opacity corrected column densities,
$\alpha_{\rm vir,\tau}$ is estimated using $\sigma_{\tau}$ and
$M_{\tau}$.  Finally, the virial parameter was estimated using the
Gaussian-fitted, opacity corrected column density spectra, i.e.,
$\alpha_{\rm vir,G}$ is estimated using $\sigma_G$ and $M_G$.
In the cases of clouds D, F, G, and H, the ``-s'' mass and velocity
dispersions were used to estimate the single component 
virial parameters of the clouds.
If cloud properties are assessed using the full $v_0\pm15\kms$
velocity range (including multiple components), then larger values of
$M$ and especially $\sigma$ are inferred, leading to estimates of
$\alpha_{\rm vir}$ that are larger by factors of about 3.
All molecular cloud virial parameters estimates are listed in Tables
\ref{tab2} and \ref{tabirdc}. We note that for CE-defined GMCs, the virial analysis is carried out
using cloud centers that are the center-of-mass of all the CE gas
within the 30~pc radius region.  The cloud masses are those that are
estimated from the material that is within the radii $R_M$, $R_A$ and
$R_{1/2}$. For SE-defined GMCs, the virial analysis uses the extracted
cloud radius as the characteristic size of the cloud.

For the same cloud definition cases, the virial parameter was also
estimated for each IRDC.  Since these clouds are elliptical, we used
the geometric mean observed radius, $R=\sqrt{(R_{\rm maj}R_{\rm
    min})}$, as an estimate of the cloud's equivalent spherical shape
(but note that, as discussed above, effects of asphericity are minor).

\begin{figure}[t!]
\begin{center}$
\begin{array}{c}
\includegraphics[width=6in, angle=0]{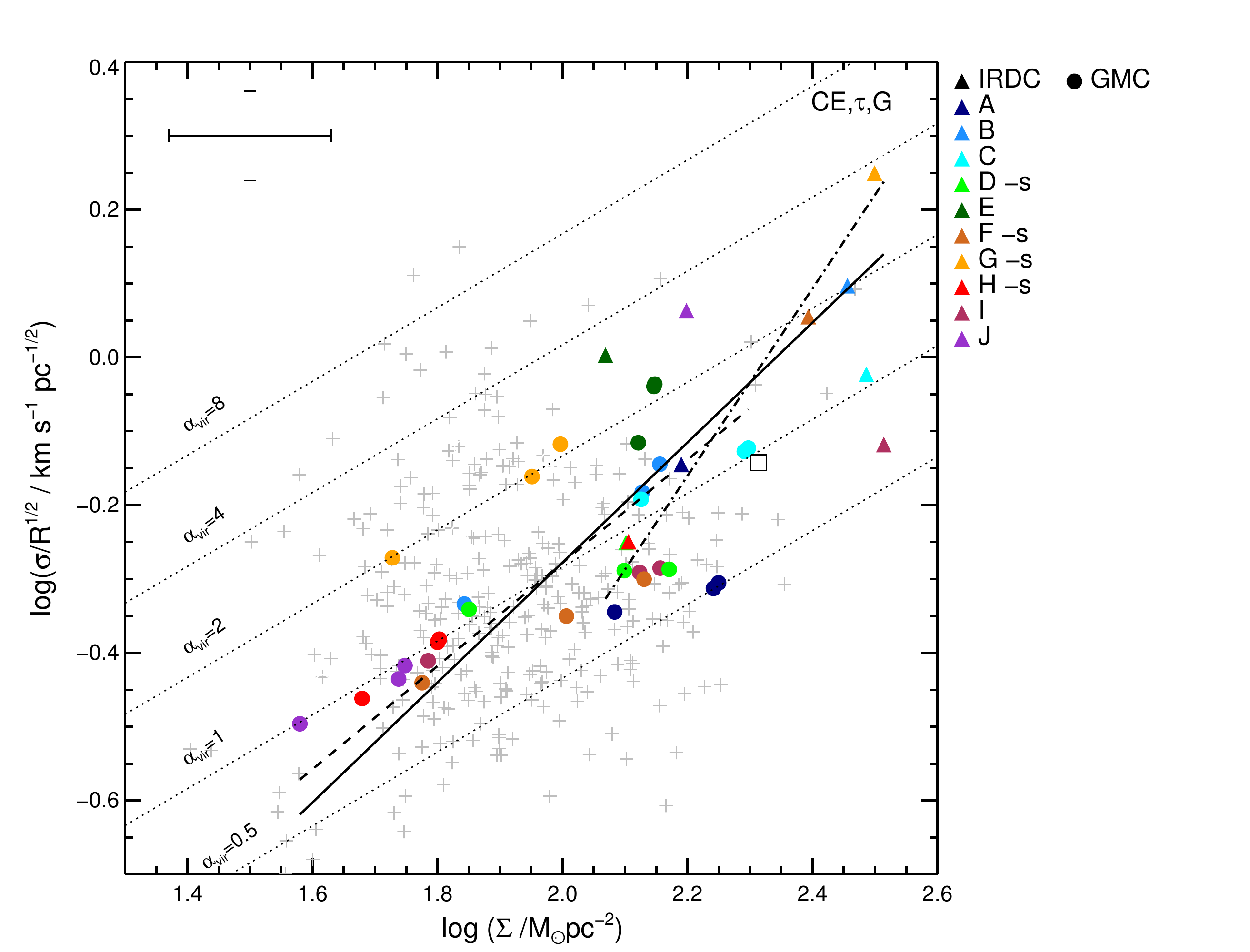}  
\end{array}$
\end{center}
\caption[]{
Dependence of $\sigma/R^{1/2}$ with mass surface density, $\Sigma$,
for the all clouds defined by CE, opacity corrected column densities,
and gaussian fitted velocity dispersions. As in Figure~\ref{heyer},
IRDCs are shown by the filled triangles and the GMCs by the filled
circles (see legend).
The error bar in
the upper-left corner represents the mean uncertainties of $\sim14\%$
in $\sigma/R^{1/2}$ and $\sim30\%$ in $\Sigma$.  The results of RD10,
a total of 329 clouds with $M>10^{4}~\sm$, are shown by the gray
crosses. The mean results of
\citet{Solomon1987} are also shown (open black square).  
The best-fit power-law relation ($\sigma/R^{1/2} \propto \Sigma^n$) is
shown for both the IRDCs
($n=1.27\pm0.52$; dot-dashed line) and the GMCs ($n=0.70\pm0.12$;
dashed line).  Collectively, the IRDCs and GMCs have a best-fit
power-law relation of $n=0.81\pm0.10$ (solid line).  The dotted lines
represents the expected scaling relation for virialized clouds with
different values of $\alpha_{\rm vir}=0.5$, 1, 2, 4, and 8.}
\label{CEtg}
\end{figure}

Figure \ref{alpha1} displays the virial parameters as a function of
mass for each cloud definition case.  
The overall range of $\alpha_{\rm vir}$ is nearly the same for both
IRDCs and GMCs.  We estimate a typical random uncertainty in
$\alpha_{\rm vir}$ of
$\sim40\%$ after accounting for uncertainties in the mass surface
densities, velocity dispersions, and the cloud kinematic distances.
However, systematic uncertainties in mass surface density of a factor
of $\sim2$ would lead to similar levels of uncertainty in $\alpha_{\rm
  vir}$.
Considering the different cloud definition methods, the virial
parameters estimated using the CE are lower than those estimated using
SE. For both cloud extraction methods, $\alpha_{\rm vir}$ is generally
smaller for the clouds using opacity corrected column densities
compared to those assuming optically thin gas conditions.  This trend
results from larger masses and slightly narrower line profiles in the
opacity-corrected cases. Similarly, the smallest $\alpha_{\rm vir}$
values were estimated in the cases of using the Gaussian fitted
spectra to measure both the velocity dispersion and mass, $\sigma_G$
and $M_G$.

Thus, the virial parameter is sensitive to the method used to extract
the $\thco$ emission associated with a GMC. Overall, the lowest values
of $\alpha_{\rm vir}$ were estimated for the clouds defined by CE,
opacity-corrected column densities, and Gaussian fitted masses and
velocity dispersions. The mean and median values of these estimates of
$\alpha_{\rm vir,G}$ are $1.12$ and $0.99$, respectively, with a
standard deviation of $0.51$ (similar to the expected size of the
random errors).  Similarly, for the IRDCs under the same definitions,
the mean and median values of $\alpha_{\rm vir,G}$ are $1.88$ and
$1.93$, respectively, with a standard deviation of $1.11$. 

We see that IRDCs have virial parameters that are about a factor of
two larger than GMCs. Again, this can be interpreted as evidence for
more disturbed kinematics. Alternatively, it could indicate that
$\thco$-derived masses for IRDCs have been underestimated, potentially
due to CO depletion.

A population of virialized clouds with similar degrees of pressure
confinement may be expected to have a relatively small dispersion in
their values of $\alpha_{\rm vir}$, although some variation is likely
due to variation in internal density structure and level of support by
large scale magnetic fields. If we denote the mean virial equilibrium
value of the virial parameters of such a cloud population as
$\bar{\alpha}_{\rm vir,eq}$, then from eq.~(\ref{eqn:alpha}) we can
derive
\begin{equation}
\frac{\sigma}{R^{1/2}} = \left(\frac{G\pi}{5}\right)^{1/2} \alpha_{\rm vir,eq}^{1/2} \Sigma^{1/2}
\label{eq:heyer}
\end{equation}
\citep[see also][]{Heyer2009,McKee2010}. By examining the observables
$\sigma/R^{1/2}$ and $\Sigma$ of our IRDC and GMC samples, we can see
if they follow this correlation that is expected if the clouds are
virialized. We estimate an uncertainty in $\sigma/R^{1/2}$ of
$\sim14\%$ based on $10\%$ uncertainties in both $\sigma$ and $R$.

Thus we considered the above scaling relationship for the results of
our ten IRDCs and GMCs (Fig. \ref{heyer}). For IRDCs, $\Sigma$ was
estimated for each ellipsoidal cloud using the geometric mean observed
radius $R$ and the total $\thco$-derived mass.  For GMCs, $\Sigma$ was
estimated using the circular area defined by the extraction radius $R$
and its corresponding $\thco$-derived mass. We estimate a random
uncertainty in $\Sigma$ of $\sim30\%$ based on the uncertainty in
$T_{\rm ex}$ and potential variations in our assumed
abundances. 
However, systematic uncertainties could be present at the
factor of $\sim2$ level.

For the ten GMCs, we find that the clouds analyzed using gaussian
profile fitting on the opacity corrected column density spectra to
estimate the cloud mass and velocity dispersion, yield results most
consistent with the scaling expected for gravitational virial
equilibrium. Specifically, we find the best-fit power-law relation
$\sigma/R^{1/2} \propto \Sigma^n$ has $n=0.54\pm0.05$ and
$n=0.70\pm0.12$, for the SE and CE extraction, respectively. We gauge
the significance of these fits by estimating the Pearson's correlation
coefficient, $\rho$, and the one-sided $p$-value. For all the GMCs
defined by CE and gaussian profile fitting on the opacity corrected
column density spectra, we estimate $\rho=0.71$ and a $p$-value of
$6.0\times 10^{-6}$, indicating a significant linearity in the
data. 

On the other hand, for the ten IRDCs analyzed using gaussian
profile fitting on the opacity corrected column density spectra to
estimate the cloud mass and velocity dispersion, yield steeper
best-fit power-law relations, with $n=1.57\pm0.80$ and
$n=1.27\pm0.52$, for the SE and CE extraction, respectively.  These
power-law fits are less significant than those seen for the GMCs, with
$\rho=0.56$ and a $p$-value of $0.047$ for the ``CE,$\tau$,G'' method.

Focusing on the ``CE,$\tau$,G'' extraction method,
these IRDC and GMC results are displayed together in Fig.~\ref{CEtg}.
Overall the IRDCs and GMCs together yield a best-fit power-law
relation of $n=0.81\pm0.10$, somewhat steeper than the $n=1/2$
expected for virialized clouds. If a power-law relation with $n=0.5$
is fit, then the normalization implies $\bar{\alpha}_{\rm vir}=1.60$
and 1.03 for IRDCs and GMCs, respectively, and $\bar{\alpha}_{\rm
  vir}=1.15$ for the whole sample (with a Pearson's correlation
coefficient of $\rho=0.78$ with a $p$-value of $1.7\times 10^{-9}$).
We note that some of the dispersion in the data is due to
cloud-to-cloud variation, which might be expected if there is
systematic variation in degree of surface pressure confinement,
magnetization, distance uncertainty (affecting $R$) or $\thco$
abundance.

We conclude that the relatively tight correlation shown by the data in
Fig.~\ref{CEtg} is evidence of cloud virialization over a wide range
of scales within GMCs, perhaps representing a self-similar hierarchy
of self-gravitating virialized structures. There is tentative evidence
that IRDCs deviate from this virialized scaling relation in the sense
of being more kinematically disturbed.
However, the level of deviation of the IRDCs is quite modest and their
best-fit $\bar{\alpha}_{\rm vir}\simeq 1.6$ is a value that is quite
compatible with models of virialized singular polytropic
($k_\rho\simeq 1.5$) spherical clouds embedded in a pressure confining
ambient medium. Such a model would yield $\bar{\alpha}_{\rm vir}=3.75$
(but quite sensitive to $k_\rho$). Including support from large-scale
magnetic fields that lead to Alfv\'en Mach numbers of unity would
lower this to $\bar{\alpha}_{\rm vir}=1.34$. However, given these
uncertainties in $k_\rho$, together with those from $\thco$ abundance,
using the absolute value of $\bar{\alpha}_{\rm vir}$ to assess the
degree of virialization is problematic, and we argue it is better to
focus on the degree of correlation in the $\sigma/R^{1/2}$ versus
$\Sigma$ diagram.

For comparison, the results of RD10 for a total of 329 clouds with
$M>10^{4}~\sm$ and with their $\Sigma$ values scaled by a factor of
0.49 to reflect our adopted $^{13}$CO abundance \citep[see][]{Tan2013}
are also shown in Fig.~\ref{CEtg}. The RD10 GMCs also follow the
expected virialized scaling, but with larger scatter. This is to be
expected given that we also see larger scatter with our simpler cloud
extraction methods. 
Consistent with our GMC results, the RD10 GMCs also have values of
$\bar{\alpha}_{\rm vir}\simeq 1.1$.

We also note here that the absolute values of $\Sigma$ that we derive
for GMCs are similar to those of H09 and RD10. H09 found an average
mass surface density of $\bar{\Sigma}\simeq 42\:\sm\:{\rm pc}^{-2}$
using 162 of the \citet{Solomon1987} $^{12}$CO-defined molecular
clouds (but they noted that the expected values are likely
underestimated by a factor of $\sim 2$ for these larger scale GMCs due
to sub-thermal excitation of $\thco$). Additionally, the 329 RD10
clouds with $M>10^{4}~\sm$ yield $\bar{\Sigma}\simeq 88\:\sm\:{\rm
  pc}^{-2}$.  For the 10
GMCs studied here, we find
$\bar{\Sigma}_{G}\simeq 80\:\sm\:{\rm pc}^{-2}$ assuming CE, opacity
corrected column densities, fitted gaussian line profile to derive
mass, and circular cloud radii defined by $R_A$.  For the 10 IRDCs,
defined by the Simon et al. (2006) ellipses, we find
$\bar{\Sigma}_{G}\simeq 217\:\sm\:{\rm pc}^{-2}$.
However, it is important to note that there is a real spread in
$\Sigma$ values between and among the considered IRDC and GMC scales
of greater than a factor of 10.

\section{Conclusions}

We have presented a detailed study of the $\thco$ gas on various size
scales out to radii of 30~pc around 10 well-studied IRDCs (i.e., the
sample of BT09 and BT12). We find that all the IRDCs are embedded in a
$\thco$-defined GMC, i.e., a molecular cloud with $M\gtrsim
10^4\:M_\odot$. We have taken care to assess the effects of cloud
definition, specifically comparing simple extraction of a slab in
position-velocity space with extraction only of connected emission
features. We have also taken notice of the effects of multiple
Gaussian components when they are present. We have considered column
density estimates via assuming, first, optically thin $\thco$(1-0)
emission (with excitation temperatures estimated from associated
$^{12}$CO emission) and then correcting for line optical depth
effects. Our preferred method of ``CE,$\tau$,G'' cloud definition
involves connected extraction, opacity-corrected column densities and
fitting a single Gaussian profile to derive velocity dispersion and
mass.

We discussed the particular morphologies of the individual clouds in
both position-velocity space and in relation to plane of sky imaging
of the IRDCs. We find most IRDCs are offset from their 30-pc-scale GMC
center by $\sim 2-10$~pc, but have only small velocity offsets of
$\sim 1\:{\rm km\:s^{-1}}$. This suggests IRDCs have been formed by
contraction/compression of structures that were already pre-existing
with the GMCs rather than being formed from material swept-up by a shock. Six of
the ten clouds have relatively simple morphologies in
position-velocity space, while the other four exhibit more complex,
multi-component structures.

We then focussed on the kinematic and dynamical properties of the
clouds. We measured velocity gradients as a function of size scale,
$s$, finding they decrease approximately as $s^{-1/2}$, i.e., consistent
with observations that IRDCs show larger velocity gradients than
GMCs. Overall, this is also consistent with the gradients being caused
by turbulence with a velocity dispersion versus size relation of the
form $\sigma \propto s^{1/2}$. Such a scaling is seen in our sample of
GMCs, but a somewhat shallower relation is apparent when considering
IRDCs and GMCs together. Using velocity gradients to define projected
rotation axes we do not see correlations with Galactic plane
orientation and have approximately equal fractions of pro- and
retrograde rotators with respect to Galactic orbital rotation (though
this analysis is limited by the small sample of only 10
clouds). We measure rotational energy to gravitational binding energy ratios of
$\beta\simeq 0.045$, with a slight increase seen on going from GMC to
IRDC scales.

Assessing virial equilibrium, we find cloud definition has a
significant impact on the virial parameter, i.e., the ratio of
internal kinetic to gravitational energy. CE,$\tau$,G-defined GMCs
have $\bar{\alpha}_{\rm vir}\simeq 1.0$, while IRDCs have
$\bar{\alpha}_{\rm vir}\simeq 1.6$. We argue that it is difficult to
assess the dynamical state, i.e., how close to virial equilibrium,
from the absolute values of $\bar{\alpha}_{\rm vir}$, especially given
the uncertain effects of pressure confinement, large-scale magnetic
field support and mass surface density estimation from
$\thco$. However, these values $\bar{\alpha}_{\rm vir}\sim 1.0$ to 1.6
are quite compatible with Turbulent Core/Clump/Cloud models of
singular polytropic spherical clouds that are pressure-confined by an
ambient medium and have moderate large scale $B$-field support that
yields Alf\'enic Mach numbers of order unity. Note, that
$\thco$-defined clouds are likely to have significant surrounding
molecular material that could provide pressure confinement.

Rather than focus on absolute values of $\bar{\alpha}_{\rm vir}$, we
follow H09 in examining the predicted scaling in the $\sigma/R^{1/2}$
versus $\Sigma$ plane. The results can be sensitive to the method of cloud definition. For
our preferred method of CE,$\tau$,G-defined clouds, the GMCs, analyzed
over a range of scales, show a strong correlation in this diagram,
with a power law slope of $\sigma/R^{1/2} \propto \Sigma^n$ with
$n=0.70\pm0.12$, similar to the $n=1/2$ value expected under virial
equilibrium. Potentially this represents a self-similar hierarchy of
self-gravitating virialized structures.

However, the IRDCs exhibit a somewhat steeper scaling with
$n=1.27\pm0.52$ (or $n=0.81\pm0.10$ for the combined GMC and IRDC
sample). This may indicate that IRDCs have moderately more disturbed
kinematics than GMCs, which is qualitative expectation of models in
which star formation is triggered by GMC-GMC collisions 
\citep[e.g.,][]{Scoville1986,Tan2000}. However, it is also possible that
systematic uncertainties affecting mass surface density estimation
from $\thco$, such as increased CO freeze-out in IRDCs \citep[e.g.,][]{Hernandez2011,
Hernandez2012}, are playing a role. Improved methods of
measuring mass and assessing degree of CO freeze-out in these clouds
are needed to help distinguish these possibilities and thus reliably
measure the dynamical state of gas that is likely to form star
clusters.

Additional future work will involve applying the techniques developed
here to larger samples of clouds to better study connections with
Galactic dynamics and Galactic environment, and potential systematic
differences of GMCs that are forming IRDCs or that have relatively
high dense gas mass fractions.

\acknowledgements A.K.H acknowledges support from grants awarded 
to Bart Wakker at the University of Wisconsin-Madison.  
We thank Brian Babler for contributing to the GLIMPSE data
processing. This publication makes use of molecular line data from the
Boston University-FCRAO Galactic Ring Survey (GRS). The GRS is a joint
project of Boston University and Five College Radio Astronomy
Observatory, funded by the National Science Foundation under grants
AST-9800334, AST-0098562, and AST-0100793.
J.C.T. acknowledges support from NASA grants ATP09-0094 and
ADAP10-0110.

\newpage
\appendix
\section{CLOUD MAPS: B-J}
\begin{figure}[ht!]
\begin{center}$
\begin{array}{c}
\includegraphics[width=7.5in, angle=0]{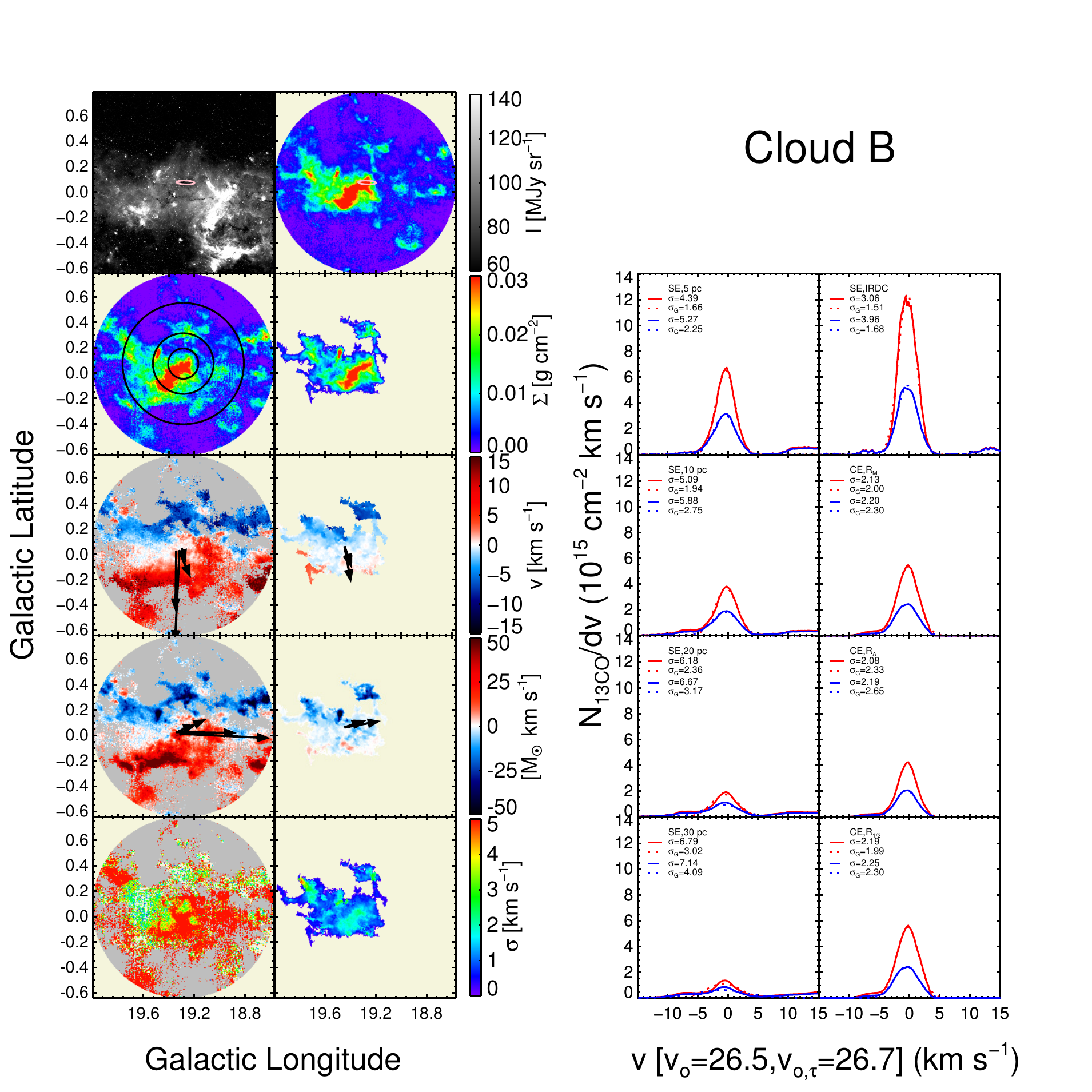}  
\end{array}$
\end{center}
\caption{
Same as Fig. \ref{mapA}, but for Cloud B.  The $\thco$(1-0) GRS
integrated intensity map of the IRDC emission profile is over the
velocity range: $v_{\rm LSR}=20.7-31.6~\kms$.}
\label{mapB}
\end{figure}
\newpage

\begin{figure}[ht!]
\begin{center}$
\begin{array}{c}
\includegraphics[width=7.5in, angle=0]{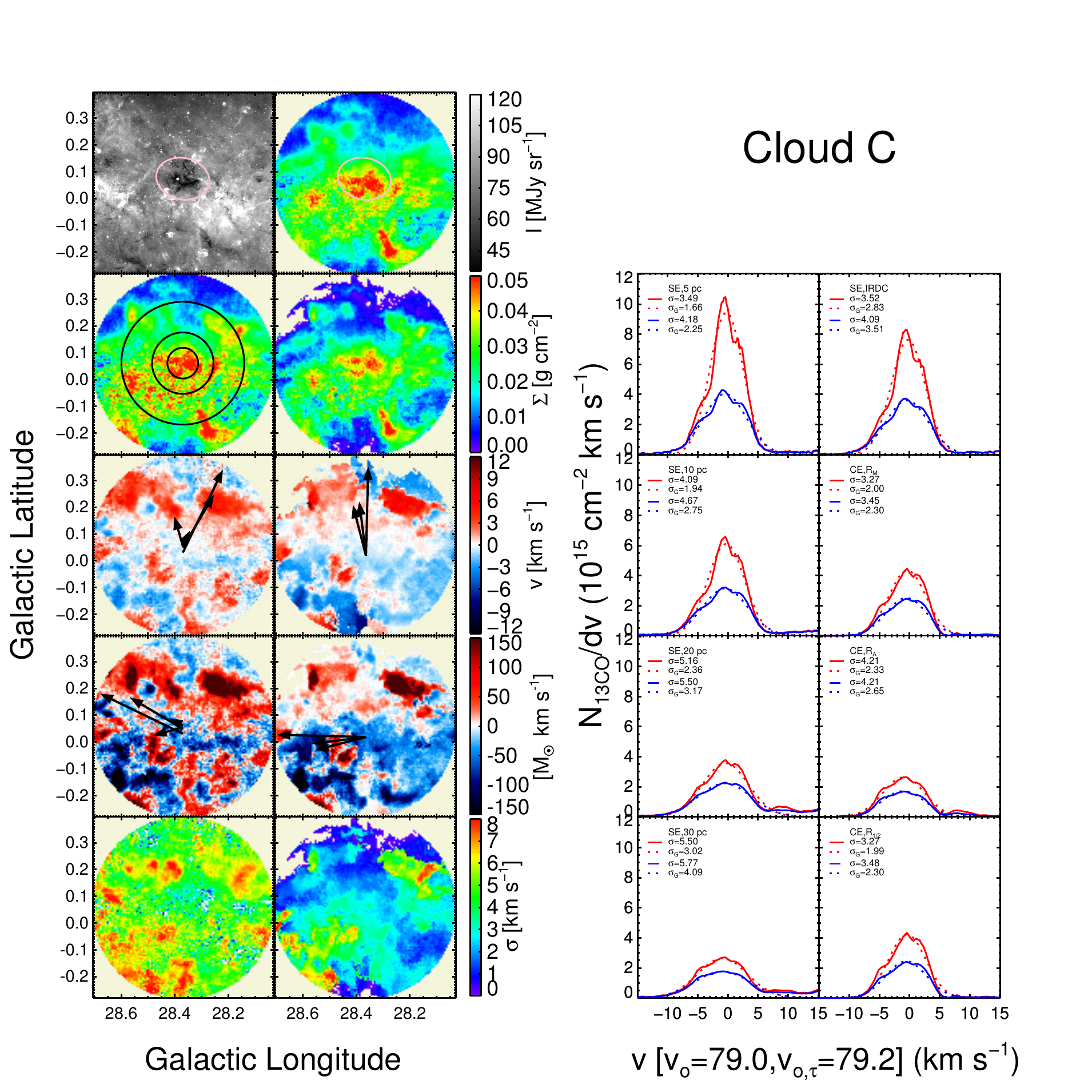}  
\end{array}$
\end{center}
\caption{
Same as Fig. \ref{mapA}, but for Cloud C. The $\thco$(1-0) GRS
integrated intensity map of the IRDC emission profile is over the
velocity range: $v_{\rm LSR}=66.3-85.7~\kms$.}
\label{mapC}
\end{figure}
\newpage

\begin{figure}[ht!]
\begin{center}$
\begin{array}{c}
\includegraphics[width=7.5in, angle=0]{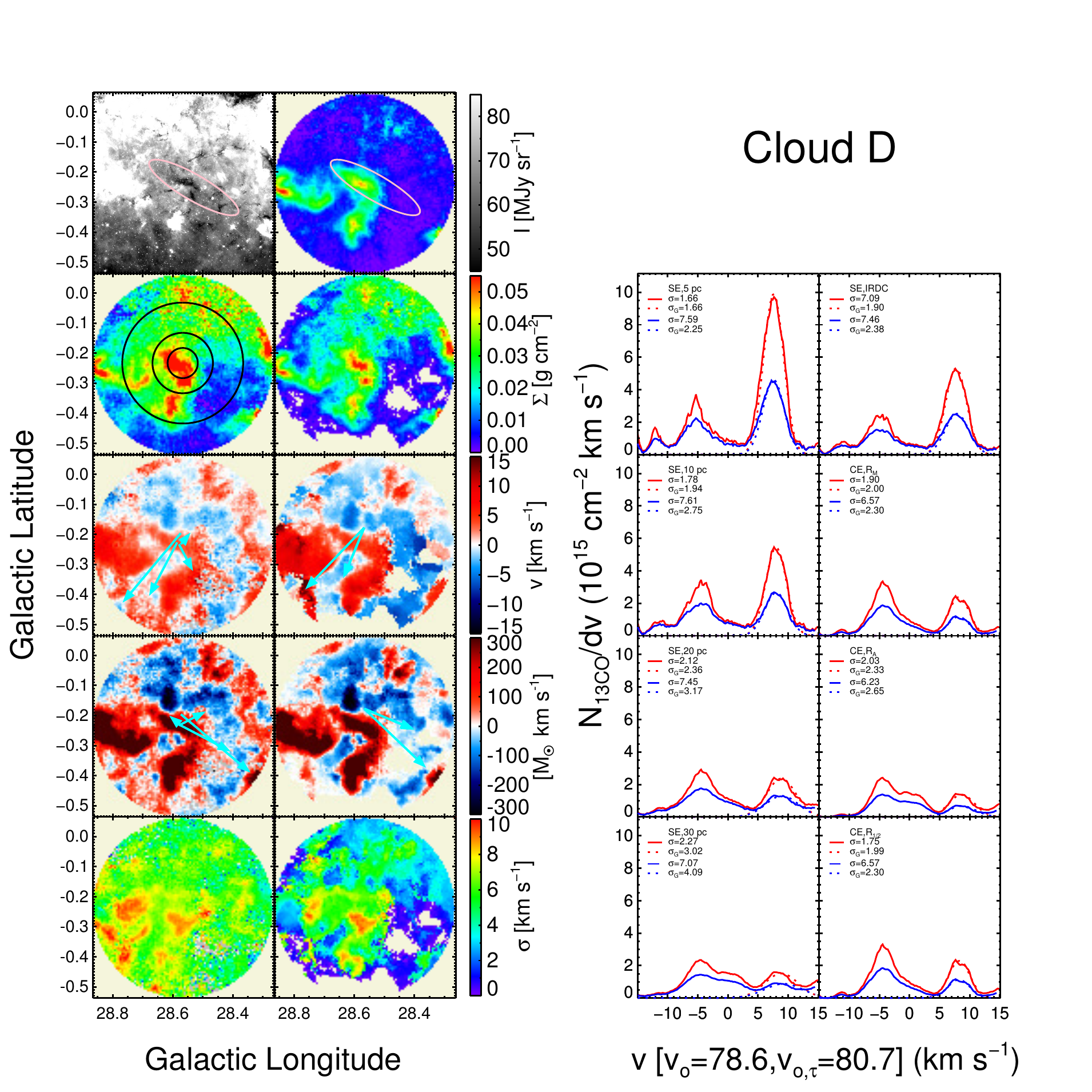}  
\end{array}$
\end{center}
\caption{
Same as Fig. \ref{mapA}, but for Cloud D. The $\thco$(1-0) GRS
integrated intensity map of the IRDC emission profile is over the
velocity range: $v_{\rm LSR}=84.1-93.7~\kms$.}
\label{mapD}
\end{figure}
\newpage

\begin{figure}[ht!]
\begin{center}$
\begin{array}{c}
\includegraphics[width=7.5in, angle=0]{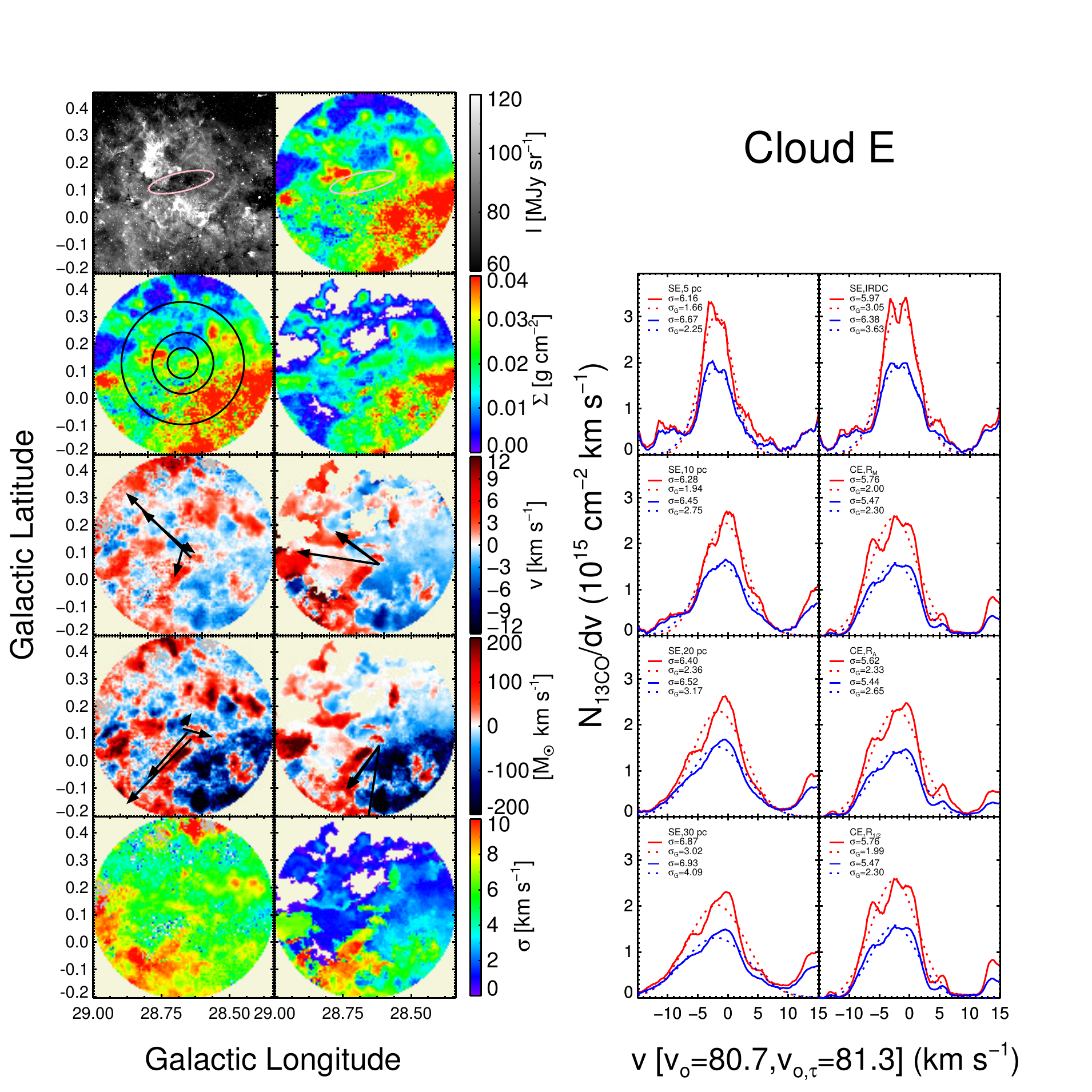}  
\end{array}$
\end{center}
\caption{
Same as Fig. \ref{mapA}, but for Cloud E. The $\thco$(1-0) GRS
integrated intensity map of the IRDC emission profile is over the
velocity range: $v_{\rm LSR}=66.5-89.8~\kms$.}
\label{mapE}
\end{figure}
\newpage

\begin{figure}[ht!]
\begin{center}$
\begin{array}{c}
\includegraphics[width=7.5in, angle=0]{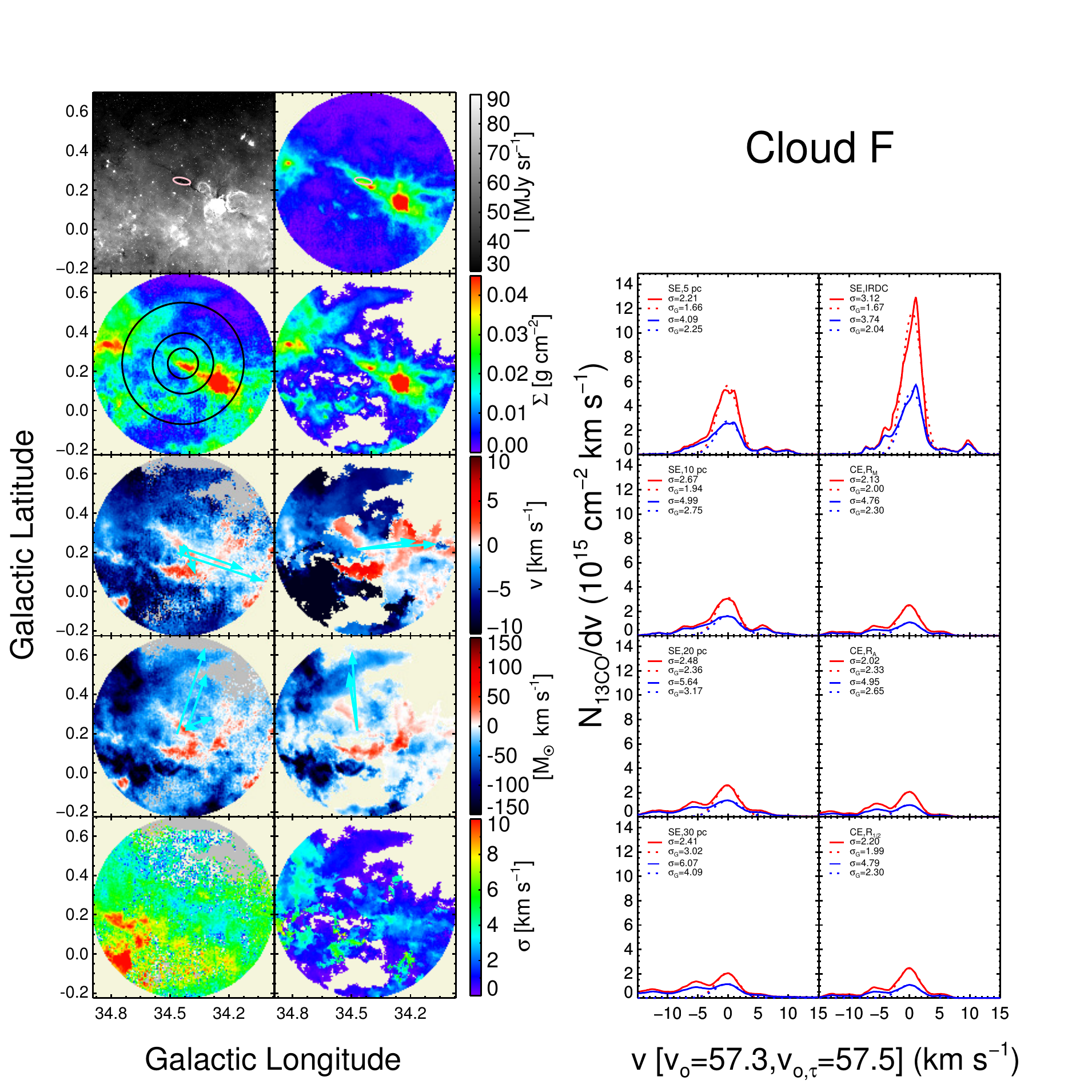}  
\end{array}$
\end{center}
\caption{
Same as Fig. \ref{mapA}, but for Cloud F. The $\thco$(1-0) GRS
integrated intensity map of the IRDC emission profile is over the
velocity range: $v_{\rm LSR}=54.3-65.6~\kms$.}
\label{mapF}
\end{figure}
\newpage

\begin{figure}[ht!]
\begin{center}$
\begin{array}{c}
\includegraphics[width=7.5in, angle=0]{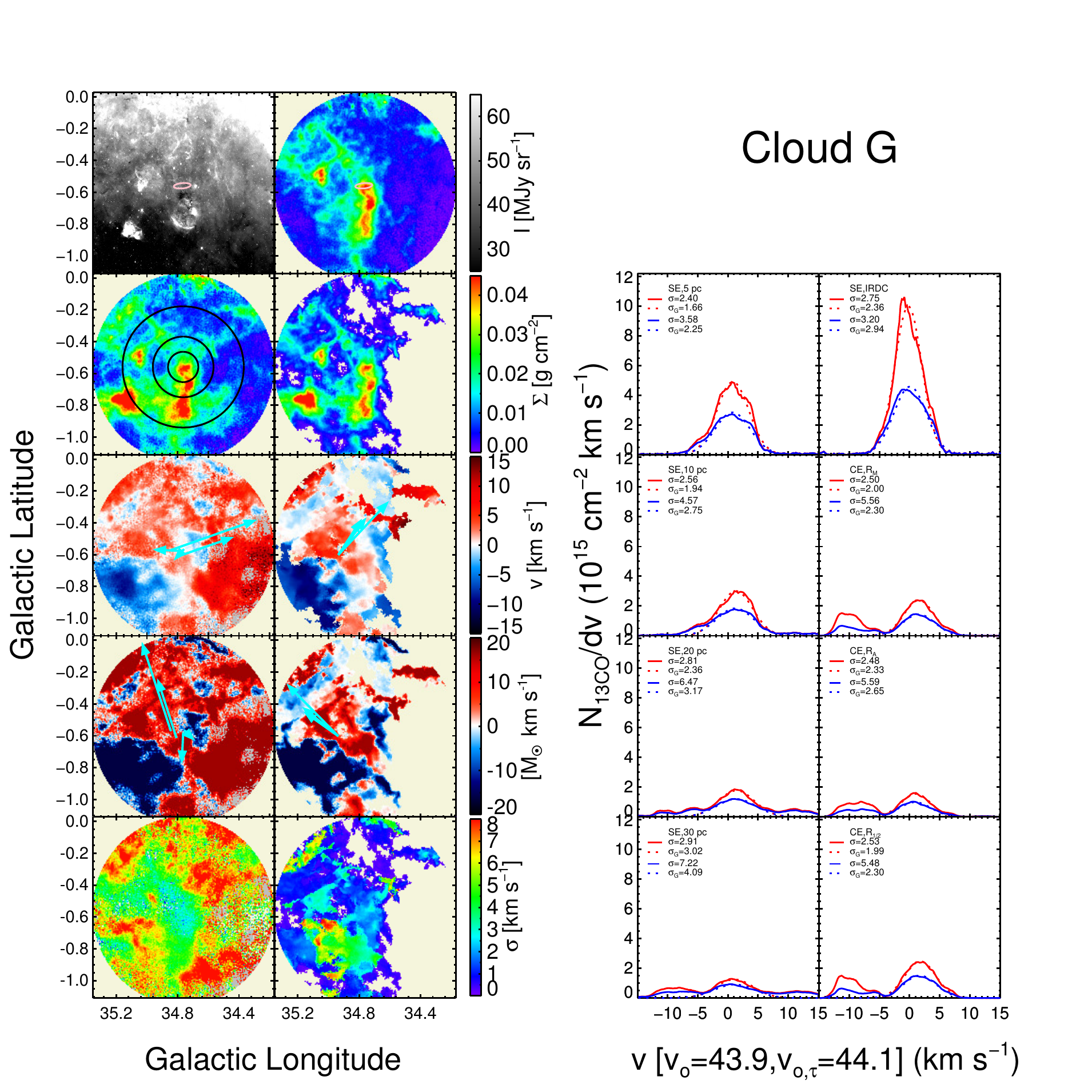}  
\end{array}$
\end{center}
\caption{
Same as Fig. \ref{mapA}, but for Cloud G.  The $\thco$(1-0) GRS
integrated intensity map of the IRDC emission profile is over the
velocity range: $v_{\rm LSR}= 39.7-51.8~\kms$.}
\label{mapG}
\end{figure}
\newpage

\begin{figure}[ht!]
\begin{center}$
\begin{array}{c}
\includegraphics[width=7.5in, angle=0]{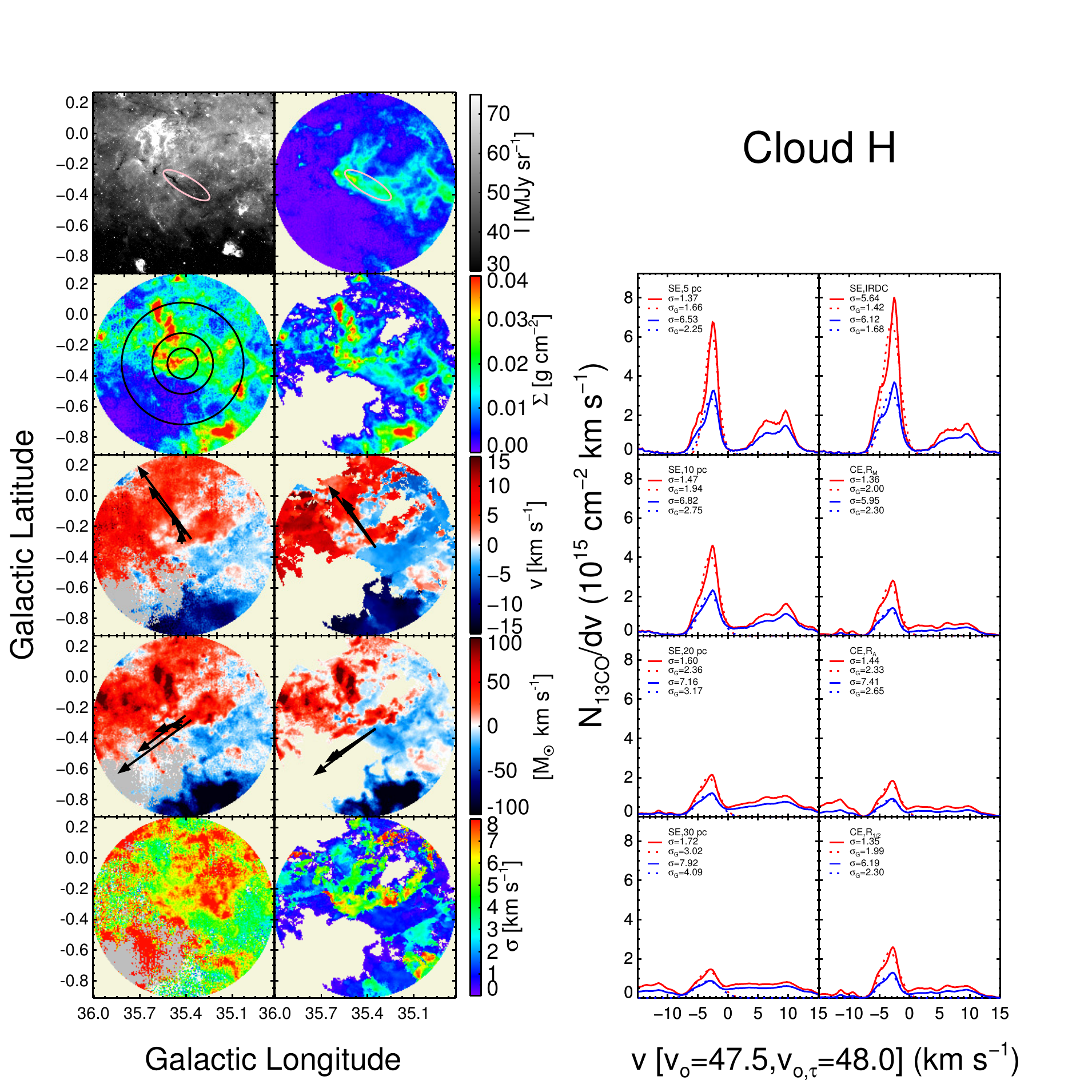}  
\end{array}$
\end{center}
\caption{
Same as Fig. \ref{mapA}, but for Cloud H.  The $\thco$(1-0) GRS
integrated intensity map of the IRDC emission profile is over the
velocity range: $v_{\rm LSR}=39.7-47.1~\kms$.}
\label{mapH}
\end{figure}
\newpage

\begin{figure}[ht!]
\begin{center}$
\begin{array}{c}
\includegraphics[width=7.5in, angle=0]{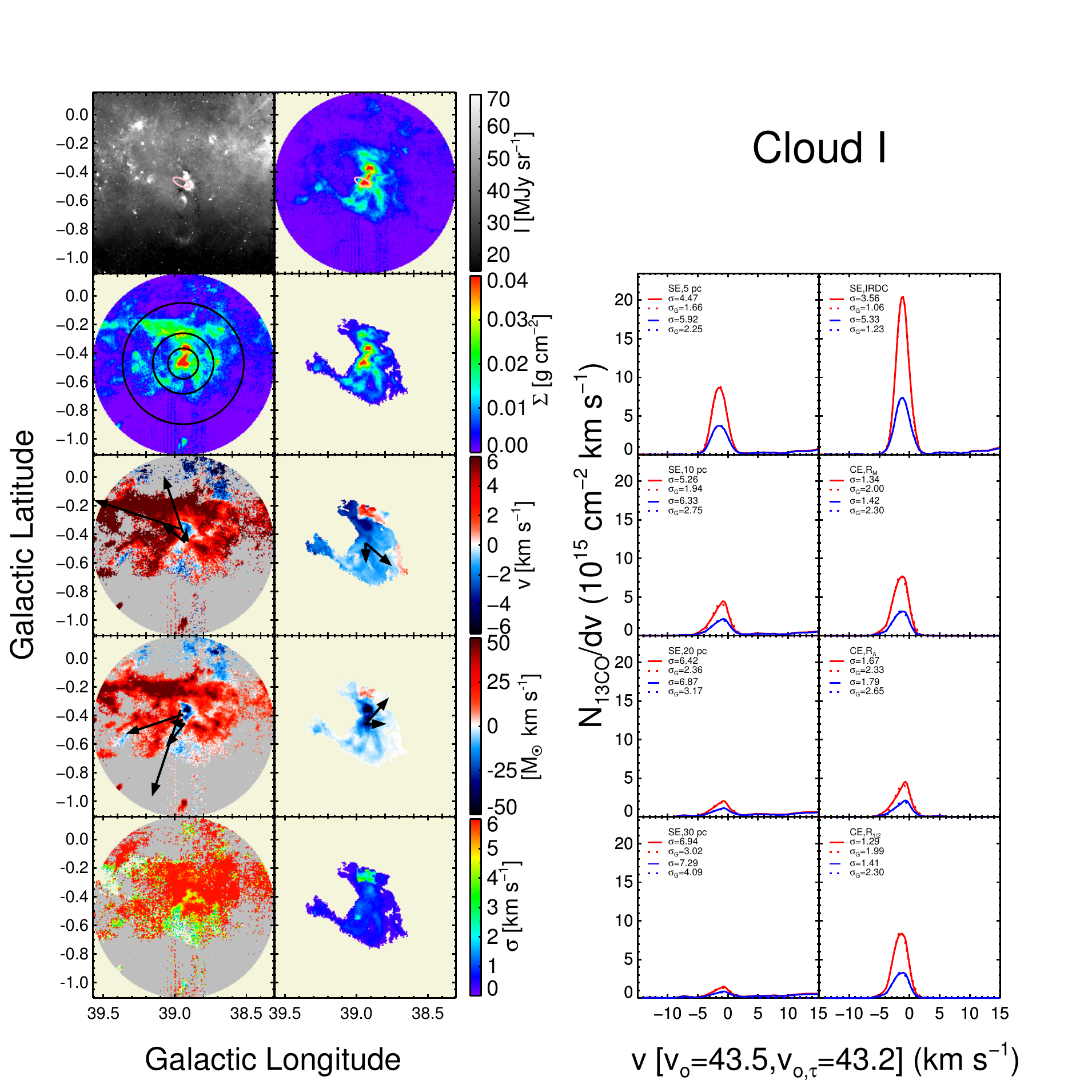}  
\end{array}$
\end{center}
\caption{
Same as Fig. \ref{mapA}, but for Cloud I.  The $\thco$(1-0) GRS
integrated intensity map of the IRDC emission profile is over the
velocity range: $v_{\rm LSR}=37.0-45.1~\kms$.}
\label{mapI}
\end{figure}
\newpage

\begin{figure}[ht!]
\begin{center}$
\begin{array}{c}
\includegraphics[width=7.5in, angle=0]{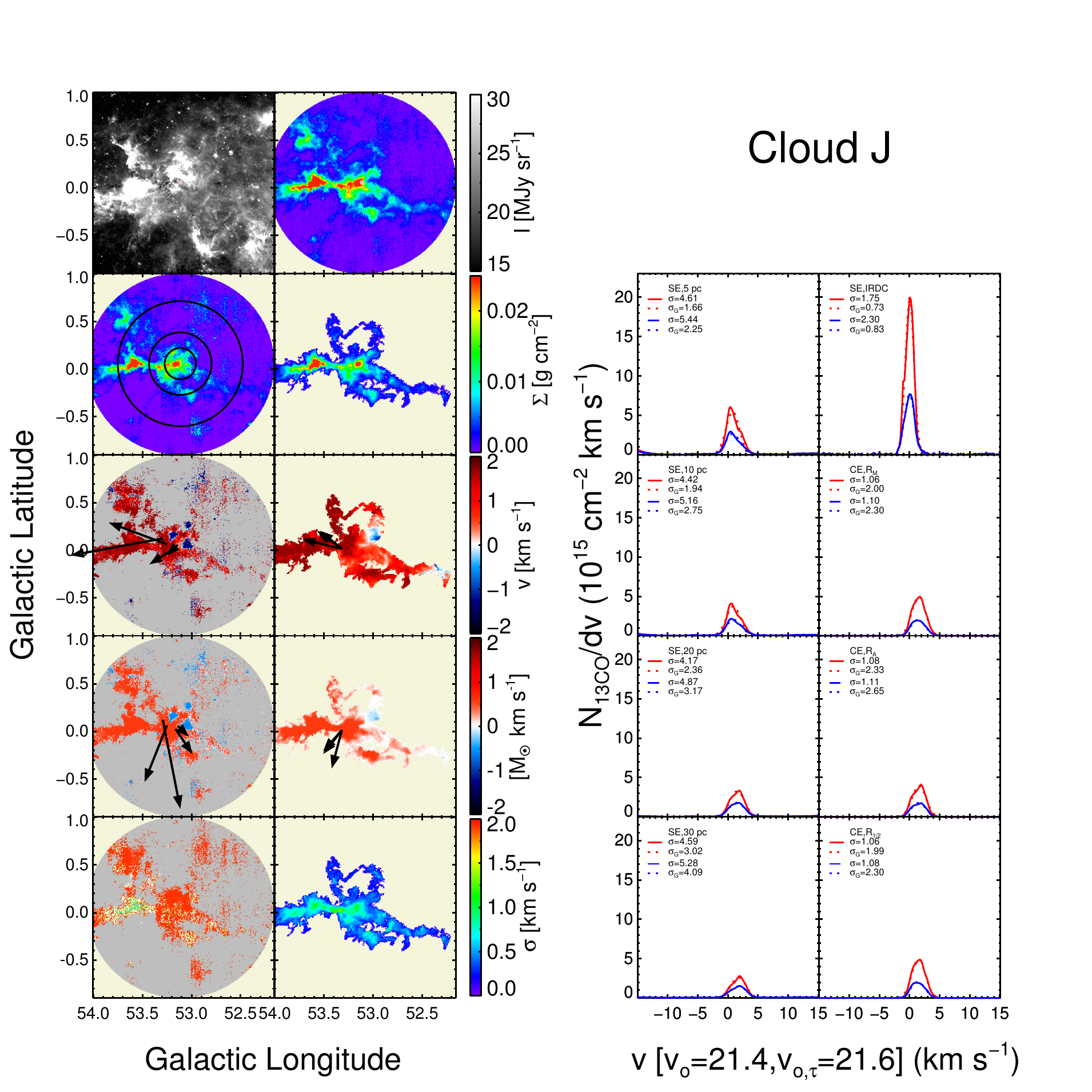}  
\end{array}$
\end{center}
\caption{
Same as Fig. \ref{mapA}, but for Cloud J.  The $\thco$(1-0) GRS
integrated intensity map of the IRDC emission profile is over the
velocity range: $v_{\rm LSR}=18.7-27.1~\kms$.}
\label{mapJ}
\end{figure}
\clearpage

\bibliography{irdc}

\end{document}